\documentclass[fleqn,usenatbib]{mnras}

\usepackage{etoolbox}
\makeatletter
\newcount\c@additionalboxlevel
\newcount\c@maxboxlevel
\patchcmd\@combinedblfloats{\box\@outputbox}{%
  \stepcounter{additionalboxlevel}%
  \box\@outputbox
}{}{\errmessage{\noexpand\@combinedblfloats could not be patched}}

\AtBeginShipout{%
  \ifnum\value{additionalboxlevel}>\value{maxboxlevel}%
    \typeout{Warning: maxboxlevel might be too small, increase to %
      \the\value{additionalboxlevel}%
    }%
  \fi 
  \@whilenum\value{additionalboxlevel}<\value{maxboxlevel}\do{%
    \typeout{* Additional boxing of page `\thepage'}%
    \setbox\AtBeginShipoutBox=\hbox{\copy\AtBeginShipoutBox}%
    \stepcounter{additionalboxlevel}%
  }%
  \setcounter{additionalboxlevel}{0}%
}
\makeatother

\usepackage[T1]{fontenc}
\usepackage{ae,aecompl}

\usepackage{caption}
\usepackage{tabularx}
\usepackage{rotating}
\usepackage{tablefootnote}
\usepackage{graphicx}	% Including figure files
\usepackage{amsmath}	% Advanced maths commands
\usepackage{amssymb}	% Extra maths symbols
\usepackage{bbm}
\usepackage{booktabs,supertabular}
\usepackage{longtable}

\usepackage{lscape}

\title[JINGLE -- IV. Dust, HI gas and metal scaling laws]{JINGLE -- IV. Dust, H{\sc{i}} gas and metal scaling laws in the local Universe}

\author[Ilse De Looze et al.]{I. De Looze$^{1,2,}$\thanks{E-mail: ilse.delooze@ugent.be, idelooze@star.ucl.ac.uk}\thanks{Institutional affiliations are shown at the end of the paper.},
I. Lamperti$^{2}$,
A. Saintonge$^{2}$,
M.~Rela{\~n}o$^{3,4}$,
M.~W.~L. Smith$^{5}$,
\newauthor C.~J.~R. Clark$^{6}$,
C.~D. Wilson$^{7}$, 
M.~Decleir$^{6}$,
A.~P. Jones$^{8}$,
R.~C.~Kennicutt$^{9,10}$,
\newauthor G. Accurso$^{2}$,
E. Brinks$^{11}$, 
M. Bureau$^{12,13}$, 
P. Cigan$^{5}$, 
D L. Clements$^{14}$, 
P.~De Vis$^{5}$, 
\newauthor L. Fanciullo$^{15}$,
Y. Gao$^{16,17}$,
W. K. Gear$^{18}$, 
L.~C.~Ho$^{19,20}$,  
H. S. Hwang$^{21}$,
\newauthor M.~J. Micha{\l}owski$^{22}$,
J. C. Lee$^{21}$, 
C. Li$^{23}$,
L. Lin$^{15}$, 
T. Liu$^{24}$, 
M. Lomaeva$^{2}$,
\newauthor H.-A. Pan$^{15,25}$,
M. Sargent$^{26}$,
T. Williams$^{25}$,
T. Xiao$^{15,27}$,
M. Zhu$^{28}$ }
\date{Accepted 2020 April 9. Received 2020 March 6; in original form 2020 January 10.}
\pubyear{2020}
\begin{document}
\label{firstpage}
\pagerange{\pageref{firstpage}--\pageref{lastpage}}
\maketitle

\begin{abstract}
Scaling laws of dust, H{\sc{i}} gas and metal mass with stellar mass, specific star formation rate and metallicity are crucial to our understanding of the buildup of galaxies through their enrichment with metals and dust. In this work, we analyse how the dust and metal content varies with specific gas mass ($M_{\text{HI}}$/$M_{\star}$) across a diverse sample of 423 nearby galaxies. The observed trends are interpreted with a set of Dust and Element evolUtion modelS (DEUS) -- including stellar dust production, grain growth, and dust destruction -- within a Bayesian framework to enable a rigorous search of the multi-dimensional parameter space. We find that these scaling laws for galaxies with $-1.0\lesssim\log M_{\text{HI}}/M_{\star}\lesssim0$ can be reproduced using closed-box models with high fractions (37-89$\%$) of supernova dust surviving a reverse shock, relatively low grain growth efficiencies ($\epsilon$=30-40), and long dust lifetimes (1-2\,Gyr). The models have present-day dust masses with similar contributions from stellar sources (50-80$\%$) and grain growth (20-50$\%$). Over the entire lifetime of these galaxies, the contribution from stardust ($>$90\,$\%$) outweighs the fraction of dust grown in the interstellar medium ($<$10\,$\%$). Our results provide an alternative for the chemical evolution models that require extremely low supernova dust production efficiencies and short grain growth timescales to reproduce local scaling laws, and could help solving the conundrum on whether or not grains can grow efficiently in the interstellar medium. \\ 
\end{abstract}
% compared to single temperature models which are biased to the luminous warm dust component. 


%%%%%%%%%%%%%%%%%%%%%%%%%%%%%%%%%%%%%%%%%%%%%%%%%%

%%%%%%%%%%%%%%%%% BODY OF PAPER %%%%%%%%%%%%%%%%%%

%\section{Introduction}
\section{Setting the scene}

Dust grains make up only a small fraction ($\sim$1$\%$ on average) of the interstellar mass in galaxies. Nonetheless, these dust particles play a crucial role in balancing local gas heating and cooling processes. Chemical reactions on the surfaces of dust grains result in the formation of a large variety of molecules, especially in regions of the interstellar medium (ISM) where gas phase chemistry is inefficient. The processing of about 30 to 50$\%$ of all stellar light in the Universe by dust grains (e.g., \citealt{2007MNRAS.379.1022D,2018A&A...620A.112B}) makes observations of the infrared (IR) dust emission furthermore essential for all studies of star formation to recover the bright ultraviolet (UV) and optical light emitted by young stellar populations. %nowadays and in the early days of the Universe. 

Although the ubiquitous presence of interstellar gas \citep{1904ApJ....19..268H} and dust \citep{1930PASP...42..214T} has been recognised for nearly a century, the origin and main formation channels for interstellar dust grains remain an open question. It is commonly accepted that dust grains can form through the condensation of metals in the cool envelopes of asymptotic giant branch (AGB) stars (e.g., \citealt{2006A&A...447..553F,2013MNRAS.434.2390N}) and in the expanding ejecta of core-collapse supernovae (e.g., \citealt{2010A&A...518L.138B,2012ApJ...760...96G,2015ApJ...800...50M,2017MNRAS.465.3309D,2017ApJ...836..129T,2019MNRAS.488..164D,2019ApJ...886...51C}), but these two stellar dust production sources appear not able to account for the bulk of the dust mass observed in galaxies at high redshift \citep{2010A&A...522A..15M,2011MNRAS.416.1916V,2014MNRAS.441.1040R,2015A&A...577A..80M,2015MNRAS.451L..70M,2019arXiv190907388G} and in the nearby Universe \citep{2013MNRAS.429.2527M,2016MNRAS.459.3900D,2016MNRAS.457.1842S,2018MNRAS.473.4538G,2020MNRAS.tmp..415T}. The reformation of dust grains through the accretion of metals in dense ISM clouds is thought to provide the key to explaining the large amounts of interstellar dust observed in galaxies (e.g., \citealt{2014MNRAS.441.1040R,2014A&A...562A..76Z,2016MNRAS.457.1842S,2016ApJ...831..147Z,2017MNRAS.471.1743D,2017MNRAS.471.3152P}), but the exact physical processes that enable this type of ``grain growth" in the interstellar medium remain poorly understood \citep{1978MNRAS.183..417B,2016MNRAS.463L.112F,2018MNRAS.476.1371C}. 

To better understand the main dust formation mechanisms in galaxies, and whether or not grain growth can dominate the dust production, we require substantial progress on two independent fronts. First of all, we need reliable estimates of the dust content in galaxies. In this work, we rely on a set of carefully determined dust masses (see Appendix \ref{DustMasses.sec}) inferred from fitting the mid-infrared to sub-millimetre dust spectral energy distribution (SED) with a Bayesian method that builds upon the grain mix and dust properties from the THEMIS dust model \citep{2017A&A...602A..46J} and a multi-component interstellar radiation field heating these dust grains \citep{2001ApJ...549..215D}. Secondly, we require measurements of how the dust, metal and gas content in galaxies scales with respect to other global galaxy properties (i.e., stellar mass, specific star formation rate, metallicity) through scaling relations to infer how a galaxy's dust content evolves with time and to shed light on the main sources of dust production in the ISM. Understanding how the amount of dust, metals and gas evolves for a large ensemble of galaxies, at different stages of their evolution, will allow us to pin down the importance of various dust production and destruction mechanisms. Tracking how metals and dust are built up throughout a galaxy's lifetime necessitates simultaneously quantifying dust and gas reservoirs. The JINGLE (JCMT dust and gas In Nearby Galaxies Legacy Exploration) galaxy sample (\citealt{2018MNRAS.481.3497S}, hereafter JINGLE Paper I) was designed to acquire dust mass measurements from \textit{Herschel} and SCUBA-2 data, in addition to ancillary H{\sc{i}} observations, and molecular gas mass measurements currently available for 63 JINGLE galaxies.

In this paper, we present dust, gas and metal scaling relations for a sample of 423 nearby galaxies, including JINGLE, HRS, HAPLESS, HiGH and KINGFISH samples\footnote{The combined galaxy sample consists of 568 galaxies. We consider the subsample of those galaxies: (1.) with available H{\sc{i}} gas measurements, and (2.) classified as non-H{\sc{i}}-deficient galaxies.}. We split up this local galaxy sample into six subsamples according to their stage of evolution. We assume in this paper that the evolutionary stage of a galaxy is relatively well approximated by their $M_{\text{HI}}$/$M_{\star}$ ratios and infer representative star formation histories according to the evolutionary stage of these galaxies. We compare the average dust, gas and metal mass fractions along these evolutionary sequences with a set of Dust and Element evolUtion modelS (DEUS) in a Bayesian framework in order to cover a large range of input parameters and to elucidate what processes drive these scaling laws. This is the first study (to our knowledge) where such a rigorous search of the full parameter space has been pursued.

Section \ref{Samples.sec} discusses the main characteristics of our five nearby galaxy samples (JINGLE, HRS, KINGFISH, HAPLESS, HIGH). In Section \ref{ScalingRelations.sec}, we analyse the observed scaling laws for the dust, gas and metal content of these five galaxy samples. In Section \ref{EvolutionModels.sec}, we subdivide our local galaxy sample into six bins according to their specific H{\sc{i}} gas masses, and compare their average scaling laws with DEUS to infer how their dust and metal content has been built up across cosmic time. In Section \ref{Conclusions.sec}, we summarise our conclusions. In the Appendices, we outline the method used to model the dust masses (Appendix \ref{DustMasses.sec}), detail the datasets and methods used to infer galaxy specific properties (Appendix \ref{OtherSamples.sec}), describe how we infer customised star formation histories (SFH) for galaxies at different evolutionary stages (Appendix \ref{Sec_SFH}), discuss the specifics of DEUS (Appendix \ref{DEUS.sec}), while a list of acronyms and symbols is presented in Appendix \ref{Acronyms.sec}, and additional Tables and Figures are presented in Appendices \ref{Tables.sec} and \ref{Figures.sec}. %In Section \ref{DustTracer.sec}, we discuss the applicability of dust mass as a tracer of the interstellar gas reservoir. 

\begin{figure*}
	\includegraphics[width=17.5cm]{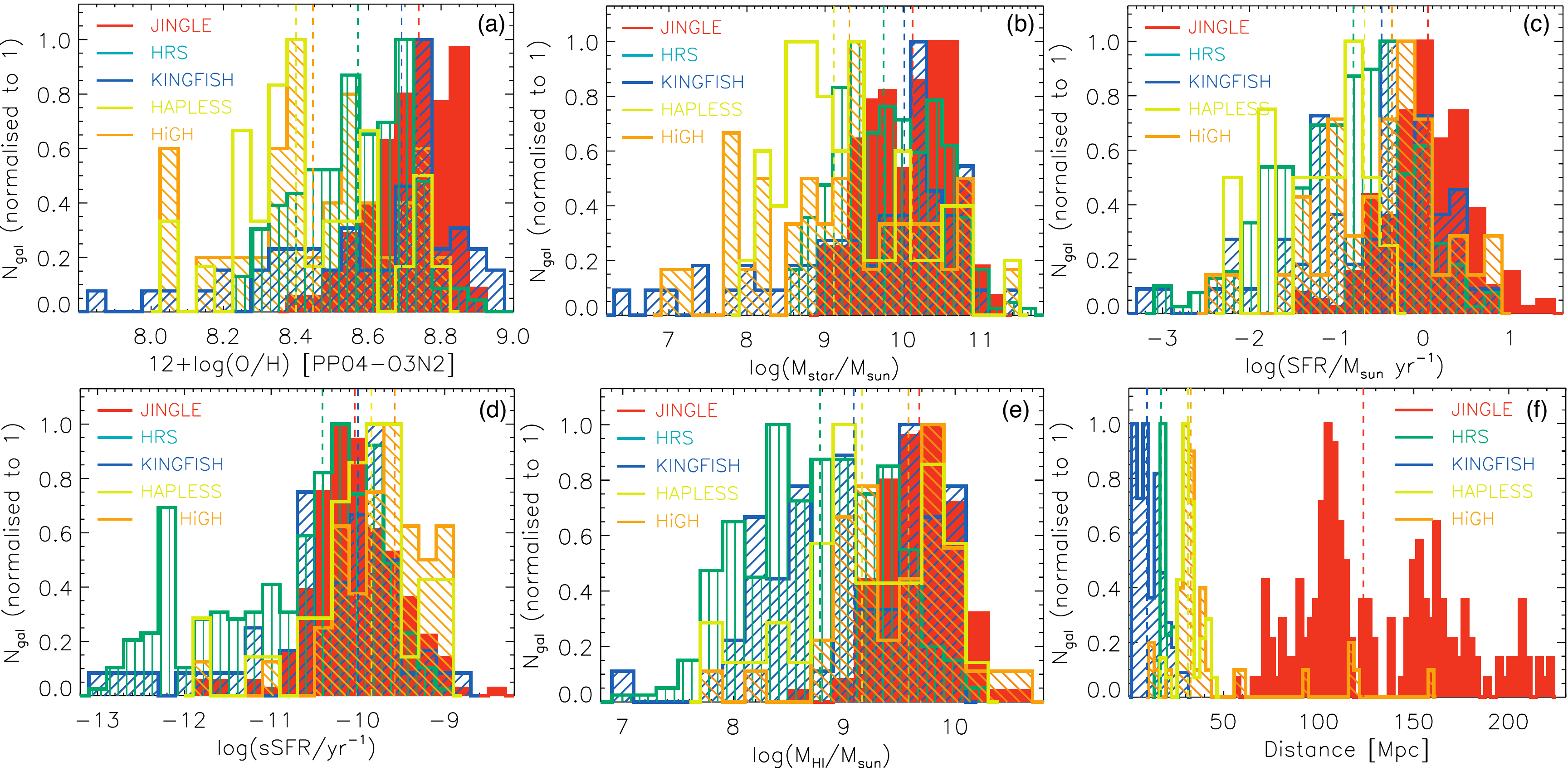} 
    \caption{From left to right, and top to bottom: histograms of the metallicities (as traced by the oxygen abundance), stellar masses, star formation rates (SFRs), specific star formation rates (sSFRs), H{\sc{i}} masses ($M_{\text{HI}}$) and distances for the JINGLE (red filled histograms), HRS (green vertical lines), KINGFISH (blue diagonal lines), HAPLESS (yellow lines) and HiGH (orange diagonal lines) galaxies. Median sample values are indicated with vertical dashed lines using the same colour-coding.}
    \label{Compare_nearbysamples}
\end{figure*}

\section{Sample description}
\label{Samples.sec}
\subsection{An introduction to JINGLE}
\label{JINGLE.sec}
JINGLE is a large program on the James Clerk Maxwell Telescope (JCMT) aiming to assemble dust mass measurements for a sample of 193 local galaxies and molecular gas masses for part of this sample. The JINGLE sample populates the redshift range between z=0.01 and z=0.05, and was drawn from the MaNGA (Mapping Nearby Galaxies at Apache Point Observatory, \citealt{2015ApJ...798....7B}) sample with optical integral-field spectroscopy data. In brief, JINGLE galaxies were selected to homogeneously sample the SFR--$M_{\star}$ plane between 10$^{9}$ and 10$^{11}$\,M$_{\sun}$.  As part of the sample selection procedure, JINGLE galaxies were required to have detections in the \textit{Herschel} SPIRE\,250 and 350\,$\mu$m bands. New JCMT SCUBA-2 850\,$\mu$m (and 450\,$\mu$m) observations probe the dust emission spectrum along the Rayleigh-Jeans tail (\citealt{2019MNRAS.486.4166S}, hereafter JINGLE Paper II), while RxA CO J=2-1 observations provide measurements of the molecular gas content (currently) for 63 JINGLE galaxies (Xiao et al.\,in prep., hereafter JINGLE Paper III). As a consequence of the sample selection, most JINGLE galaxies are classified as late-type spirals or irregular galaxies with a subset of only 7 early-type galaxies.%, and to show $\gtrsim$3$\sigma$ detections in the \textit{Herschel} SPIRE\,250 and 350\,$\mu$m wavebands. 

The sample selection and main science goals of the JINGLE survey are described in JINGLE Paper I, with specific details about the observational setup and data reduction of the RxA CO J=2-1 line spectroscopy and SCUBA-2 450 and 850\,$\mu$m dust continuum observations presented in JINGLE Papers III and II, respectively. In \citet{2019MNRAS.489.4389L} (hereafter JINGLE Paper V), a hierarchical Bayesian fitting algorithm has been used to infer dust temperatures, dust emissivity indices, and dust masses for the ensemble of JINGLE (and HRS) galaxies. In this paper, we rely on the dust masses for JINGLE and the other nearby galaxies inferred from an alternative modelling method using a non-hierarchical Bayesian implementation of the THEMIS dust model, that enables us to constrain the small grain size distribution, dust masses and starlight intensity distribution responsible for the dust heating (see Appendix \ref{DustMasses.sec}). We note that the dust masses inferred here and in JINGLE Papers V are in excellent agreement after considering the differences in the assumed dust mass absorption coefficients: JINGLE Paper V assumes $\kappa_{\text{500}}$=0.051\,m$^{2}$\,kg$^{-1}$ \citep{2016MNRAS.459.1646C}, while here we adopt $\kappa_{\text{500}}$=0.185\,m$^{2}$\,kg$^{-1}$ from the THEMIS dust model \citep{2013A&A...558A..62J,2017A&A...602A..46J}. Due to growing evidence (both from observations and laboratory experiments) indicating that interstellar dust is more emissive than considered in the previous generation of dust models (e.g., \citealt{2016A&A...586A.132P,2017A&A...600A.123D,2017A&A...606A..50D,2019MNRAS.489.5256C}), we base our analysis upon the dust masses inferred with the THEMIS dust model to account for this increased dust emissivity and to avoid overestimating the dust masses for a set of observed flux densities (compared to the previous generation of dust models).

\subsection{Nearby galaxy comparison samples}
\label{Nearby.sec}
%To compare JINGLE dust properties and their dust content to the average nearby galaxy population
In addition to JINGLE, we have selected four nearby galaxy samples with well-studied dust characteristics and general galaxy properties. The combination of samples, whilst not statistical, allows the scaling relations in this paper to be explored over the widest possible extent of the parameter spaces in question.

The first sample consists of the galaxies from the \textit{Herschel} Reference Survey (HRS, \citealt{2010PASP..122..261B}) which is a volume-limited, K-band selected sample of 322 nearby galaxies with distances between 15 and 25\,Mpc. More than half of the HRS sample consists of cluster galaxies (resi\-ding in the Virgo and Ursa Major cluster), with the remaining galaxies located in massive groups surrounding these clusters. The second sample is composed of galaxies from the \textit{Herschel} program KINGFISH (Key Insights on Nearby Galaxies: A Far-Infrared Survey with Herschel, \citealt{2011PASP..123.1347K}) which consists of 61 nearby galaxies with distances $D\leq30$\,Mpc, covering a variety of different morphological classifications, star formation activity and galaxy environments. The third and fourth sample, HAPLESS and HiGH, were selected from the \textit{Herschel} Astrophysical Terahertz Large Area Survey (H-ATLAS, \citealt{2010PASP..122..499E}) based on their SPIRE\,250\,$\mu$m (HAPLESS, \citealt{2015MNRAS.452..397C}) and H{\sc{i}} (HiGH, \citealt{2017MNRAS.464.4680D}) detections, respectively. Since dusty galaxies often contain a considerable amount of gas, and vice versa, it is not surprising that the HAPLESS (42 galaxies) and HiGH (40 galaxies) samples have 22 sources in common. Average sample properties are summarised in Table \ref{AverageProperties}, and are briefly discussed in Section \ref{Characteristics.sec}. To compare properties of different galaxy samples, we have performed Mann--Whitney U--tests using the \texttt{RS}$\_$\texttt{TEST} procedure in IDL (see Table \ref{Table_MannWhitneyU} for the test results). This procedure tests the hypothesis that two samples have the same median of distribution at a significance level of 5$\%$, with probabilities higher than this value indicative of both samples not being significantly different.

\begin{table*}
%\centering
\caption{Overview of the median values for a set of galaxy properties, with the error bars reflecting the dispersion observed for galaxies within a specific galaxy sample. For the HRS sample, we report the sample characteristics for the entire set of HRS galaxies, and the subsamples of H{\sc{i}}-deficient and non-deficient (H{\sc{i}}$_{\text{def}}\leq$0.5) HRS galaxies.}
\label{AverageProperties}
\begin{tabular}{lccccccc} % four columns, alignment for each
\hline
Quantity & JINGLE & HRS (all) & HRS (H{\sc{i}} def$<$0.5) & HRS (H{\sc{i}} def$\ge$0.5) & KINGFISH & HAPLESS & HIGH \\ 
\hline
 12+$\log$(O/H) & 8.74$\pm$0.10 & 8.58$\pm$0.15 & 8.57$\pm$0.15 & 8.64$\pm$0.13 & 8.69$\pm$0.22 & 8.44$\pm$0.17 & 8.50$\pm$0.20 \\
$\log$\,$M_{\star}$ [M$_{\odot}$] & 10.13$\pm$0.55 & 9.67$\pm$0.63 & 9.53$\pm$0.59 & 9.95$\pm$0.64 & 9.95$\pm$0.98 & 9.06$\pm$0.64 & 9.39$\pm$0.86 \\
$\log$\,SFR [M$_{\odot}$ yr$^{-1}$] & 0.052$\pm$0.48 & -0.70$\pm$0.67 & -0.47$\pm$0.56 & -1.18$\pm$0.65 & -0.48$\pm$0.85 & -0.83$\pm$0.31 & -0.24$\pm$0.54 \\
$\log$\,sSFR [yr$^{-1}$] & -10.03$\pm$0.49 & -10.30$\pm$0.80 & -10.08$\pm$0.51 & -10.98$\pm$0.82 & -10.0$\pm$0.63 & -9.92$\pm$0.68 & -9.72$\pm$0.50 \\
$\log$\,$M_{\text{HI}}$ [M$_{\odot}$] & 9.66$\pm$0.39 & 8.92$\pm$0.60 & 9.20$\pm$0.44 & 8.37$\pm$0.45 & 9.08$\pm$0.71 & 8.90$\pm$0.52 & 9.74$\pm$0.48 \\
D [Mpc] & 123.4$\pm$41.6 & 17.0$\pm$1.2 & 17.3$\pm$2.8 & 17.0$\pm$2.4 & 9.8$\pm$6.8 & 31.1$\pm$5.3 & 32.4$\pm$5.0 \\
\hline
$\log$\,$M_{\text{dust}}$/$M_{\star}$ & -2.71$\pm$0.36 & -2.90$\pm$0.43 & -2.76$\pm$0.29 & -3.19$\pm$0.55 & -2.86$\pm$0.48 & -2.82$\pm$0.46 & -2.78$\pm$0.44 \\
$\log$\,$M_{\text{HI}}$/$M_{\star}$ & -0.43$\pm$0.48 & -0.76$\pm$0.75 & -0.50$\pm$0.50 & -1.49$\pm$0.66 & -0.60$\pm$0.91 & -0.35$\pm$0.69 & 0.02$\pm$0.61 \\
$\log$\,$M_{\text{dust}}$/$M_{\text{metals}}$ & -0.67$\pm$0.23 & -0.60$\pm$0.21 & -0.62$\pm$0.21 & -0.44$\pm$0.08 & -0.63$\pm$0.38 & -0.65$\pm$0.15 & -0.78$\pm$0.30 \\
$\log$\,$M_{\text{dust}}$/$M_{\text{HI}}$ & -2.25$\pm$0.31 & -2.17$\pm$0.47 & -2.28$\pm$0.35 & -1.80$\pm$0.44 & -2.30$\pm$0.69 & -2.59$\pm$0.23 & -2.61$\pm$0.45 \\
\hline 
\end{tabular}
\end{table*}

\subsection{Sample characteristics}
\label{Characteristics.sec}
JINGLE and KINGFISH galaxies are more metal-rich as compared to other nearby galaxy samples (see Fig.\,\ref{Compare_nearbysamples}a), whereas the oxygen abundance distributions for HRS, HAPLESS and HiGH samples are not considered to be significantly different. The JINGLE sample has a relatively flat stellar mass distribution (which was by selection, see \citealt{2018MNRAS.481.3497S}) with values ranging from 10$^{9}$ to 10$^{11}$ M$_{\odot}$ (see Fig.\,\ref{Compare_nearbysamples}b), significantly different from the other four nearby galaxy samples. The HRS, KINGFISH and HiGH samples extend towards low stellar masses with several galaxies in the 10$^{6}$-10$^{9}$\,M$_{\odot}$ stellar mass range. HAPLESS does not contain galaxies with stellar masses below 10$^{8}$\,M$_{\odot}$, nor does it contain many $M_{\star}\,>\,10^{10}$\,M$_{\odot}$ galaxies like JINGLE. Based on the mass-metallicity relation (e.g., \citealt{2004ApJ...613..898T,2013A&A...550A.115H,2017MNRAS.469.2121S}), it is thus not surprising that JINGLE galaxies are characterised by the highest metal abundances among our local galaxy sample.

The median star formation rate (SFR) of JINGLE galaxies (1\,M$_{\odot}$ yr$^{-1}$, see Fig.\,\ref{Compare_nearbysamples}c) is similar to the average present-day star formation activity in our own Galaxy \citep{2010ApJ...710L..11R}. SFRs are a factor of three lower in KINGFISH and HiGH galaxies, and lower by a factor of six in HRS and HAPLESS galaxies, than for JINGLE galaxies. The low SFRs and specific star formation rates (sSFRs) imply that the majority of HRS galaxies are undergoing a period of low star formation activity, and have built up the majority of their stellar mass content during earlier epochs. The subsample of more evolved HRS galaxies is also evident from the long tail in the sSFR diagram at the low sSFR end (see Fig.\,\ref{Compare_nearbysamples}d). Although HiGH, KINGFISH and HAPLESS galaxies have a median SFR two, three and eight times lower than JINGLE, respectively, the similarity in their median sSFRs suggests that these samples contain several galaxies with elevated levels of recent star formation activity. 

The H{\sc{i}} mass content of JINGLE galaxies is similar to the median H{\sc{i}} reservoirs present in the H{\sc{i}}-selected HiGH sample, but the specific H{\sc{i}} gas mass of HiGH galaxies ($\log$\,$M_{\text{HI}}$/$M_{\star}$=0.02$\pm$0.61) is higher than for JINGLE ($\log$\,$M_{\text{HI}}$/$M_{\star}$=-0.43$\pm$0.48). HiGH galaxies are therefore considered to be in a very early stage of galaxy evolution \citep{2017MNRAS.464.4680D}. Nonetheless, JINGLE H{\sc{i}} masses are clearly higher than those of KINGFISH, HAPLESS and HRS galaxies, suggesting that JINGLE galaxies have retained a non-negligible part of their H{\sc{i}} reservoir for future star formation, and are also at an earlier stage of galaxy evolution. It is worth noting that the spatial extent of the H{\sc{i}} reservoir has not been taken into consideration in the comparison of these H{\sc{i}} masses (due to the availability of single-dish measurements only), and that, in particular, low-mass metal-poor galaxies can have a large H{\sc{i}} reservoir that extends well beyond the stellar body (e.g., \citealt{2011AJ....142..173H}). HRS galaxies have a median H{\sc{i}} mass almost an order of magnitude below the median for JINGLE, which supports the interpretation of the HRS sample consisting of more evolved galaxies. A subset of the HRS galaxies have been characterised to be H{\sc{i}}-deficient\footnote{The H{\sc{i}}-deficiency is calculated as the logarithmic difference between the expected and observed H{\sc{i}} mass, i.e. H{\sc{i}} def = $\log\,M_{\text{HI,ref}}$\,-\,$\log\,M_{\text{HI,obs}}$, following the definition in \citet{1984AJ.....89..758H}.}, and their reduced star formation activity has been attributed to the removal of part of their H{\sc{i}} gas reservoir due to environmental processes that inhibit new stars from forming (e.g., \citealt{2011MNRAS.415.1797C}). 

The median distance of JINGLE galaxies ($D$=123.4\,Mpc) is higher than the median distances ($D$=10-30\,Mpc) for the other samples, which will likely bias the JINGLE sample selection to only include the dustiest galaxies at those distances.

Similar to the low fraction of early-type galaxies in the JINGLE sample (i.e., 3.6$\%$), the HAPLESS and HiGH samples consist of late-type star-forming galaxies with a range of different morphologies (ranging from early-type spirals to bulgeless highly flocculent galaxies), with the exception of 2 early-type HAPLESS galaxies. The KINGFISH sample contains 10 early-type galaxies (E/S0/S0a), 22 early-type spirals (Sa/Sb/Sbc), 16 late-type spirals (Sc/Sd/Scd) and 13 irregular galaxies (I/Sm). The HRS sample contains a significant subpopulation of 23 elliptical and 39 spheroidal galaxies \citep{2012ApJ...748..123S}, with the remaining 261 galaxies classified as late-type galaxies.

\section{Dust, gas and metal scaling laws}
\label{ScalingRelations.sec}
The main goal of this part of the paper is to analyse local dust, H{\sc{i}} gas and metal scaling laws, to understand how the dust content and metallicity evolves over time, and what processes drive this evolution.   %, and evolutionary stage of JINGLE galaxies compares with other local galaxy samples. 

\subsection{Dust scaling relations}
%\label{ScalingRelations.sec}

\begin{figure*}
	\includegraphics[width=8.7cm]{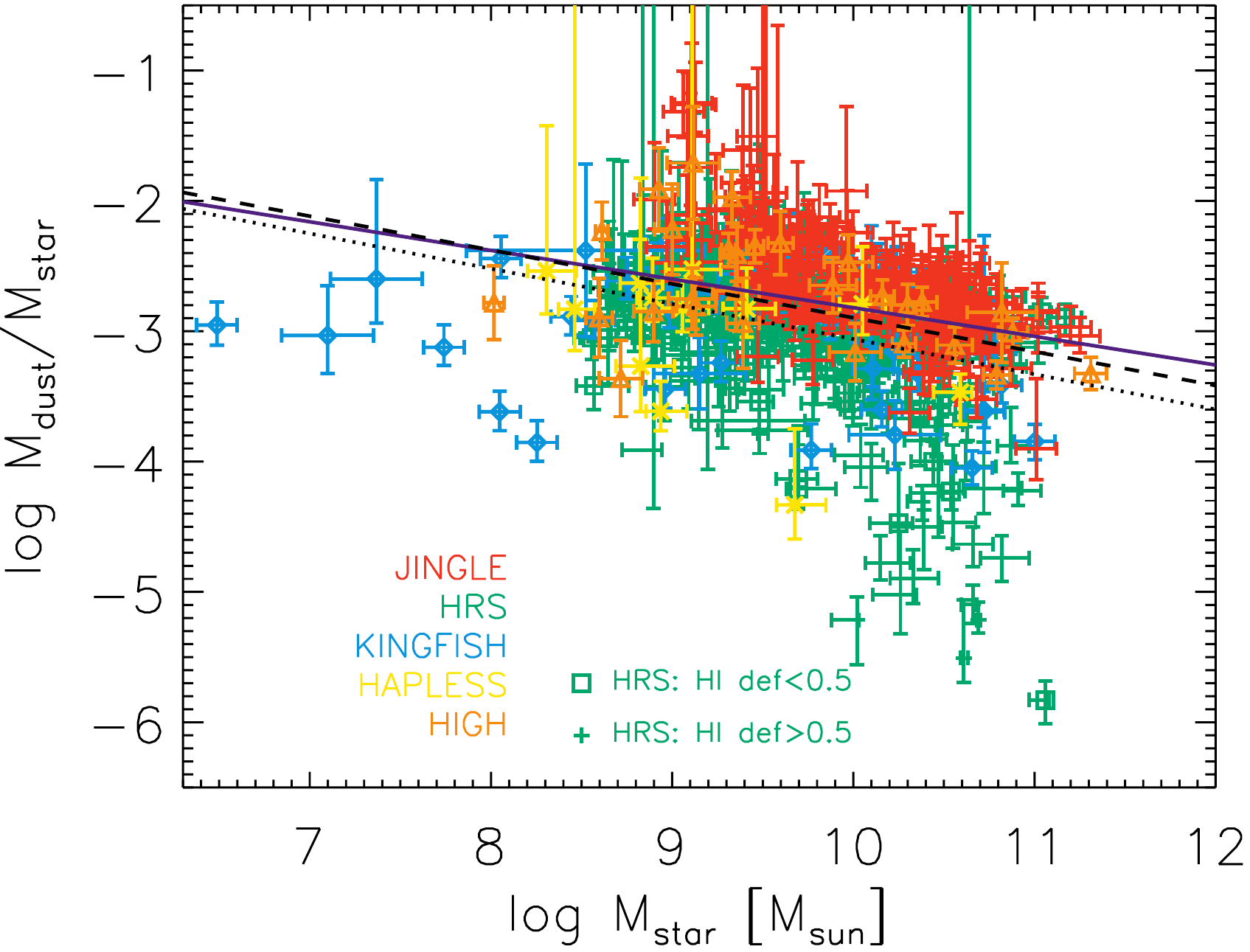}
	\includegraphics[width=8.7cm]{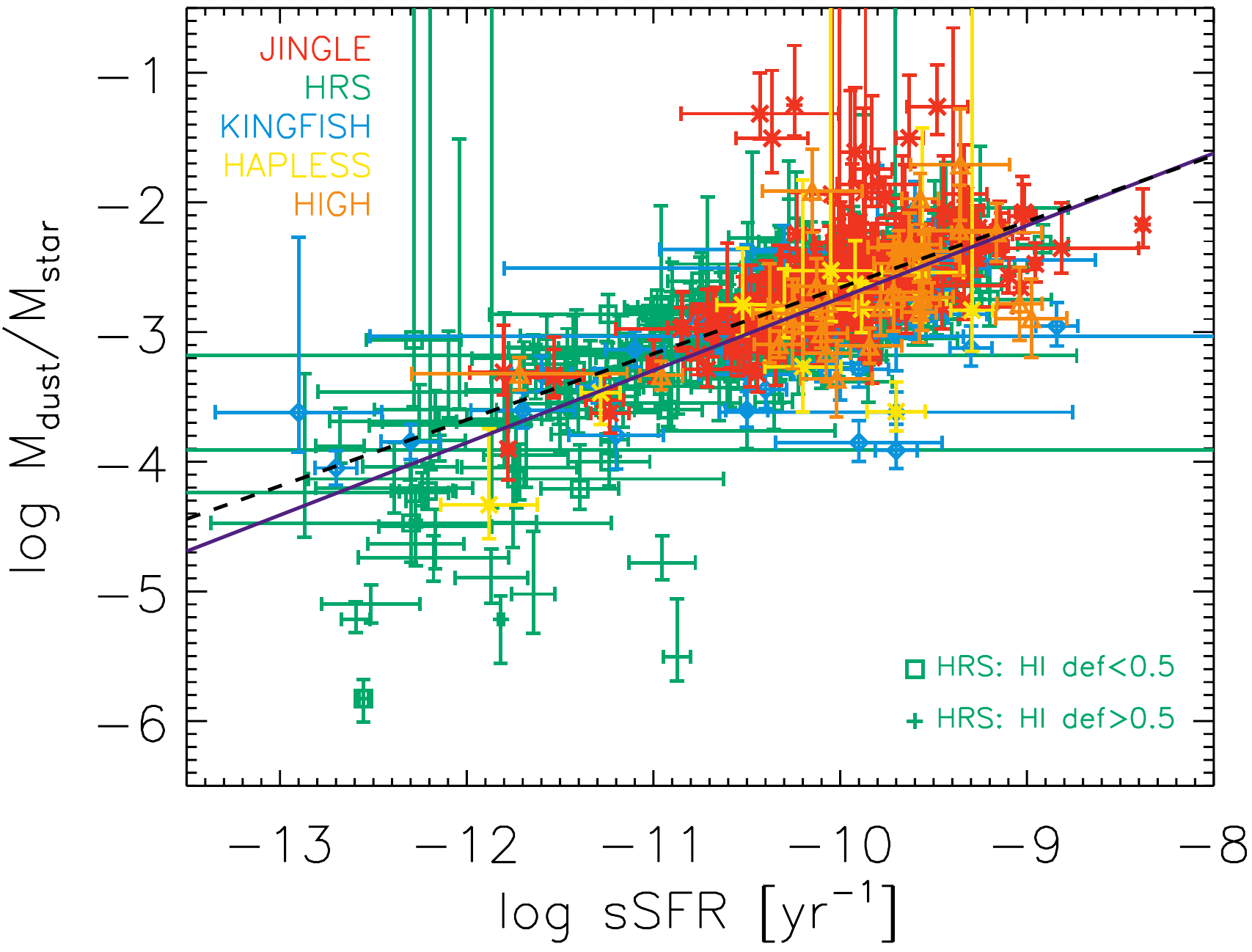} 
    \caption{The scaling of the dust-to-stellar mass ratio (i.e., $M_{\text{dust}}$/$M_{\star}$) with stellar mass ($M_{\star}$, left panel) and specific star formation rate (sSFR with sSFR=SFR/$M_{\star}$, right panel) is shown for JINGLE (red cross), HRS (green square/cross), KINGFISH (blue diamond), HAPLESS (yellow cross) and HiGH (orange triangle) galaxies. A distinction is made between H{\sc{i}}-deficient and non-deficient HRS galaxies (H\,{\sc{i}}$_{\text{def}}\leq$0.5, green square). Best-fit relations (as inferred for the entire nearby galaxy sample, with the exception of H{\sc{i}}-deficient HRS galaxies) have been overlaid as a purple solid line, and are compared (where possible) to local galaxy scaling laws from \citet{2017MNRAS.464.4680D} (black dashed curve) and from \citet{2019arXiv191109187C} (black dotted curve). The scaling relation from \citet{2017MNRAS.464.4680D} was adjusted to account for the difference in the assumed dust opacities.}     \label{DustScaling}  %Upper (5$\sigma$) limits have been indicated using downward arrows.
\end{figure*}

With dust being formed through the condensation of metals synthesised in recent generations of stars, the dust content is closely linked to the stellar mass and star formation activity in galaxies. Since the stellar mass typically scales with the metal richness of the interstellar medium (through the stellar mass-metallicity relation, e.g., \citealt{2004ApJ...613..898T}), the dust-to-stellar mass ratio can be interpreted as the ratio of metals locked into dust grains versus the metals in the gas phase. It is known that specific dust masses ($M_{\text{dust}}$/$M_{\star}$) decrease towards high stellar masses (see Fig. \ref{DustScaling}, left panel) due to dust destruction dominating over dust production processes in more massive systems. The latter trend can also be understood in view of the downsizing of galaxies (e.g., \citealt{1996AJ....112..839C}), where most of the massive galaxies already converted most of their gas into stars, and the bulk of dust mass was formed during these main star formation episodes. The wide spread in $\log\,M_{\text{dust}}$/$M_{\star}$ ratio from -2.5 to -5 that we find for galaxies with stellar masses $M_{\star}$=10$^{10}$-10$^{11}$\,M$_{\odot}$ is a reflection of galaxies with similar stellar masses but at different stages of evolution. 

JINGLE galaxies populate the high end of the $M_{\text{dust}}$/$M_{\star}$ range at a given stellar mass. Their high $\log\,M_{\text{dust}}$/$M_{\star}$ ratios (-2.71$\pm$0.36) are not surprising considering that JINGLE galaxies were selected from their detections in the \textit{Herschel} SPIRE bands \citep{2018MNRAS.481.3497S}. The JINGLE galaxies have $M_{\text{dust}}$/$M_{\star}$ ratios similar to (or even slightly higher than) the majority of dust- and H{\sc{i}}-selected HAPLESS/HiGH galaxies in the stellar mass range that those samples have in common. Several KINGFISH and HAPLESS/HiGH galaxies with stellar masses $M_{\star}\,\leq$10$^{9}$\,M$_{\sun}$ are characterised by low $M_{\text{dust}}$/$M_{\star}$ ratios, and deviate from the general trend for more massive galaxies. The HAPLESS/HiGH galaxies with low specific dust masses were identified by \citet{2017MNRAS.464.4680D} as a unique population of galaxies, at an extremely early phase of evolution where most of the dust still needs to be formed. H\,{\sc{i}}-deficient HRS galaxies populate the bottom part of the diagram with systematically lower $M_{\text{dust}}$/$M_{\star}$ ratios in comparison to other nearby galaxies. The lower $M_{\text{dust}}$/$M_{\star}$ for H\,{\sc{i}}-deficient galaxies suggests that these galaxies have had part of their dust content stripped along with their H\,{\sc{i}} gas content (see also \citealt{2014MNRAS.440..942C}), or that star formation has ceased in these objects a long time ago, resulting in a lack of recent dust replenishment, with dust destruction processes further diminishing their dust content. Our best-fit relation is very similar compared to the best-fit relation from \citet{2017MNRAS.464.4680D} (inferred for HRS, HAPLESS and HiGH late-type galaxies). The relation inferred by \citet{2019arXiv191109187C} for a sample of 436 late-type local DustPedia galaxies is lower by up to 0.2\,dex, which can likely be attributed to a selection effect. Our sample includes dust-selected galaxies at larger distances (see Fig. \ref{Compare_nearbysamples}f), which are likely to be more dusty on average compared to a local galaxy sample.
 
The importance of recent star formation activity to determine a galaxy's dust content is evidently shown from the scaling of $M_{\text{dust}}$/$M_{\star}$ with sSFR (see Fig. \ref{DustScaling}, right panel). Independent of their morphological classification, all galaxies follow a similar trend of decreasing $M_{\text{dust}}$/$M_{\star}$ towards low sSFR over three orders of magnitude in both quantities. The tight correlation ($\rho$=$0.63$) between $M_{\text{dust}}$/$M_{\star}$ and sSFR was first shown by \citet{2010MNRAS.403.1894D} for a sample of nearby galaxies. The fact that dust-selected samples such as JINGLE and HAPLESS follow the same trend as the stellar-mass selected HRS sample indicates that sSFR is a more fundamental parameter than $M_{\star}$ to determine the specific dust mass of a galaxy (either directly or through a secondary correlation).

The present-day dust mass of a galaxy is set by the balance between the sources producing dust (i.e., evolved stars, supernovae, grain growth) and the sinks destroying dust grains (i.e., astration, supernova shocks). The observed correlation between $M_{\text{dust}}$/$M_{\star}$ and sSFR could be a reflection of an equilibrium process where the amount of dust grains formed/destroyed scales with the recent star formation activity in a galaxy. Alternatively, the relation of the $M_{\text{dust}}$/$M_{\star}$ with sSFR can be interpreted as an indirect measure of the total gas mass in a galaxy which is known to scale with the star formation rate through the Kennicutt-Schmidt relation \citep{1959ApJ...129..243S,1998ApJ...498..541K}. Given that the $M_{\text{HI}}$/$M_{\star}$ ratio dominates the scatter in local scaling relations with $M_{\star}$ and sSFR (see also Section \ref{HIrelations.sec}), and correlates strongly with the observed $M_{\text{dust}}$/$M_{\star}$, $M_{\text{dust}}$/$M_{\text{metals}}$ and $M_{\text{dust}}$/$M_{\text{HI}}$ ratios in our local galaxy samples, we favour the latter interpretation (see below). % for the observed trend between $M_{\text{dust}}$/$M_{\star}$ and sSFR.}

\subsection{H{\sc{i}} gas scaling relations}
\label{HIrelations.sec}

\begin{figure*}
	\includegraphics[width=8.7cm]{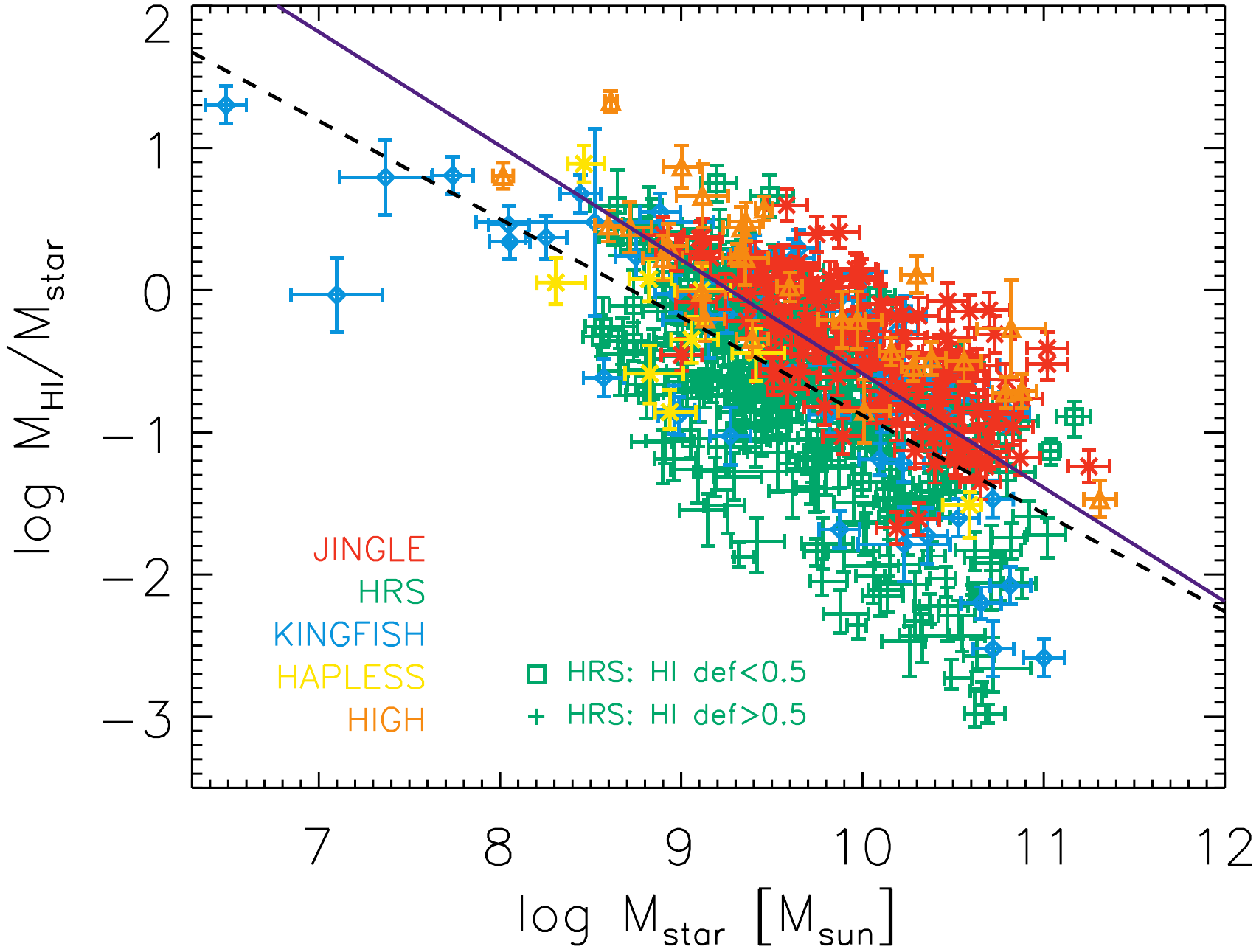}
	\includegraphics[width=8.7cm]{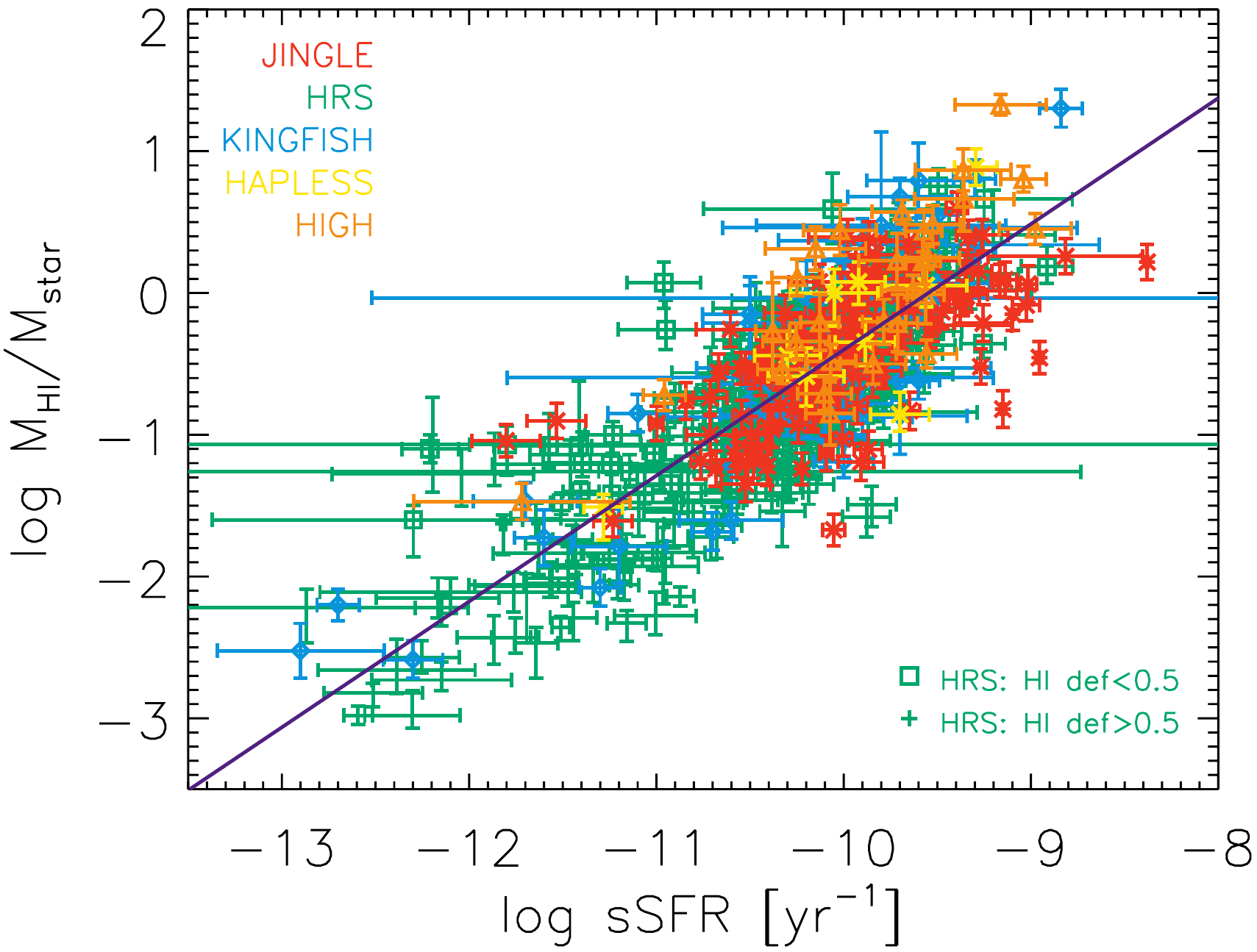} 
    \caption{The scaling of the H{\sc{i}}-to-stellar mass ratio (i.e., $M_{\text{HI}}$/$M_{\star}$) with stellar mass ($M_{\star}$, left panel) and specific star formation rate (sSFR with sSFR=SFR/$M_{\star}$, right panel). See caption of Fig. \ref{DustScaling} for more details on the symbols and plotted curves.}
    \label{HIScaling}
\end{figure*}

%of hydrogen (either in atomic or molecular form) and a small fraction of heavier elements
With dust grains making up about 1$\%$ of the ISM in mass, the gas reservoir dominates the ISM budget of a galaxy. In this paper, we will make the assumption that galaxies with massive H{\sc{i}} reservoirs (compared to their $M_{\star}$) are considered to be at an early stage of evolution, while a low gas content is indicative of an evolved galaxy which had most of its gas reservoir turned into stars already. The $M_{\text{HI}}$/$M_{\star}$ ratio in Fig. \ref{HIScaling} (left panel) shows a similar anti-correlation ($\rho$=$-0.64$) with stellar mass as the $M_{\text{dust}}$/$M_{\star}$ ratio in Fig. \ref{DustScaling} (left panel) which is consistent with the least massive galaxies having the largest atomic gas reservoir proportional to their stellar mass. 

\begin{figure*}
	\includegraphics[width=8.7cm]{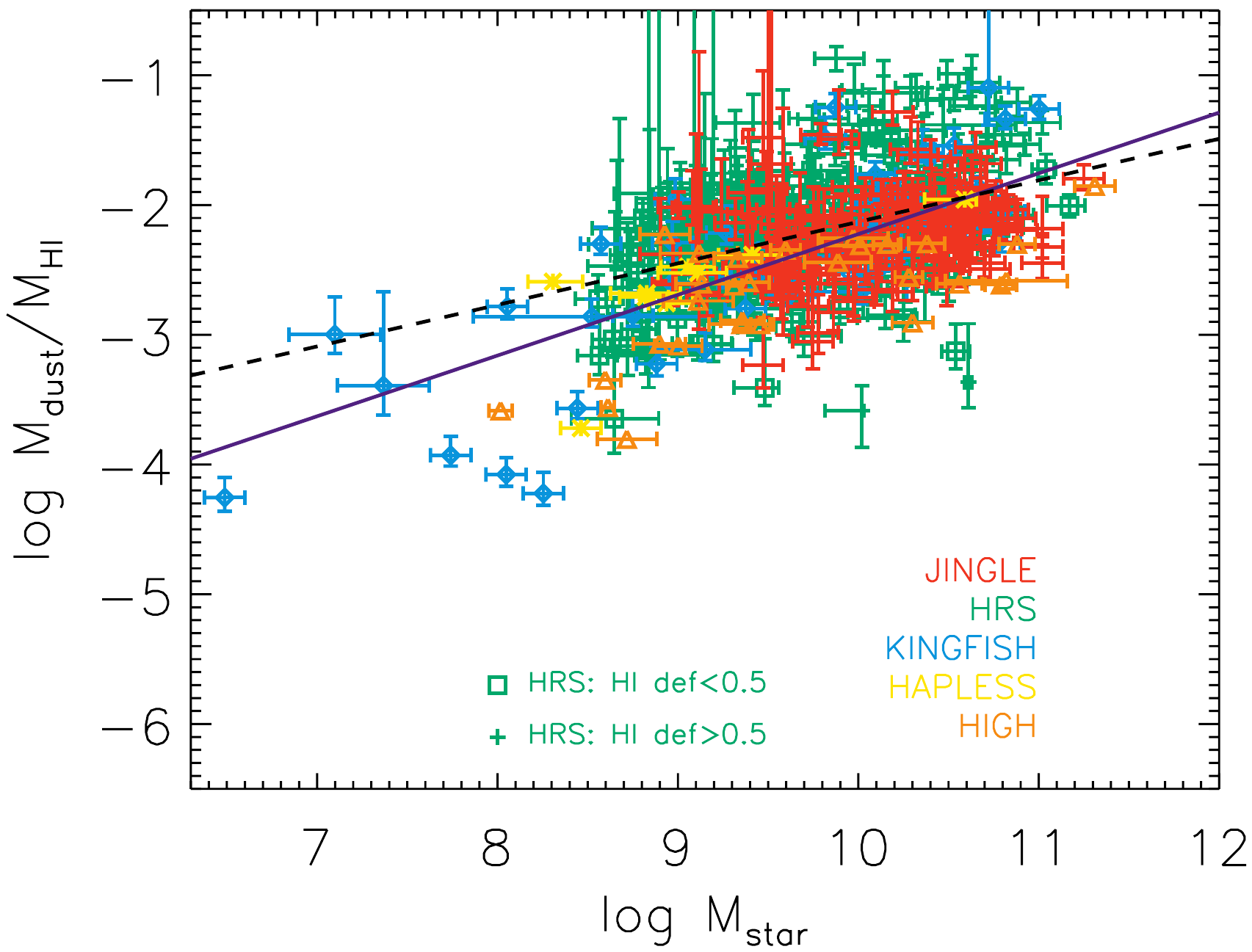}
	\includegraphics[width=8.7cm]{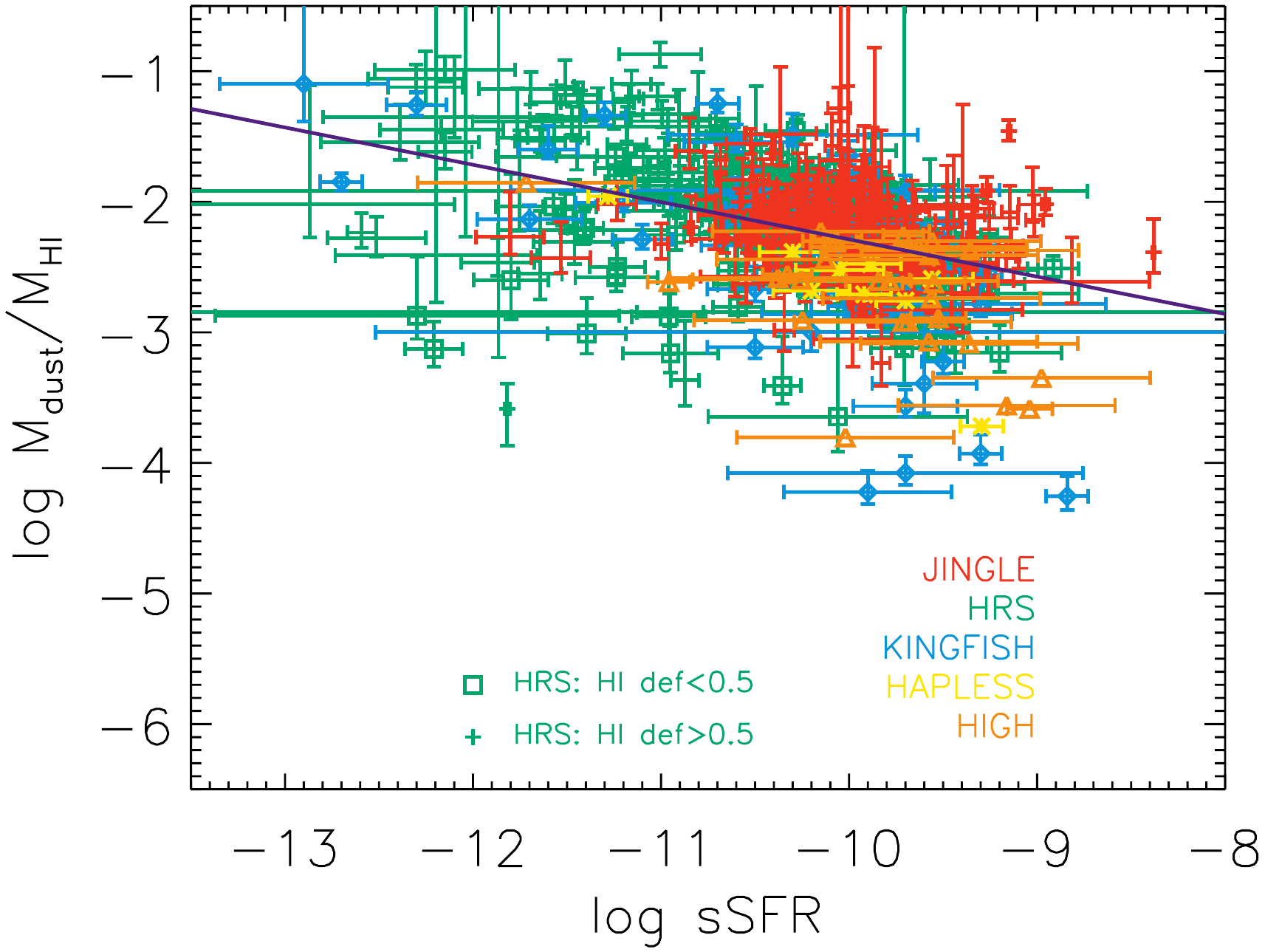}  \\
	    \caption{The scaling of the dust-to-H{\sc{i}} mass ratio (i.e., $M_{\text{dust}}$/$M_{\text{HI}}$) with stellar mass ($M_{\star}$, left panel) and specific star formation rate (sSFR with $sSFR=SFR$/$M_{\star}$, right panel). See caption of Fig. \ref{DustScaling} for more details on the symbols and plotted curves.}
    \label{DusttoHIScaling}
\end{figure*}

Our best-fit relation for the specific H{\sc{i}} gas mass as a function of stellar mass is shifted upwards by 0.3 to 0.4\,dex compared to the trend from \citet{2017MNRAS.464.4680D} due to the high specific H{\sc{i}} gas masses of JINGLE galaxies, and due to our omission of H{\sc{i}}-deficient HRS galaxies to determine the best-fit relation. The scatter observed in the relations of $M_{\text{HI}}$/$M_{\star}$ with $M_{\star}$ ($\sigma$=$0.57$) and sSFR ($\sigma$=$0.40$) dominates over the dispersion in the respective trends of $M_{\text{dust}}$/$M_{\star}$ with $M_{\star}$ ($\sigma$=$0.39$) and sSFR ($\sigma$=$0.29$), which suggests that the scatter in the trends of $M_{\text{dust}}$/$M_{\star}$ with $M_{\star}$ and $sSFR$ are likely dominated by variations in the galaxy's specific H{\sc{i}} gas masses, and not necessarily directly influenced by the various dust production and destruction mechanisms at work in these galaxies. This scenario is also supported by the scatter observed in the scaling laws (see Fig. \ref{DusttoHIScaling}) for $M_{\text{dust}}$/$M_{\text{HI}}$ with $M_{\star}$ ($\sigma$=$0.33$) and sSFR ($\sigma$=$0.37$), which is lower than similar relations for $M_{\text{HI}}$/$M_{\star}$ and suggests that the specific H{\sc{i}} gas mass dominates the scatter in these scaling relations. The trends between $M_{\text{HI}}$/$M_{\star}$, and $M_{\text{dust}}$/$M_{\star}$ ($\rho=0.67$), $M_{\text{dust}}$/$M_{\text{metals}}$ ($\rho=-0.61$) and $M_{\text{dust}}$/$M_{\text{HI}}$ ($\rho=-0.72$) furthermore show strong correlations (see Fig. \ref{Scaling_MHIMstar_trends} and Table \ref{BestFitScalingLaws}) compared to the relations of the latter ratios with $M_{\star}$ or $sSFR$, reinforcing the above reasoning. To study what processes drive the observed trends and scatter in local scaling laws, we therefore verify how the dust and metal content of these galaxies varies as a function of $M_{\text{HI}}$/$M_{\star}$ in Section \ref{EvolutionModels.sec}.  

\begin{table*}
%\centering %
\caption{The best-fit relations (of the form $y=a\,\times\,x+b$) have been inferred based on linear regression fits using the IDL procedure \texttt{MPFITEXY}, which is based on the non-linear least-squares fitting package MPFIT \citep{2009ASPC..411..251M}. In addition to JINGLE, KINGFISH, HAPLESS and HiGH galaxies, only HRS galaxies with an H{\sc{i}}-deficiency lower than 0.5 (i.e., classified as non-deficient galaxies) have been considered. The observed scatter ($\sigma$) around each of the best-fit relations has been inferred. The Spearman rank correlation coefficient, $\rho$, and corresponding p value have been inferred from the IDL procedure \texttt{r$\_$correlate} to quantify the degree of (non-)linear correlation between the various quantities.} %  h
\label{BestFitScalingLaws}
\begin{tabular}{llcccrr} % four columns, alignment for each
\hline
x & y & a & b & $\sigma$ & $\rho$ & p value\\ 
\hline
$\log$\,$M_{\star}$ & $\log$\,$M_{\text{dust}}$/$M_{\star}$ & -0.22$\pm$0.01 & -0.62$\pm$0.11 & 0.39 & -0.39 & $<10^{-6}$ \\
$\log$\,$M_{\star}$ & $\log$\,$M_{\text{HI}}$/$M_{\star}$ & -0.80$\pm$0.01 & 7.42$\pm$0.12 & 0.57 & -0.64 & $<10^{-6}$ \\
$\log$\,$M_{\star}$ & $\log$\,$M_{\text{dust}}$/$M_{\text{HI}}$ & 0.47$\pm$0.01 & -6.90$\pm$0.11 & 0.33 & 0.51 & $<10^{-6}$ \\
$\log$\,$M_{\star}$ & $\log$\,$M_{\text{dust}}$/$M_{\text{metals}}$ & 0.19$\pm$0.01 & -2.54$\pm$0.09 & 0.24 & 0.26 & $<10^{-6}$ \\
\hline
$\log$\,sSFR & $\log$\,$M_{\text{dust}}$/$M_{\star}$ & 0.56$\pm$0.01 & 2.85$\pm$0.15 & 0.29 & 0.63 & $<10^{-6}$ \\
$\log$\,sSFR & $\log$\,$M_{\text{HI}}$/$M_{\star}$ & 0.89$\pm$0.02 & 8.47$\pm$0.15 & 0.40 & 0.72 & $<10^{-6}$ \\
$\log$\,sSFR & $\log$\,$M_{\text{dust}}$/$M_{\text{HI}}$ & -0.28$\pm$0.01 & -5.08$\pm$0.10 & 0.37 & -0.41 & $<10^{-6}$ \\
$\log$\,sSFR & $\log$\,$M_{\text{dust}}$/$M_{\text{metals}}$ & -0.29$\pm$0.01 & -3.54$\pm$0.11 & 0.24 & -0.32 & $<10^{-6}$ \\
\hline 
Metallicity & $\log$\,$M_{\text{dust}}$/$M_{\text{HI}}$ & 2.27$\pm$0.06 & -21.89$\pm$0.55 & 0.34 & 0.53 & $<10^{-6}$ \\
Metallicity & $\log$\,$M_{\text{dust}}$/$M_{\text{metals}}$ & 0.40$\pm$0.14 & -4.10$\pm$1.18 & 0.26 & 0.11 & $4\times10^{-2}$ \\
%$Z$ & $M_{\text{dust}}$/$M_{\text{metals}}$ & & & & & \\
\hline 
$\log$\,$M_{\text{HI}}$/$M_{\star}$ & $\log$\,$M_{\star}$ & -1.25$\pm$0.01 & 9.27$\pm$0.01 & 0.51 & -0.65 & $<10^{-6}$ \\ %still to be changed!!!
$\log$\,$M_{\text{HI}}$/$M_{\star}$ & $\log$\,$M_{\text{dust}}$/$M_{\star}$ & 0.49$\pm$0.02 & -2.55$\pm$0.01 & 0.28 & 0.67 & $<10^{-6}$ \\
$\log$\,$M_{\text{HI}}$/$M_{\star}$ & $\log$\,$M_{\text{dust}}$/$M_{\text{metals}}$ & -0.37$\pm$0.01 & -0.84$\pm$0.01 & 0.20 & -0.61 & $<10^{-6}$ \\
$\log$\,$M_{\text{HI}}$/$M_{\star}$ & Metallicity & -0.25$\pm$0.01 & 8.54$\pm$0.01 & 0.15 & -0.55 & $<10^{-6}$ \\
$\log$\,$M_{\text{HI}}$/$M_{\star}$ & $\log$\,$M_{\text{dust}}$/$M_{\text{HI}}$ & -0.63$\pm$0.01 & -2.63$\pm$0.01 & 0.28 & -0.72 & $<10^{-6}$ \\
\hline
$\log$\,$M_{\text{dust}}$/$M_{\text{HI}}$ & $\log$\,$M_{\text{dust}}$/$M_{\text{metals}}$ & 0.61$\pm$0.02 & 0.75$\pm$0.04 & 0.11 & 0.88 & 0.0 \\
\hline 
\end{tabular}
\end{table*}

\begin{figure*}
	\includegraphics[width=8.7cm]{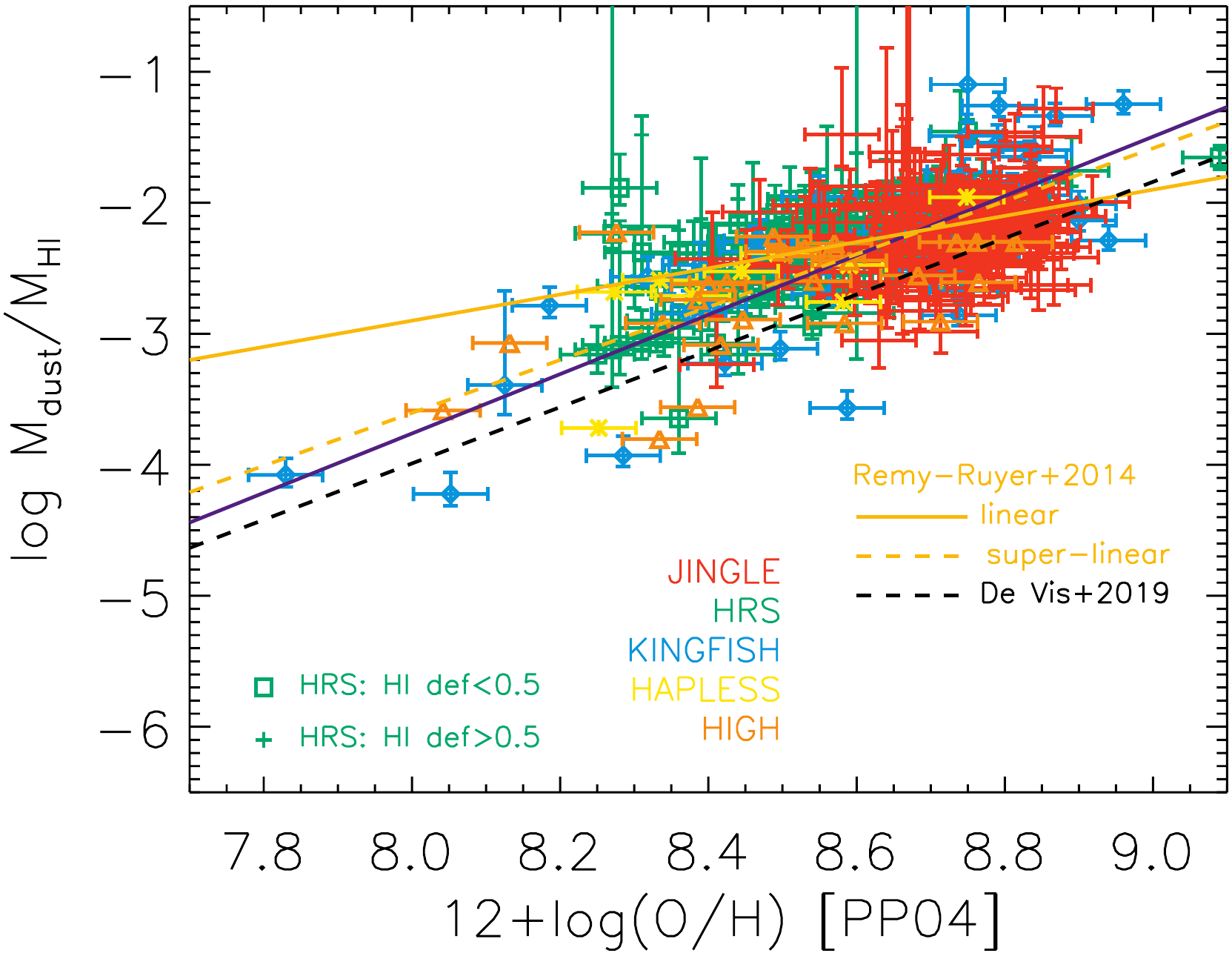}
	\includegraphics[width=8.7cm]{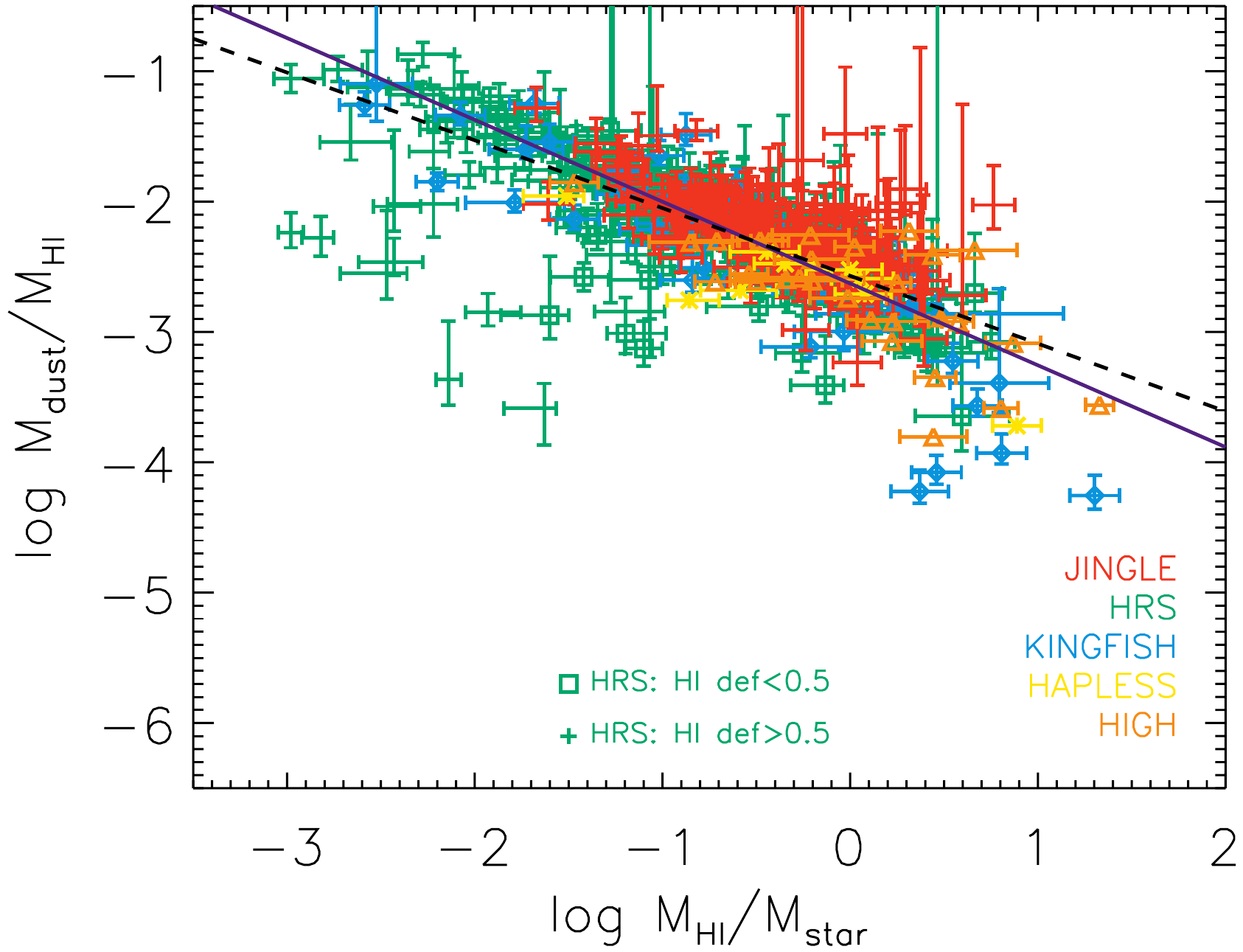}
    \caption{The scaling of the dust-to-H{\sc{i}} mass ratio (i.e., $M_{\text{dust}}$/$M_{\text{HI}}$) with metallicity (as traced by the oxygen abundance 12+$\log$(O/H)) and the specific H{\sc{i}} gas mass (i.e., $M_{\text{HI}}$/$M_{\star}$). See caption of Fig. \ref{DustScaling} for more details on the symbols and plotted curves.}
    \label{Plot_Met_DusttoHI}
\end{figure*}

\subsection{Dust-to-H{\sc{i}} ratios}
\begin{figure*}
	\includegraphics[width=8.7cm]{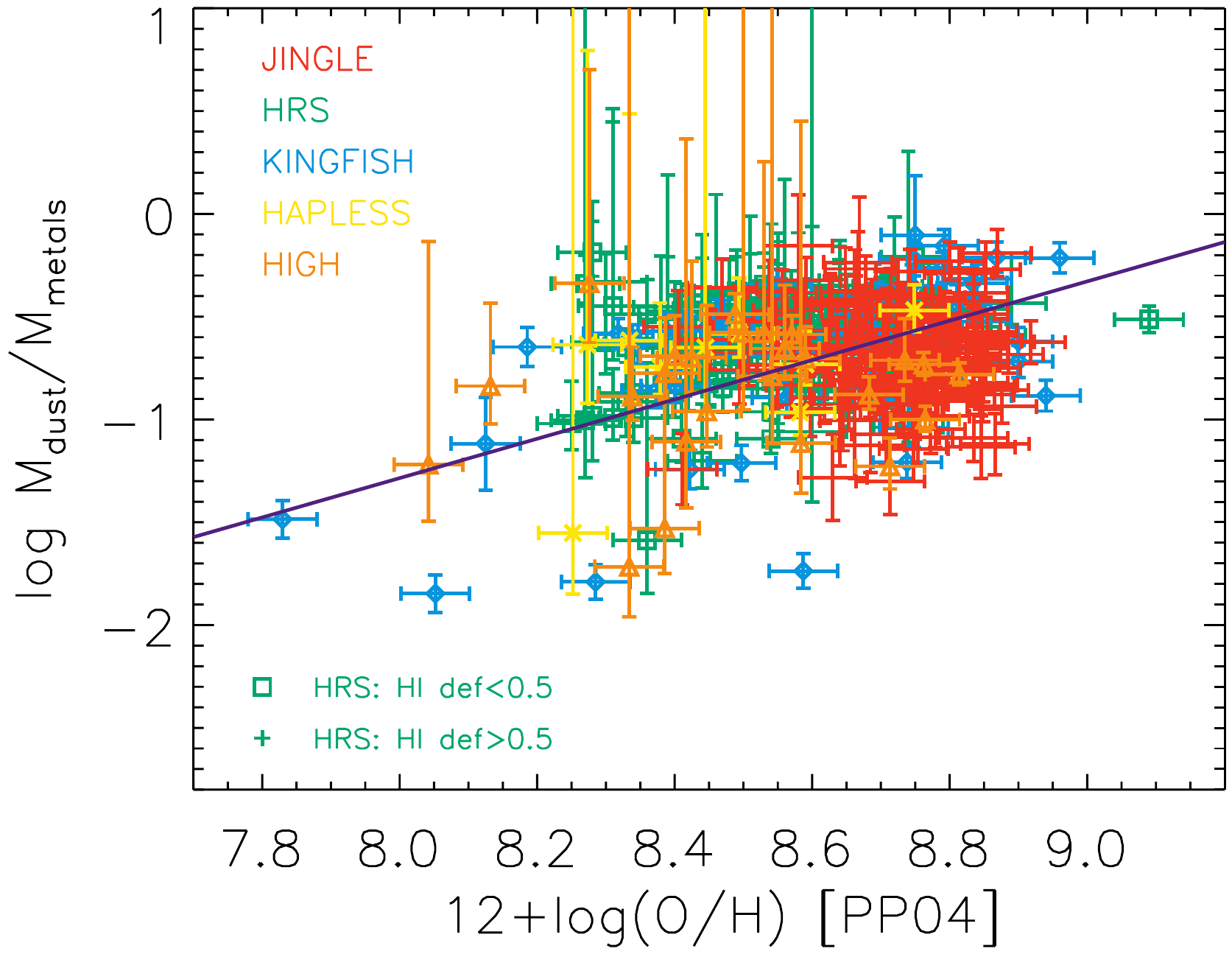}
	\includegraphics[width=8.7cm]{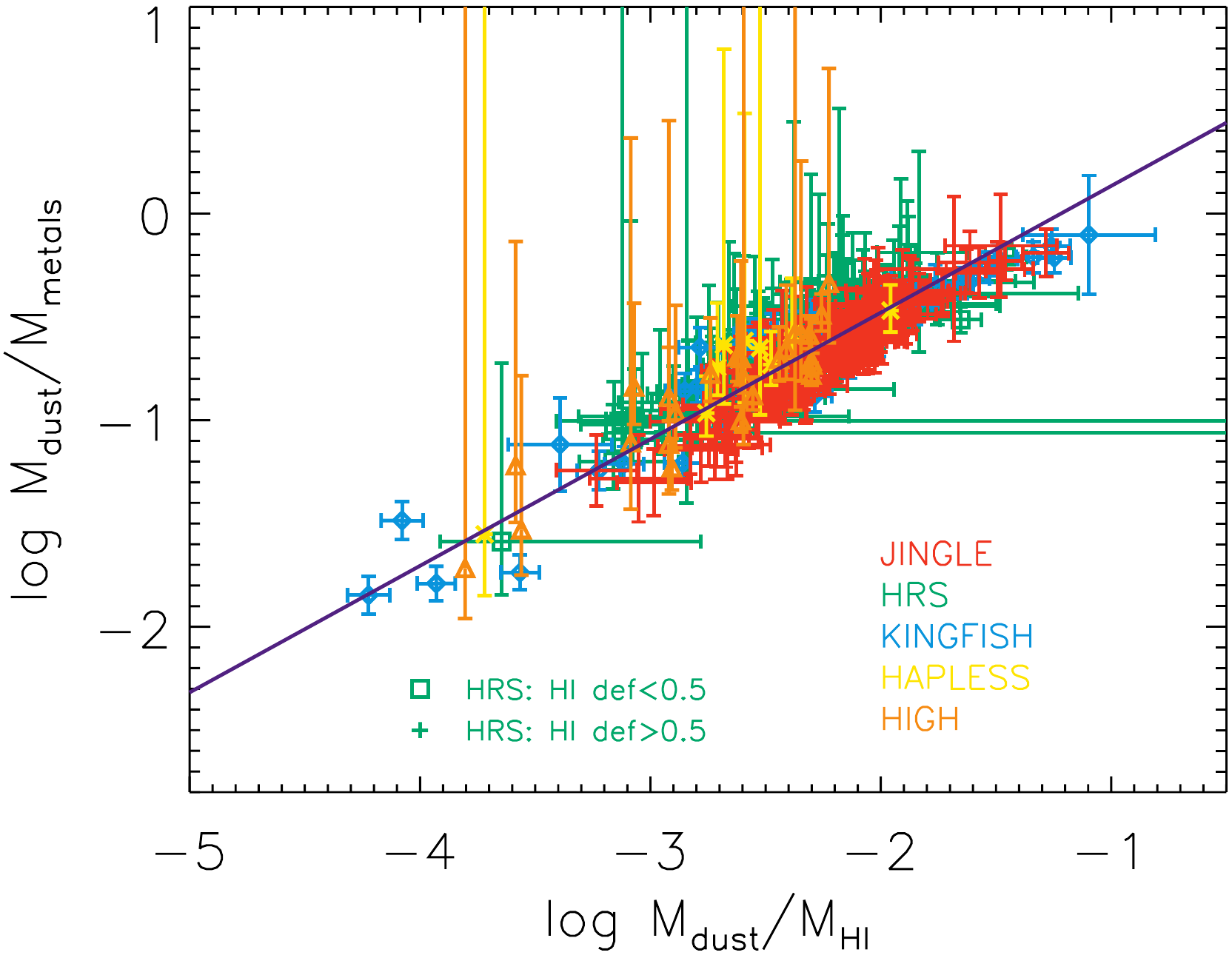}  \\
	    \caption{The scaling of the dust-to-metal mass ratio (i.e., $M_{\text{dust}}$/$M_{\text{metals}}$) with oxygen abundance (12+$\log$(O/H), left panel) and dust-to-H{\sc{i}} mass ratio (i.e., $M_{\text{dust}}$/$M_{\text{HI}}$, right panel). See caption of Fig. \ref{DustScaling} for more details on the symbols and plotted curves.}
    \label{DusttoMetalScalingII}
\end{figure*}

The $M_{\text{dust}}$/$M_{\text{HI}}$ ratio (or dust-to-gas ratio, if the contribution from molecular gas can be marginalised\footnote{We note that for 44 KINGFISH and 81 HRS galaxies with CO data the median H$_{\text{2}}$/H{\sc{i}} ratio is equal to 0.62 and 0.31, respectively, assuming a Galactic $X_{\text{CO}}$ factor. For a metallicity- and luminosity-dependent $X_{\text{CO}}$ factor, the median H$_{\text{2}}$/H{\sc{i}} ratio for KINGFISH and HRS galaxies changes to 1.90 and 0.30, respectively. For HRS galaxies, these values are in line with the average $M_{\text{H}_{2}}$/$M_{\text{HI}}$ ratio of 0.3 for xGASS galaxies with stellar masses above 10$^{10}$\,M$_{\odot}$ \citep{2018MNRAS.476..875C}.}) of a galaxy measures how many metals have been locked up in dust grains compared to the metals in the gas phase. To verify the reliability of this proxy, we plot the $M_{\text{dust}}$/$M_{\text{metals}}$ ratio (see Section \ref{DTM.sec}) as a function of the $M_{\text{dust}}$/$M_{\text{HI}}$ ratio in Figure \ref{DusttoMetalScalingII} (right panel), which shows a strong correlation ($\rho=0.88$) with little scatter ($\sigma=0.11$) around the best-fit trend. 

The $M_{\text{dust}}$/$M_{\text{HI}}$ ratios of our nearby galaxy samples range between 10$^{-1.1}$ and 10$^{-4.3}$ with a median 10$^{-2.3\pm0.4}$ (see Figure \ref{DusttoHIScaling}) which is roughly consistent with the Milky Way dust-to-H{\sc{i}} gas column density ratio assumed in the THEMIS dust model (1/135, \citealt{2017A&A...602A..46J}). The $M_{\text{dust}}$/$M_{\text{HI}}$ ratio decreases with decreasing stellar mass ($\rho$=$0.51$), and with increasing sSFR ($\rho$=$-0.41$), which is consistent with the consensus that less massive galaxies are currently in the process of vigorously forming stars, and that most of their metals have not been locked up in dust grains in comparison to the large reservoir of gas. In particular, HAPLESS ($\log$\,$M_{\text{dust}}$/$M_{\text{HI}}$=$-2.59\pm0.23$) and HiGH ($\log$\,$M_{\text{dust}}$/$M_{\text{HI}}$=$-2.61\pm0.45$) galaxies have median ratios at the low end of the entire nearby galaxy population, which might at first seem surprising given their ``normal" $M_{\text{dust}}$/$M_{\star}$ ratios. Similar trends were found by \citet{2010MNRAS.403.1894D} and consecutive works \citep{2012A&A...540A..52C,2015MNRAS.452..397C,2017MNRAS.464.4680D}, and attributed to galaxies with low stellar masses and high sSFRs, currently forming dust (high $M_{\text{dust}}$/$M_{\star}$), and still retaining large H{\sc{i}} gas reservoirs (low $M_{\text{dust}}$/$M_{\text{HI}}$) for future star formation. JINGLE ($\log$\,$M_{\text{dust}}$/$M_{\text{HI}}$=$-2.25\pm0.31$) and KINGFISH ($\log$\,$M_{\text{dust}}$/$M_{\text{HI}}$=$-2.30\pm0.69$) galaxies have ratios that agree well with the general trend observed for the ensemble of nearby galaxies, while the overall HRS sample median ($\log$\,$M_{\text{dust}}$/$M_{\text{HI}}$=$-2.17\pm0.47$) is increased due to the high ratios ($\log$\,$M_{\text{dust}}$/$M_{\text{HI}}$=$-1.80\pm0.44$) observed for H{\sc{i}}-deficient HRS galaxies. The latter high ratios agree with the findings of \citet{2016MNRAS.459.3574C}, and were attributed to the outside-in stripping of the interstellar medium in these H{\sc{i}}-deficient HRS galaxies (where the extended H{\sc{i}} component is affected more than the dust and molecular gas). The lowest ratios ($-4.3\leq$\,$M_{\text{dust}}$/$M_{\text{HI}}$\,$\leq-3.9$) have been observed for four irregular KINGFISH galaxies (NGC\,2915, HoII, DDO053, NGC\,5408) characterised by low stellar masses, low metal abundances, high sSFRs, high specific H{\sc{i}} gas masses and low specific dust masses, which makes them stand out from the average KINGFISH galaxy population and characterises these galaxies as being at an early stage of evolution. 

Trends of dust-to-gas ratios with metallicity reported in the literature show that the $M_{\text{dust}}$/$M_{\text{HI}}$ ratio is strongly linked to the evolutionary stage of galaxies with gradually more metals being locked up in dust grains (e.g., \citealt{2014A&A...563A..31R}). The relation between $M_{\text{dust}}$/$M_{\text{HI}}$ as a function of oxygen abundance is shown in Figure \ref{Plot_Met_DusttoHI}, and is best-fitted with a super-linear trend (slope: $2.26\pm0.07$). For reference, the linear relation (with a fixed slope of $1$) and super-linear trend (with a slope of $2.02\pm0.28$) from \citet{2014A&A...563A..31R} are overlaid as yellow solid and dashed lines\footnote{Note, that we are only interested in a comparison of their slopes, as the normalisation of these curves can not be directly compared to our values due to the differences in the assumed metallicity calibration, scaling factor for the gas mass to include heavier elements and dust opacities.}. Our trend is consistent with the super-linear relation from \citet{2014A&A...563A..31R}, which might seem surprising at first as the linear relation from \citet{2014A&A...563A..31R} was found adequate to explain the trends at metallicities $12+\log$(O/H)$\gtrsim8$ and the super-linear trend was invoked to explain the behaviour at metallicities lower than this threshold. A $\chi^{2}$ goodness-of-fit test confirms that the linear fit from \citet{2014A&A...563A..31R} does not provide a good fit to the data (p-value of 1), even when excluding galaxies below a metallicity threshold of 12+$\log$(O/H)=8.4. The p-value (0.25) inferred from our best-fit suggests that the data are neither well described by a non-linear relation, which likely results from the limited metallicity range covered by our sample, and the large degree of scatter in the relation ($\sigma$=0.34). We furthermore compare our best-fit relation to the super-linear trend (slope of 2.15$\pm$0.11) inferred by \citet{2019A&A...623A...5D} for a sample of $\sim$500 DustPedia galaxies for the same metallicity ``PP04" calibration\footnote{Note that the use of a different metallicity calibration would still yield a super-linear trend, but with a slightly different slope and/or normalisation (see Table 4 from \citealt{2019A&A...623A...5D}).}, but they included an estimation of the molecular gas content. The slope of our relation agrees well with their super-linear trend, but is offset by 0.2-0.3\,dex to higher dust-to-H{\sc{i}} ratios, which can likely be attributed to the omission of the molecular gas content in our galaxy samples and/or to the different samples under study in both works. In another DustPedia paper, a metallicity-dependent $X_{\text{CO}}$ factor is invoked to reproduce a linear relation between the dust-to-gas ratio and metallicity \citet{2019arXiv191109187C}, as frequently observed both on resolved and integrated galaxy scales in the local Universe (e.g., \citealt{1998ApJ...496..145L,2011A&A...532A..56G,2011A&A...535A..13M,2013ApJ...777....5S}). In future work, we will study the total gas scaling relations for JINGLE galaxies, and investigate the effect of different assumptions on the $X_{\text{CO}}$ conversion factor. In the next paragraphs, we discuss the applicability of dust as a gas tracer based on the H{\sc{i}} gas scaling relations of this work. 

%\section{Dust as a probe of the interstellar gas reservoir}
%\label{DustTracer.sec}
Dust mass measurements are often advocated as an alternative probe of the total ISM mass budget (e.g., \citealt{2012ApJ...761..168E,2012ApJ...760....6M,2014ApJ...783...84S,2015ApJ...799...96G,2016ApJ...820...83S,2018MNRAS.476.1390J}), due to the relative ease of obtaining infrared data and inferring dust masses, as opposed to a combination of H{\sc{i}} data (for which the sensitivity quickly drops at high redshifts) and CO observations (hampered by the notorious CO-to-H$_{\text{2}}$ conversion factor, \citealt{2013ARA&A..51..207B}). 

Figure \ref{Plot_Met_DusttoHI} shows that there is a considerable spread (0.34\,dex) in the $M_{\text{dust}}$/$M_{\text{HI}}$ ratio as a function of oxygen abundance. The use of dust as an ISM mass tracer relies on the assumption of an approximately constant dust-to-gas ratio to convert dust masses into total gas masses. Variations of the dust-to-gas ratio with metallicity have been demonstrated before (e.g., \citealt{2014A&A...563A..31R}), but the scatter around the best-fit in Figure \ref{Plot_Met_DusttoHI} implies that the dust-to-H{\sc{i}} ratio already varies by more than a factor of two at fixed metallicity. In most cases, the metal abundances of galaxies are not known a priori, and the uncertainty on the estimated ISM mass reservoir will be higher than this factor of two. Also the use of oxygen as a tracer of the total metal mass in galaxies might introduce an increased level of scatter. Some of the scatter in our relation might be caused by the missing molecular gas mass measurements; although \citet{2019arXiv191109187C} find that the H{\sc{i}} gas mass correlates more closely to the dust mass than the molecular gas. Part of the spread might furthermore be attributed to the inhomogeneous extent of dust and gas reservoirs tracing different parts of a galaxy. In particular, JINGLE galaxies may be affected by the unresolved extent of H{\sc{i}} gas observations obtained from single-dish observations. In due course, all JINGLE galaxies will be covered by future interferometric radio facilities (e.g., SKA, Apertif), which will give us a handle on the spatial extent of their H{\sc{i}} gas reservoir. JINGLE, HAPLESS and HiGH metallicities have furthermore been derived from the central 3$\arcsec$ covered by SDSS fibre optical spectroscopy data \citep{2013MNRAS.431.1383T}, which could potentially increase the uncertainty on their oxygen abundances due to the lack of a set of spatially resolved metallicity measurements, as opposed to the resolved metallicity measurements for the other nearby galaxy samples. Due to the wide spread in metallicity gradients observed in local galaxy samples (e.g., \citealt{2003ApJ...591..801K,2010ApJS..190..233M,2014A&A...570A...6S,2017MNRAS.469..151B,2018MNRAS.479.5235P,2019ApJ...878L..31A}), these central metallicity measurements will not necessarily be representative of a galaxy's average metal abundance. Metallicity measurements (in particular at low metallicity) furthermore come with large uncertainties due to the specific metallicity calibration that was applied, and its dependence on a fixed electron temperature in case of strong line calibrations. In addition, variations in the dust emissivity driven by an altered dust mineralogy or variations in carbon-to-silicate grain fractions (e.g., \citealt{2019MNRAS.489.5256C}) may be the cause of part of the scatter.

\citet{2018MNRAS.476.1390J} argued that most of the scatter in the $M_{\text{dust}}$/$M_{\text{HI}}$ relation is driven by the unknown partition between atomic and molecular gas, and variations in the H$_{2}$-to-H{\sc{i}} ratio with galaxy properties. Their study of the HRS galaxy sample suggests a dispersion of 0.22-0.25\,dex in the relation between $M_{\text{dust}}$/$M_{\text{HI}}$ and metallicity, which is somewhat lower than the 0.34\,dex scatter inferred for the sample of nearby galaxies in this paper.

\begin{figure*}
	\includegraphics[width=8.7cm]{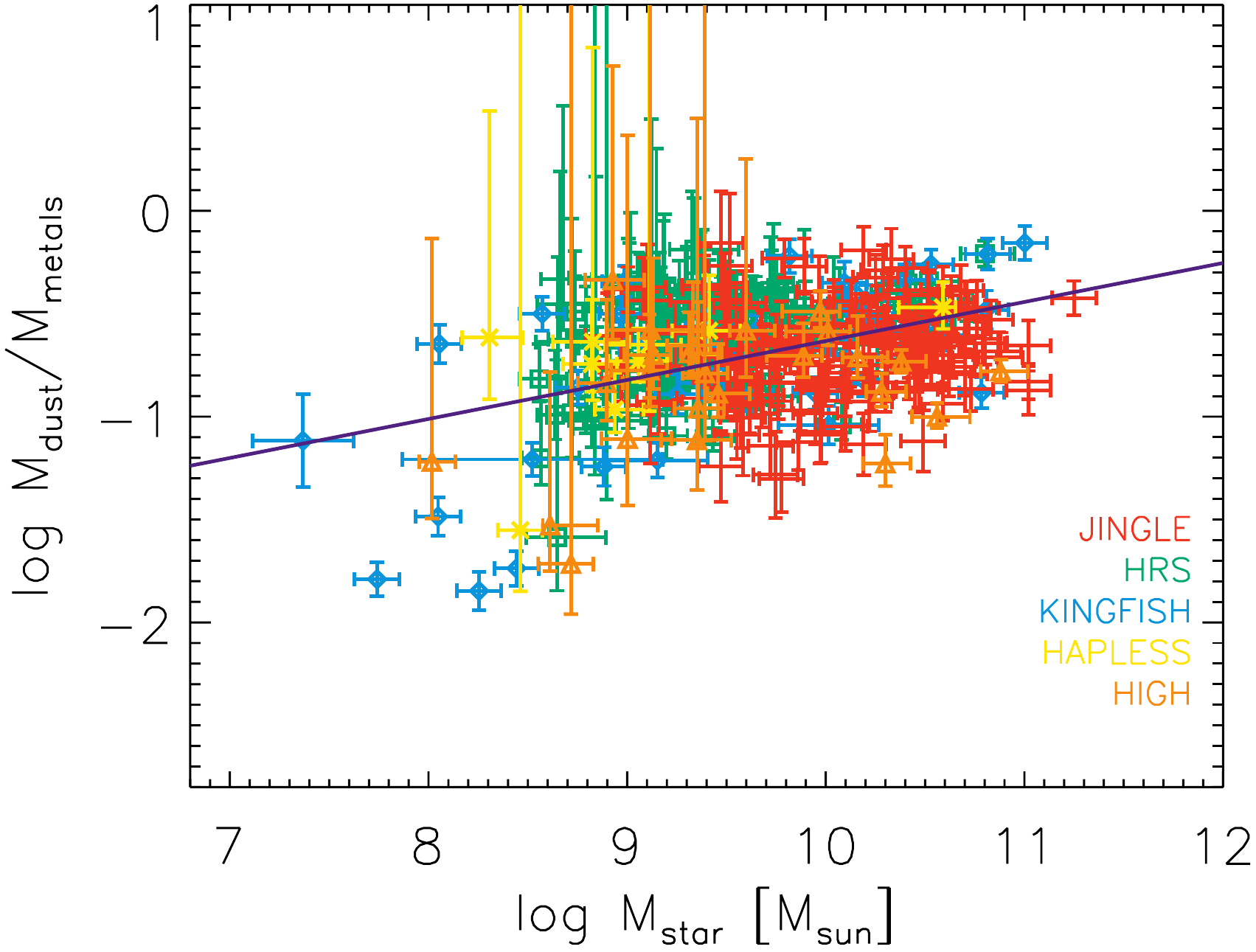}
	\includegraphics[width=8.7cm]{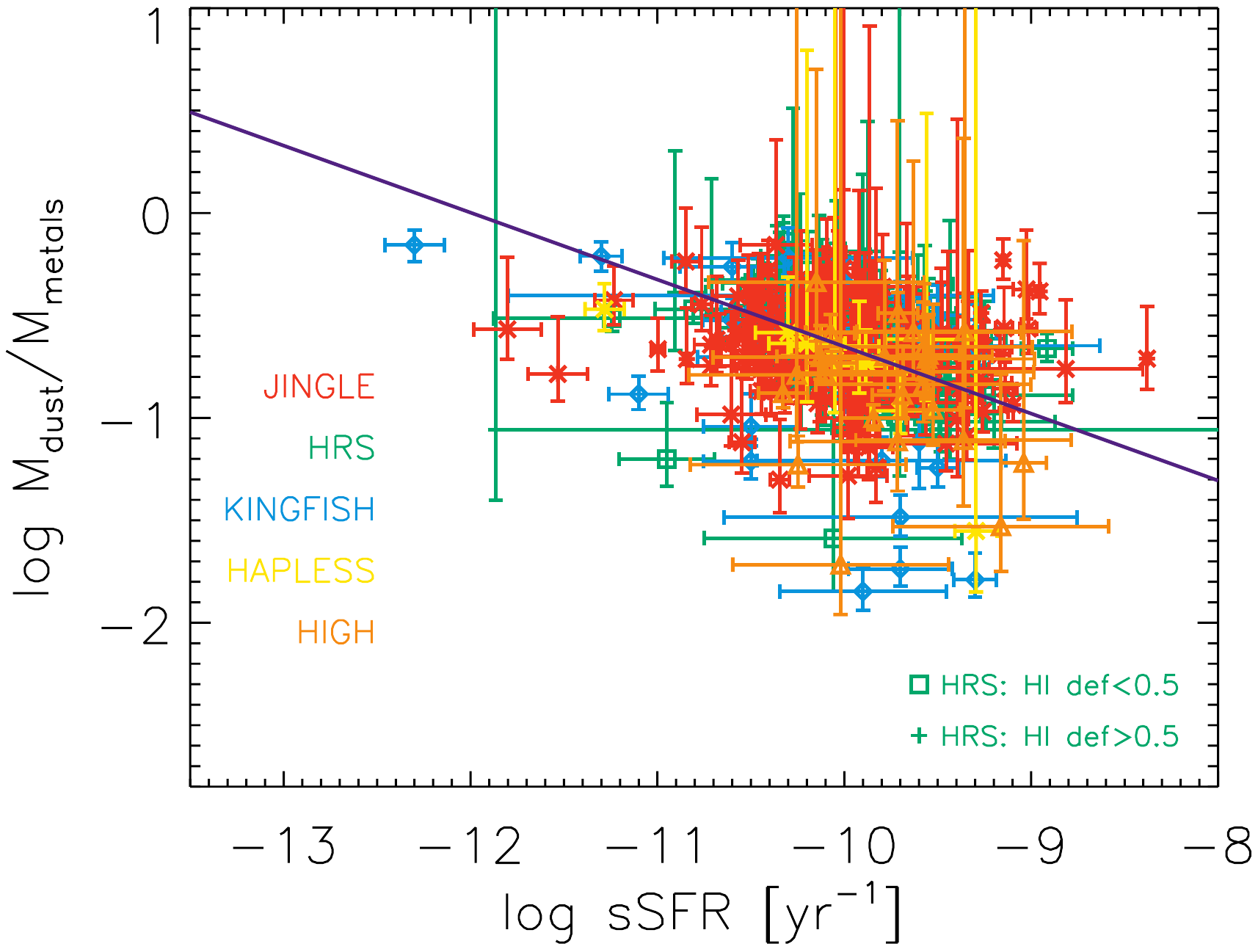}  \\
    \caption{The scaling of the dust-to-metal mass ratio (i.e., $M_{\text{dust}}$/$M_{\text{metals}}$) with stellar mass ($M_{\star}$, left panel) and specific star formation rate (sSFR with $sSFR=SFR$/$M_{\star}$, right panel). See caption of Fig. \ref{DustScaling} for more details on the symbols and plotted curves.}
    \label{DusttoMetalScaling}
\end{figure*}

\subsection{Dust-to-metal ratios}
\label{DTM.sec}
We have calculated the dust-to-metal ratios (DTM) as the ratio of the dust mass and the total amount of metals (and thus accounting for metals in the gas phase and locked up in dust grains) similar to other literature works (e.g., \citealt{2019A&A...623A...5D}):
\begin{equation}
DTM = M_{\text{dust}}/M_{\text{metals}}(gas+dust),
\end{equation}
with $M_{\text{metals}}$(gas+dust)=$f_{\text{Z}}\times M_{\text{gas}}$+$M_{\text{dust}}$. This prescription allows for a direct comparison with the measurements of dust depletion in damped Lyman\,$\alpha$ absorbers  out to large redshifts (e.g., \citealt{2016A&A...596A..97D}). The metal mass fraction $f_{\text{Z}}$ is calculated based on a galaxy's oxygen abundance, and the values of the metal mass fraction ($f_{Z_{\odot}}$=0.0134) and oxygen abundance (12+$\log$(O/H)$_{\odot}$=$8.69$) inferred for the Sun from \citet{2009ARA&A..47..481A}, which results in $f_{\text{Z}}$=27.36$\times$10$^{12+\log(O/H)-12}$. Due to the lack of molecular gas mass estimates, we have used the H{\sc{i}} gas mass (corrected for the contribution from elements heavier than hydrogen, see Eq. \ref{Eq_heavyelcorr}) to calculate the metal mass fractions. We inferred that the dust-to-metal ratios are lower by -0.11\,dex and -0.19\,dex for 81 HRS and 44 KINGFISH galaxies, respectively, if we account for molecular gas masses assuming a Galactic $X_{\text{CO}}$ conversion factor. A metallicity-dependent $X_{\text{CO}}$ conversion factor would lower the dust-to-metal ratios by -0.46\,dex for the KINGFISH sample (which contain the lowest metallicity galaxies in our local galaxy sample). It is worth noting that the metal mass furthermore relies on measurements of the oxygen abundance, which does not necessarily scale linearly to the total mass of metals in galaxies at different stages of evolution. 

The DTM ratio provides a measure of the relative fraction of metals in the interstellar medium that have been locked up in dust grains, and therefore sensitively depends on the efficiency of various dust production and destruction mechanisms. It is assumed that the dust-to-metal ratio remains more or less constant if dust is predominantly produced via stellar sources\footnote{This statement relies on the assumption that stellar dust yields, dust condensation efficiencies and reverse shock destruction rates do not have a strong metallicity dependence.}. If grain growth dominates the dust production, the DTM ratio is thought to increase as galaxies evolve and their interstellar medium is enriched with metals\footnote{This inference is somewhat model dependent, and is also influenced by grain destruction efficiencies.}, with grain growth believed to be more efficient than stellar dust production sources once a critical metallicity threshold has been reached \citep{2013EP&S...65..213A}. Dust destruction through supernova shocks (where metals locked up in dust grains are returned to the interstellar medium) have the opposite effect and will lower the DTM ratio.

The majority of nearby galaxies fall within the same range of DTM ratios ($-0.90\leq\log DTM\leq-0.40$), with little variation among the different galaxy populations (see Table \ref{AverageProperties} and Fig. \ref{DusttoMetalScaling}). The Milky Way is situated on the high end of this range with $\log$\,DTM=-0.45, if we have assume a total gas mass $M_{\text{gas}}$=12.5$\times$10$^{9}$\,M$_{\odot}$ \citep{2009ARA&A..47...27K}, solar metallicity and dust-to-H{\sc{i}} ratio of 1/135 as inferred from the Milky Way THEMIS model \citep{2017A&A...602A..46J}. The median ratio for HiGH galaxies ($\log$\,DTM=-0.78$\pm$0.30) is slightly lower than the other galaxy populations (but not significantly different, see Table \ref{Table_MannWhitneyU}) and confirms their early stage of evolution. The second lowest ratio ($\log$\,DTM=-0.67$\pm$0.23) is observed for JINGLE galaxies, but similarly does not differ significantly from HAPLESS ($\log$\,DTM=-0.65$\pm$0.15) and KINGFISH ($\log$\,DTM=-0.63$\pm$0.38) galaxies. The median ratio for HRS galaxies ($\log$\,DTM=-0.60$\pm$0.21) is significantly higher than for the other four samples due to the contribution from H{\sc{i}}-deficient HRS galaxies, with the latter being characterised by significantly higher ratios ($\log$\,DTM=-0.44$\pm$0.08). This DTM is more than 60$\%$ higher than the median DTM observed in our sample of nearby galaxies ($\log$\,DTM=-0.66$\pm$0.24, excluding the H{\sc{i}}-deficient HRS galaxies). This high DTM ratio appears consistent with the high $M_{\text{H}_{2}}$/$M_{\text{dust}}$ ratios observed in H{\sc{i}}-deficient HRS galaxies \citep{2016MNRAS.459.3574C}, and a picture of outside-in stripping of interstellar material where metals and H{\sc{i}} are more easily stripped compared to the more centrally concentrated dust and molecular gas content.  
The median ratio for our nearby galaxy sample is higher than the average $\log$\,DTM=-0.82$\pm$0.23 from \citet{2019A&A...623A...5D}, which we attribute to the fact that we did not consider molecular hydrogen measurements. Indeed, we discussed earlier that neglecting the molecular gas content will overestimate the DTM ratios by 0.11\,dex up to 0.46\,dex.

\begin{figure*}
	\includegraphics[width=8.7cm]{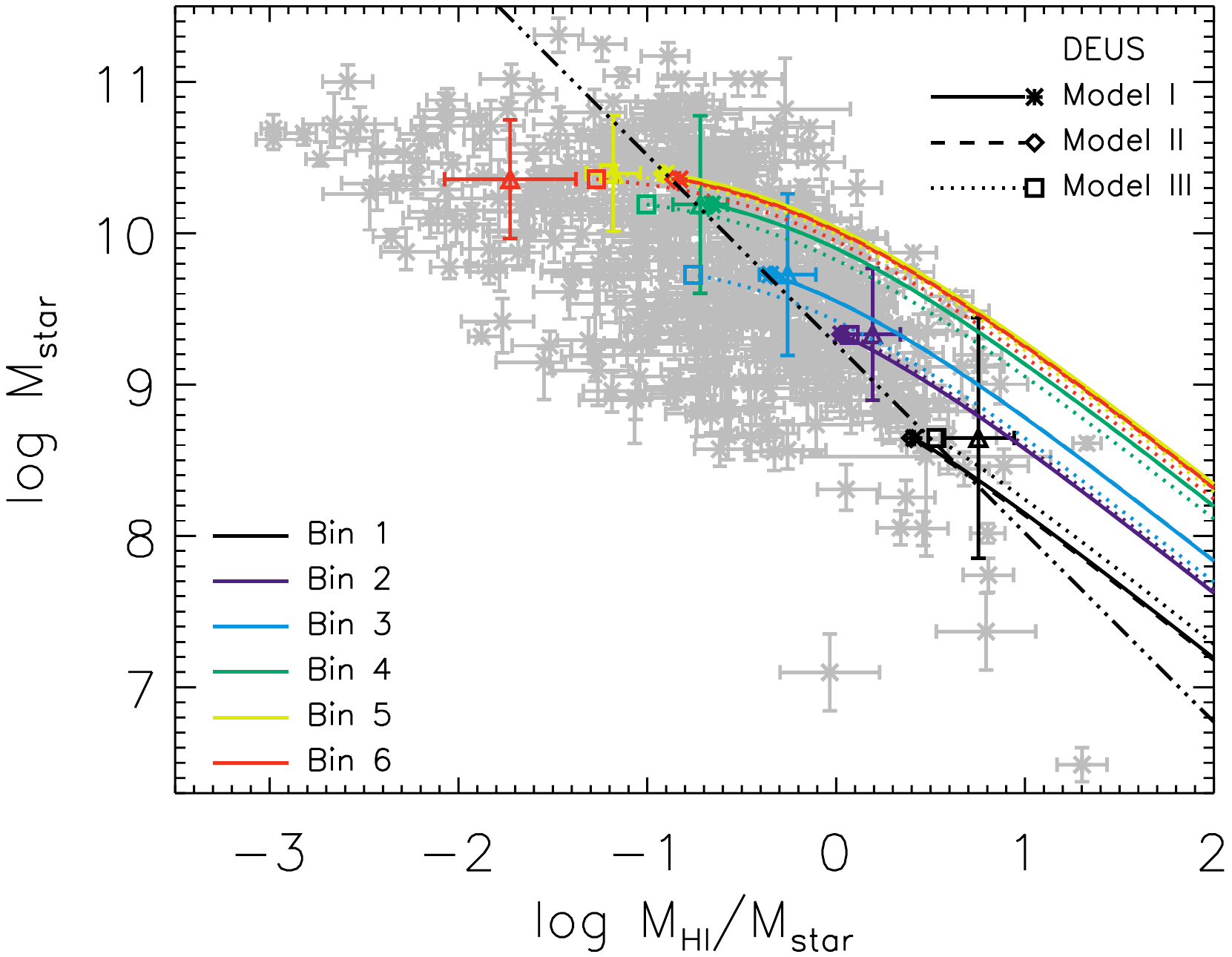}  
	\includegraphics[width=8.7cm]{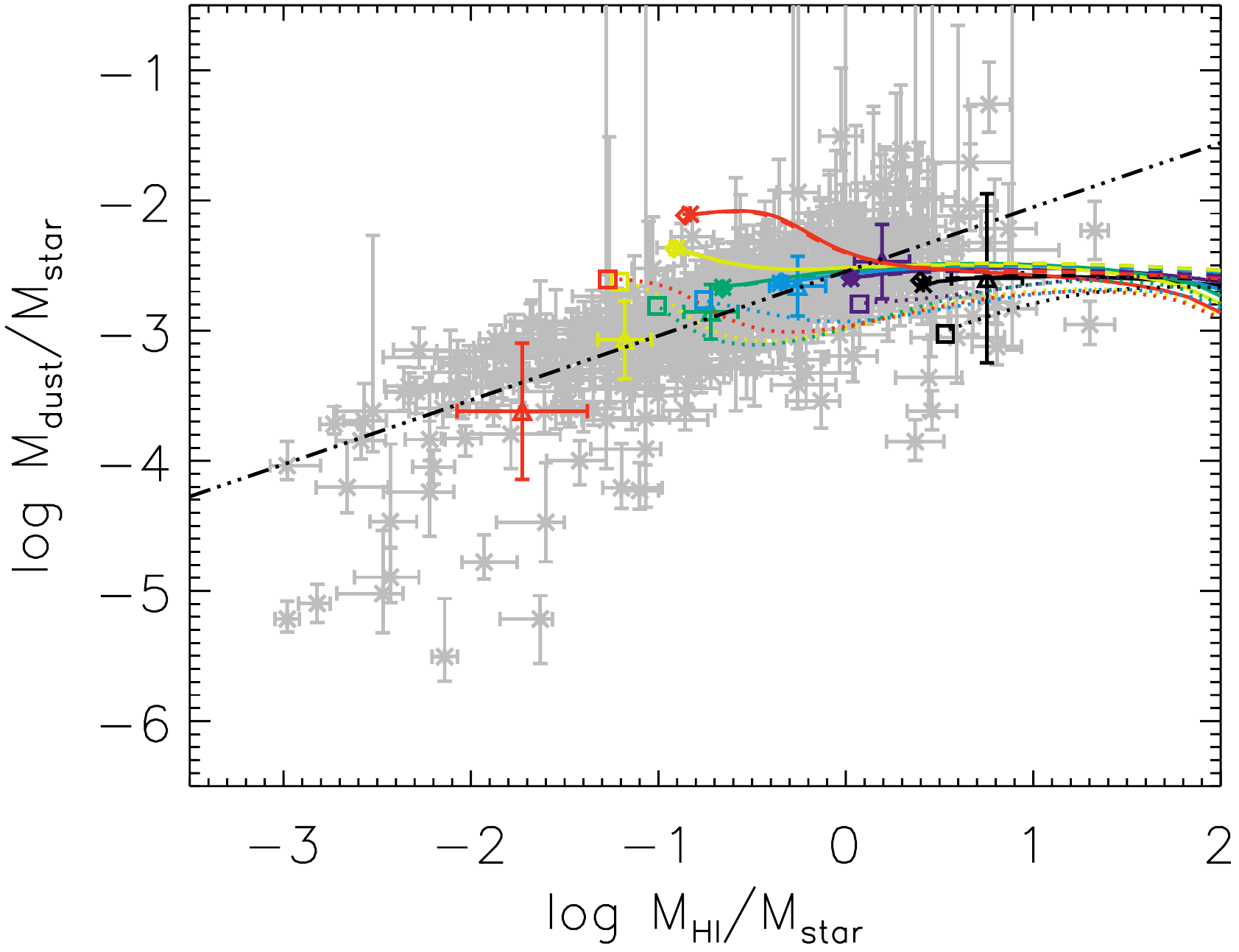} \\
	\includegraphics[width=8.7cm]{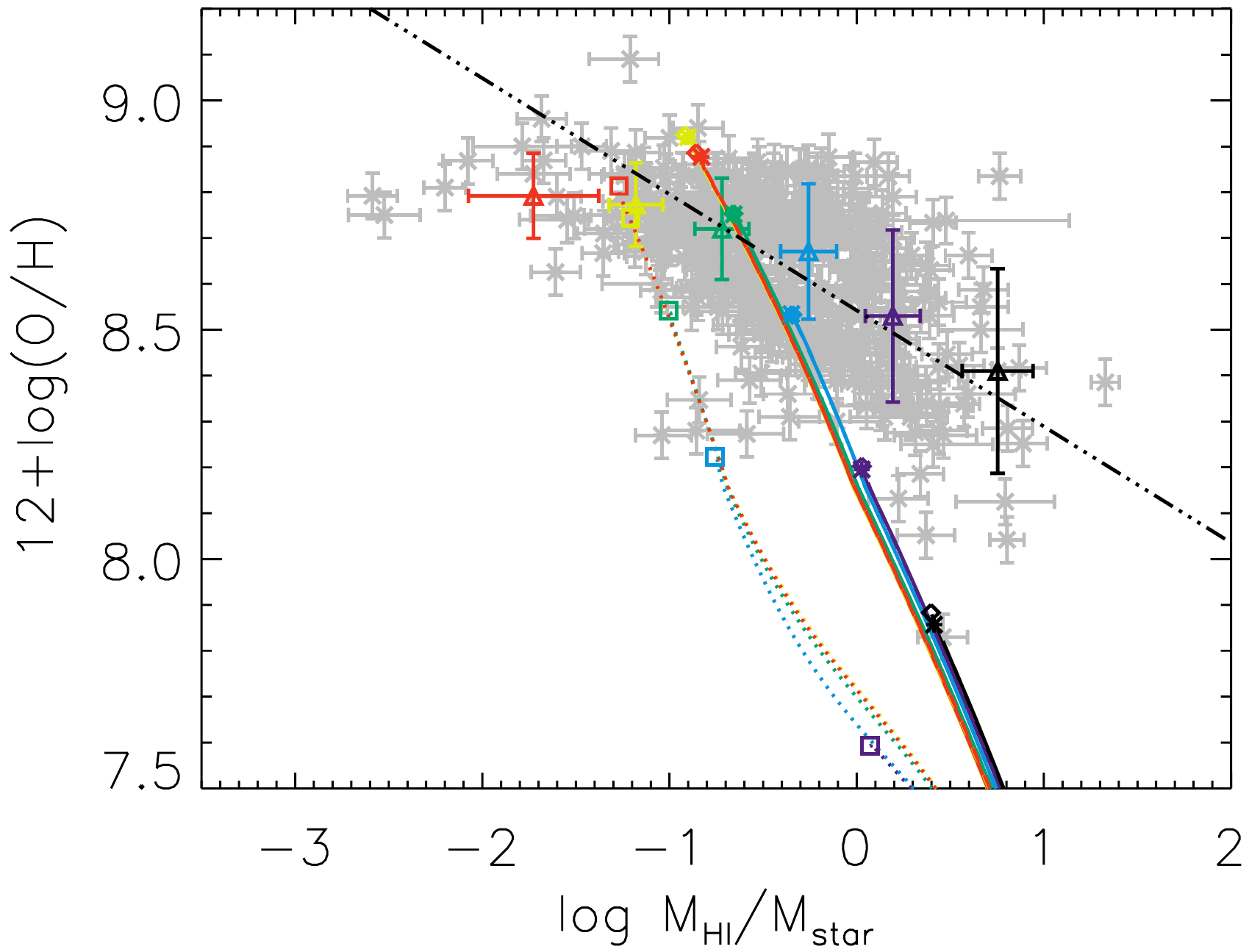}  
		\includegraphics[width=8.7cm]{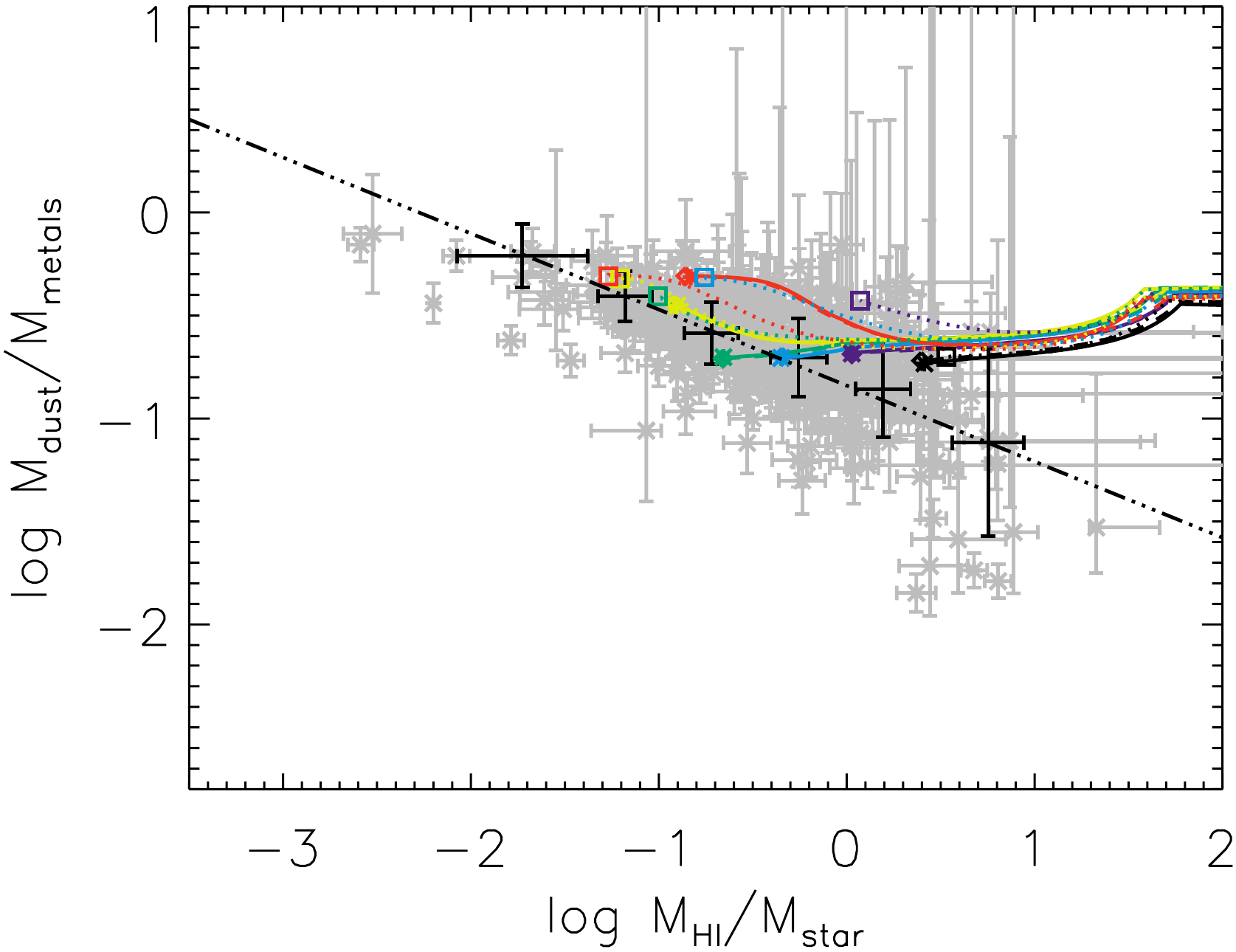}  
    \caption{The scaling of the stellar mass ($M_{\star}$, top left), $M_{\text{dust}}$/$M_{\star}$ (top right), oxygen abundance (bottom left) and $M_{\text{dust}}$/$M_{\text{metals}}$(gas+dust) (bottom right) with specific H{\sc{i}} gas mass ($M_{\text{HI}}$/$M_{\star}$) shown for our entire sample of nearby galaxies (grey symbols), with the best-fit trends overlaid as black dashed-triple dotted lines. The median values for each of the six galaxy bins are overlaid with coloured triangles and errors that correspond to the spread in each bin. The $M_{\text{HI}}$/$M_{\star}$ ratio is highest for galaxies in Bin\,1 (which are thought to correspond to the least evolved galaxies), while the more evolved galaxies with the lowest $M_{\text{HI}}$/$M_{\star}$ ratios populate Bin\,6. The evolutionary trends (over a period of 12\,Gyr) that were inferred from the median parameters for Models I, II and III are overlaid as solid, dashed and dotted lines, respectively, and the final present-day model values are indicated with coloured asterisks, diamonds and squares, respectively. The colour coding of model trends and observed median properties range from black to red corresponding to Bins 1 through to 6 (see legend in the top left panel).}
    \label{Scaling_MHIMstar_trends}
\end{figure*}

We observe weak (but significant) correlations between the DTM and $M_{\star}$ ($\rho$=$0.26$), sSFR ($\rho$=$-0.32$) and $M_{\text{HI}}$/$M_{\star}$ ($\rho$=$-0.61$) (see Fig. \ref{DusttoMetalScaling} and \ref{Scaling_MHIMstar_trends}), while the relation with metallicity does not reveal a clear trend ($\rho$=$0.11$, see Fig. \ref{DusttoMetalScalingII}, left panel). These weak correlations suggest that the DTM increases as a galaxy evolves, although there is quite some scatter in these relations. In particular, galaxies with $M_{\star}\ge10^{9}$M$_{\odot}$ appear characterised by a nearly constant DTM, while the DTM drops significantly for several low mass galaxies ($M_{\star}<10^{9}$M$_{\odot}$). This sudden change in DTM becomes particularly evident for less evolved galaxies with $\log\,M_{\text{HI}}$/$M_{\star}>0.3$ (see Fig. \ref{Scaling_MHIMstar_trends}, bottom right panel), and has been attributed in the past to a critical metallicity threshold above which grain growth becomes efficient and contributes significantly to the dust production in galaxies (e.g.,  \citealt{2013EP&S...65..213A}). The absence of a clear trend with metallicity due to the large scatter in DTM ratios at low metallicities might suggest that this critical metallicity threshold can vary from one galaxy to another \citep{2013EP&S...65..213A} or, alternatively, that such a critical metallicity threshold is not relevant\footnote{We should note that the metallicity range in our local galaxy sample is limited (with only one galaxy below 12+$\log$(O/H)$<$8.0) and might not reach down to the metallicity regime where a threshold would occur.}. In Section \ref{EvolutionModels.sec}, we show that efficient grain growth is not required as a dominant dust production source to explain the current dust budgets of nearby galaxies with $-1.0\lesssim \log M_{\text{HI}}/M_{\star}\lesssim0$. With supernova shock destruction releasing elements back into the gas phase, a wide range of DTM ratios (at fixed metallicity) can also result from variations in dust destruction efficiencies and/or recent supernova rates. Also the structure of the interstellar medium, and the filling factor of different ISM phases can play an important role in determining how efficiently grains can grow in the interstellar medium, and how effectively supernova shocks can act as dust destroyers \citep{2011A&A...530A..44J}, and will add to the scatter.

To summarise our observational findings from these scaling relations, we infer that $M_{\text{HI}}$/$M_{\star}$ varies considerably at a fixed stellar mass and fixed sSFR, more so than the $M_{\text{dust}}$/$M_{\star}$ and $M_{\text{dust}}$/$M_{\text{HI}}$ ratios. This large spread can be interpreted as the specific H{\sc{i}} gas mass being the main driver of the trends and scatter observed in other scaling laws (rather than variations in the relative contributions from several dust formation and destruction processes at fixed stellar mass or sSFR). This picture is reinforced by the significant correlations between $M_{\text{HI}}$/$M_{\star}$, and $M_{\text{dust}}$/$M_{\star}$, $M_{\text{dust}}$/$M_{\text{metals}}$(gas+dust) and $M_{\text{dust}}$/$M_{\text{HI}}$ (see Figure \ref{Scaling_MHIMstar_trends}) and establishes that $M_{\text{HI}}$/$M_{\star}$ is closely linked to the enrichment of the interstellar medium with dust and metals, and the evolution of a galaxy, in general. In Section \ref{EvolutionModels.sec}, we will interpret the evolutionary trends for $M_{\text{dust}}$/$M_{\star}$, $M_{\text{dust}}$/$M_{\text{HI}}$ and $M_{\text{dust}}$/$M_{\text{metals}}$(gas+dust) using a set of chemical evolution models to infer what dust production and destruction mechanisms have contributed to the build up a galaxy's present-day dust and metal budget. 

\section{Interpreting local scaling laws with Dust and Element evolUtion modelS (DEUS)}
\label{EvolutionModels.sec}

%mention closed box!!!
\subsection{Binning the sample in an evolutionary sequence}
%\subsection{Scaling laws with specific H{\sc{i}} gas mass}
For the purpose of understanding how the dust, H{\sc{i}} gas and metal content evolves in galaxies, we have divided our local galaxy sample\footnote{We omitted H{\sc{i}}-deficient HRS galaxies, since they have experienced recent removal of large fractions of their gas content, which makes it tenuous to reproduce their current H{\sc{i}} gas, dust and metal content without detailed constraints on the timescale and the extent of their gas removal.} into six separate bins according to equally sized ranges covered by galaxies in $\log\,M_{\text{HI}}$/$M_{\star}$. This subdivision results in unequal galaxy sample sizes in each bin. We decided to take this approach as the spread in various quantities (and thus the uncertainty on our median bin values) does not depend on the number of galaxies in each bin, but rather on the intrinsic scatter for galaxies at different stages of evolution. Table \ref{MHIMstar_bins} lists the sample size, average stellar mass ($\log\,$\,$M_{\star}$), specific H{\sc{i}} gas mass ($\log$($M_{\text{HI}}$/$M_{\star}$)), specific dust mass ($\log$($M_{\text{dust}}$/$M_{\star}$)), dust depletion ($\log$($M_{\text{dust}}$/$M_{\text{metals}}$(gas+dust)) and metallicity for these six galaxy bins. The bins range from galaxies with high $M_{\text{HI}}$/$M_{\star}$ ratios and thus at an early stage of evolution (Bin\,1), down to galaxies with low $M_{\text{HI}}$/$M_{\star}$ ratios, which have converted most of their gas into stars during the course of their lifetime (Bin\,6).

\subsection{DEUS modelling framework}
To interpret what drives the evolution of the stellar mass, metal mass, H{\sc{i}} gas, and dust content as galaxies evolve, we have used a Bayesian modelling framework to find the set of parameters capable of reproducing the observed scaling relations in the local Universe. To compare dust, H{\sc{i}} gas and metal scaling relations in the local Universe to model predictions, and infer what physical processes drive the observed trends and differences between galaxy populations, we have used a chemical evolution model that tracks the buildup and evolution of dust, gas and metals throughout the lifetime of a galaxy. More specifically, we employ Dust and Element evolUtion modelS (DEUS), which account for dust production by asymptotic giant branch (AGB) stars, supernova remnants (SNRs), grain growth in the interstellar medium, and dust destruction through astration and processing by supernova shocks. Our model implementation is largely founded upon chemical evolution models presented in the literature (e.g., \citealt{1998ApJ...501..643D,2003MNRAS.343..427M,2008A&A...479..669C,2014MNRAS.441.1040R}).  An earlier version of DEUS was introduced by \citet{2017MNRAS.465.3741D}. We extended DEUS to include dust destruction by supernova shocks and dust growth in the interstellar medium. We furthermore coupled DEUS to a Bayesian Markov Chain Monte Carlo (MCMC) algorithm to study the effects of varying dust production and destruction efficiencies and to infer the set of parameters that best describes the observed scaling relations in the local Universe. In contrast to previous models (e.g., \citealt{1997nceg.book.....P,1998ApJ...501..643D}), we have accounted for the lifetime of stars, and the replenishment of the interstellar medium with metals, dust and any remaining gas after stellar death, rather than resorting to the instantaneous recycling approximation for which the enrichment is assumed to occur at stellar birth. Appendix \ref{DEUS.sec} gives a detailed overview of the DEUS code, our assumed metal and dust yields, and prescriptions for grain growth and dust destruction by supernova shocks. For this current paper, we explore three different models: \begin{itemize}
    \item Model I assumes a closed-box and predicts the amount of dust and metals produced following the customised SFHs (see next paragraph) inferred for the six galaxy bins.
    \item Model II assumes a closed-box and adopts a fixed SFH shape for all six galaxy bins. More specifically, we have adopted a scaled version of the delayed SFH from \citet{2017MNRAS.471.1743D}.  
    \item Model III deviates from the closed-box assumption, and includes gas infall and outflows (see Appendix \ref{Sec_Infall_Outflow}), and furthermore relies on the customised SFHs inferred for each of the six galaxy bins.
\end{itemize}
The amount of metals and dust produced in galaxies sensitively depends on its (recent) star formation activity. Given that the six local galaxy samples correspond to different galaxy evolutionary stages, we expect them to have gone through different levels of recent star formation activity. To account for variations in their past and recent star formation activity, we have determined a customized SFH for each of the six galaxy bins by relying on their average stellar mass, specific star formation rate, and SFR(10\,Myr)-to-SFR(100\,Myr) ratio. The latter SFRs were inferred from hybrid SFR calibrators: H$\alpha$+WISE\,22\,$\mu$m for SFR(10\,Myr) and far-ultraviolet (FUV)+total-infrared (TIR) emission for SFR(100\,Myr). The customised SFHs are presented in Appendix \ref{Sec_SFH}, where it is demonstrated that galaxies at an early stage of evolution have formed most stars during recent epochs, as opposed to more evolved galaxies which show a clear drop in their recent star formation activity. 

Due to possible degeneracies between various dust production and destruction sources, we have coupled DEUS to a Bayesian MCMC method to effectively search a large parameter space and to constrain the relative importance of stellar dust production, grain growth and dust destruction by supernova shocks. Our Bayesian model has four free parameters: (1.) the initial gas mass, $M_{\text{gas,ini}}$; (2.) the fraction of supernova dust that is able to survive the reverse shock, $f_{\text{survival}}$; (3.) the grain growth parameter, $\epsilon$ (see Eq. \ref{Eq_graingrowth}); and (4.) the interstellar mass cleared by each single supernova event, $M_{\text{cl}}$ (see Eq. \ref{Eq_dustdestruction}), which is indicative of the dust destruction efficiency through supernova shocks. We leave the initial gas mass ($M_{\text{gas,ini}}$) of the halo as a free parameter in DEUS to infer what gas mass is needed to reproduce the observed present-day specific H{\sc{i}} gas masses ($M_{\text{HI}}$/$M_{\star}$) and oxygen abundances. The initial gas mass is degenerate with the mass loading factors of infalling and outflowing gas; we therefore constrain the initial gas mass in our models at fixed in- and outflow rates (or no gaseous flows in the case of Models I and II). In a similar way, variations in the initial gas mass are hard to differentiate from merger events occurring throughout a galaxy's lifetime. To constrain the free parameters in DEUS, we have compared the present-day model output to five observational quantities: $\log$\,$M_{\star}$, $\log$\,$M_{\text{dust}}$/$M_{\star}$, $\log$\,$M_{\text{HI}}$/$M_{\star}$, $\log$\,$M_{\text{dust}}$/$M_{\text{metals}}$(gas+dust) and 12+$\log$(O/H)\footnote{We have compared the median observed values to the model predictions at the end of our simulations at a galaxy age of 12\,Gyr (assuming that these galaxies started forming stars 12\,Gyr ago).}. 

As nothing much is known about the preferred values and their expected distribution, we have assumed flat priors to avoid biasing the model output results with $\log$($M_{\text{gas,ini}}$/$M_{\odot})$ varying between 8.5 and 11, $f_{\text{survival}}$ between 0.1 and 1.0, $\log(\epsilon)$ between 0.1 and 4.0, and $\log$($M_{\text{cl}}$/$M_{\odot})$ between 0.1 and 5.0. This four-dimensional parameter space was sampled with an affine invariant ensemble sampler \citep{2010CAMCS...5...65G} as implemented in the {\tt emcee} package for MCMC in \texttt{Python} \citep{2013PASP..125..306F}. We have used a collection of 100 walkers to sample the entire parameter space, where the position of a walker is changed at each step to explore the parameter space and to look for a region with high likelihood. We assumed a likelihood function based on the commonly used $\chi^{2}$ statistic: $\chi^{2}=\Sigma_{i=1}^{N}\left(\frac{f_{\text{i}}(obs)-f_{\text{i}}(model)}{\sigma_{\text{i}(obs)}}\right)^{2}$ with $f_{\text{i}}$(obs) and $f_{\text{i}}$(model) the observed and modelled values, respectively, and $\sigma_{\text{i}(obs)}$ the observational uncertainty, for constraint i, which is equivalent to a Gaussian likelihood. The positions of the 100 walkers are recorded at each time step after a warm-up phase of $N_{\text{burn}}$=500 steps, and the simulations are run for a total of $N_{\text{steps}}$=1,500 steps. The final 1,000 time steps are used to construct the posterior probability density functions (PDFs). We furthermore verified that these steps are sufficient for each of the model parameters to converge, which requires the effective sample size $N_{\text{eff}}$ (=$N_{\text{chain}}$/$\tau_{\text{int}}$ with $N_{\text{chain}}$ the length of the chain and $\tau_{\text{int}}$ the integrated autocorrelation time of the chain) to be higher than 10 for all parameters. As an additional check, we verified that the acceptance fraction of walkers ranges between 0.2 and 0.5.

\subsection{Modelling results}
The median parameter values inferred from the 1D posterior PDFs were tabulated in Table \ref{DEUSresults_bins} for the three different models. Figures \ref{Image_cornerplot_ModelI_bin1} through to \ref{Image_cornerplot_ModelIII_bin6} present the 1D and 2D posterior PDFs for the six galaxy bins and Models I, II, and III, respectively. The evolutionary tracks -- as determined from those median parameter values and spanning a time period of 12\,Gyr -- have furthermore been overlaid on the individual panels of Figure \ref{Scaling_MHIMstar_trends}. 

The stellar mass and metal abundance gradually increases for all models as galaxies evolve. For Model III (with gaseous in/outflows), the metallicity increase is less steep compared to Models I and II due to metal-enriched outflows. Due to this slow metal enrichment, Model III is able to reproduce the low specific gas masses observed for more evolved galaxies in Bins 5 and 6. The dust-to-metal ratio (i.e., the amount of metals depleted onto dust grains) starts off at a plateau around 40$\%$ in all models, indicative of dust being produced mainly by stars, and only a minor contribution from grain growth, in the early stages of galaxy evolution. After a few 100\,Myr, the metal abundance and dust mass has increased sufficiently for grain growth to kick in. However, the dust-to-metal ratio in our models first drops due to grain destruction (i.e., supernova shocks and astration) dominating over grain growth processes. For more evolved galaxies (Bins 5 and 6), the dust-to-metal ratio continues to increase due to grain growth becoming more dominant than these dust destruction mechanisms. Similar results have been inferred from galaxy simulations (e.g., \citealt{2017MNRAS.466..105A}). %A similar trend is seen for DEUS Models III, but does not appear adequate to explain the low dust-to-metal ratios observed in galaxies with high gas fractions (Bins 1-2). 
The dust-to-stellar mass ratio shows a similar trend with a nearly flat ratio at the start due to dust forming as stars evolve, progressing to a gradual increase (if grain growth starts to become important) or decrease (if dust destruction processes dominate).

In most cases, the present-day model values (indicated with asterisks, diamonds and squares for Models I, II and III, respectively, in Fig. \ref{Scaling_MHIMstar_trends}) are capable of reproducing the observed ratios in each bin within the error bars (reflecting the dispersion observed within each $M_{\text{HI}}$/$M_{\star}$ bin) which makes us confident that the models are adequate to reproduce the dust, metal and H{\sc{i}} gas scaling relations observed for the local Universe. There are however two notable exceptions. For evolved galaxy populations (Bins 5 and 6), Models I and II are not capable of reproducing their low observed specific H{\sc{i}} gas masses ($\log M_{\text{HI}}$/$M_{\star}\lesssim-1.0$). We believe this model discrepancy is driven by the closed-box assumption in Models I and II, as Model III is capable of reproducing the $M_{\text{HI}}$/$M_{\star}$ ratios and metal abundances for these more evolved galaxies better. Due to their decrease in recent ($\lesssim$100\,Myr) star formation activity, these galaxies are likely to have experienced some type of quenching during the last stages of their evolution. The assumption of a constant star formation rate on timescales $>$100\,Myr, with a sudden drop in their recent star formation activity might therefore not be fully representative if quenching timescales are longer. However, the rapid star formation quenching inferred for several HRS galaxies \citep{2016A&A...585A..43C} suggests that at least some galaxies experience a sudden drop in their SF activity on 100\,Myr timescales. A discrepancy is also observed for galaxies at an early stage of evolution (Bins 1 and 2), for which both closed-box models and models with gaseous flows underestimate the observed metal abundances (see bottom panels in Figure \ref{Scaling_MHIMstar_trends}). We speculate that these modelled low metal abundances might be compensated for by locking fewer metals into dust grains -- either through less efficient grain growth processes or more efficient grain destruction -- which will also bring the modelled dust-to-metal ratios closer to the observed values. Other than possible model discrepancies, we should note that the oxygen abundances are missing for several galaxies at the low end of the metallicity range, which will inevitably bias our average bin measurements upwards for these less evolved galaxies as the full dynamic range of metallicity values has not been covered. 

\subsection{Dust production and destruction efficiencies}
\label{DustProductionDestruction.ref}
In the rest of the paper, we focus our discussion on the dominant dust production and destruction mechanisms for the subsample of galaxies in Bins 3 and 4 with  $-1.0\lesssim \log M_{\text{HI}}/M_{\star}\lesssim0$, which constitute the majority (266/423 or 63$\%$) of the local galaxy population. Stochastic effects will not hamper the median values inferred for the galaxies in Bins 3 and 4 as is the case for poorly sampled galaxy bins at the low and high $M_{\text{HI}}$/$M_{\star}$ end. The stellar mass range (10$^{9}$-10$^{11}$\,M$_{\odot}$) covered by Bins 3 and 4 furthermore corresponds to galaxies in which an equilibrium is reached between gaseous infall, outflow and star formation \citep{2013MNRAS.433.1425B}. Such an equilibrium implies that the choice of specific gas infall/outflow rates and mass loading factors for these galaxies will have less of an impact on the output model parameters. The galaxies outside this stellar mass range instead show a large degree of scatter, and will be more sensitive to the effect of gas infall or outflows during recent times. We prefer to focus on these galaxies, for which the effect of gas infall or outflows during recent times has been less important, having sustained star formation over several Gyr (see Figure \ref{Image_SFH}). The closed-box Models I and II result in adequate fits (as quantified by the $\chi^{2}_{\text{red}}$ statistic, see Table \ref{DEUSresults_bins}) for these galaxies at an intermediate stage of evolution. We note that the conclusions for Models III (including gaseous in/outflows) generally remain unmodified, but these models typically give rise to larger model parameter uncertainties and less well constrained fits (see Table \ref{DEUSresults_bins}) due to the increased level of model complexity. Specifically, the oxygen abundance is severely underestimated due to the recent infall of pristine gas for galaxy Bins 1 through to 4, which results in higher $\chi^{2}_{\text{red}}$ values for Model III than for Models I and II. The specific prescription adopted here to model gaseous flows might not be appropriate for the entire range of galaxies in our sample, and there result in worse model fits to the data. The assumed infall and outflow rates in Model III fit the data well for galaxies in Bins 5 and 6, resulting in better fits than for Models I and II. %Not for Bins 5 and 6

\subsubsection{Initial gas mass} The initial gas masses are well determined showing peaked 1D posterior PDFs with values that gradually increase with the evolutionary stage of galaxies (see Table \ref{DEUSresults_bins}) in line with the expectation that galaxies at an advanced stage of evolution are more massive, and thus require a larger initial mass to convert gas into stars than less evolved galaxies. It should be noted that part of this trend might be driven by merger events leading to increased gas masses at specific times throughout a galaxy lifetime rather than increased initial gas masses. Since these merger events have not been considered here, the models may have converged to large initial gas masses to reproduce present-day scaling laws for more evolved galaxies. The initial gas masses might be one of the most important parameters in DEUS as they directly influence the present-day model stellar masses and metal abundances, and play an important role in setting the posterior PDFs obtained for the other parameters. In future work, we intend to explore the importance of the initial gas mass parameter (and possible degeneracies with gaseous in- and outflows and merger events) in more detail. 

\subsubsection{Net supernova dust production rates} 
Models I and II suggest that a significant fraction (37 to 89$\%$) of freshly condensed supernova dust is able to survive the reverse shock. Dust evolution models that include the effects of sputtering and/or shattering on supernova dust grains due to the passage of a reverse shock estimate dust survival rates ranging from 1 to 100$\%$ (e.g., \citealt{2007MNRAS.378..973B,2007ApJ...666..955N,2008ApJ...682.1055N,2010ApJ...715.1575S,2015A&A...575A..95S,2016A&A...589A.132B,2016A&A...587A.157B,2016A&A...590A..65M,2019MNRAS.489.4465K}). An easy comparison between these various models is hampered by the different assumptions made to describe the ambient densities, the density contrast between dust clumps and the surrounding medium, the grain size distribution and the composition of supernova dust species. In addition, our inferred dust survival rate will account for the fact that some supernova remnants will not experience a reverse shock (e.g., the Crab Nebula) due to the low density of the surrounding medium, and should thus be considered as an ``effective" dust survival rate as it is convoluted with the probability that a reverse shock will be generated through the interaction with a dense circum- or interstellar medium, and that dust might be able to reform after the shock passage (e.g., \citealt{2019MNRAS.482.1715M}). Current observational studies tend to be biased towards interacting supernova remnants or pulsar wind nebulae which provide a heating mechanism through shock interaction or through the presence of a pulsar, respectively. It is therefore hard to estimate the fraction of SNRs that will experience a reverse shock, and at what average velocity the reverse shock will interact with the ejecta. Moreover, a non-negligible fraction of core-collapse supernovae occur ``late" (i.e., 50-200\,Myrs after birth) due to binary interactions \citep{2017A&A...601A..29Z}. On such long timescales, the birth clouds of these massive stars will have dissolved, and it will become less likely that a reverse shock is generated.

Our high dust survival fractions are in excellent agreement with recent observational constraints. Elevated dust-to-gas ratios in the shocked ejecta clumps of the Galactic supernova remnant Cassiopeia\,A suggest that a significant fraction of supernova dust is capable of surviving a reverse shock \citep{2019MNRAS.485..440P}. Several studies (e.g., \citealt{2013ApJ...774....8T,2014Natur.511..326G,2015MNRAS.446.2089W,2016MNRAS.456.1269B,2020MNRAS.491.6020P}) have also argued for rather large supernova grain sizes ($\gtrsim$0.1\,$\mu$m), which lends support to the idea that significant fractions of supernova dust are able to survive a reverse shock (with large grains being more resilient to sputtering, e.g., \citealt{2010ApJ...715.1575S}).

\subsubsection{Grain growth timescales}
The grain growth parameter has been parameterised through $\epsilon$ following \citet{2012MNRAS.423...26M} (see Eq. \ref{Eq_graingrowth} and Appendix \ref{GrainGrowth.sec} for an outline of its derivation). At a fixed gas mass, dust-to-gas ratio, metal fraction and star formation rate, the grain growth parameter $\epsilon$ is inversely proportional to the grain growth timescale, and can be considered to approximate the efficiency of grain growth processes. More specifically, large values of $\epsilon$ correspond to efficient grain growth and thus short grain growth timescales $\tau_{\text{grow}}$, while small $\epsilon$ values are indicative of long $\tau_{\text{grow}}$\footnote{Values of $\epsilon$ of 10-100 typically correspond to $\tau_{\text{grow}}>100$\,Myr, while $\epsilon\gtrsim1000$ is needed to reach down to $\tau_{\text{grow}}$ of 10\,Myr and lower (depending on the assumed SFR, gas, dust and metal mass).}.

The 1D posterior PDFs for Models I and II, and Bins 3 and 4, have $\log \epsilon$ peaking around 2.0, with a wide tail of high-likelihood models extending to lower $\log \epsilon$ values and a sudden drop in likelihood beyond values of $\log\,\epsilon\gtrsim$2.0. The 1D posterior PDFs for Model III peaks at higher values (see Figures \ref{Image_cornerplot_ModelIII_bin3} and \ref{Image_cornerplot_ModelIII}) than for closed-box models, which is not surprising given that dust and metals will be expelled from the galaxy, and thus an additional source of dust production is required in Model III. A narrow range of models with $\epsilon$ values higher than this peak seems also capable of explaining our observed scaling relations, but only if such high grain growth efficiencies are exactly compensated for by high dust destruction efficiencies (e.g., the rightmost 2D contour plot on the bottom row of Figures \ref{Image_cornerplot_ModelI}, \ref{Image_cornerplot_ModelII} and \ref{Image_cornerplot_ModelIII}). With grain growth locking refractory elements into grains, and dust destruction releasing these same elements back into the gas phase, our observational constraints are not capable of distinguishing between both mechanisms. The model prescriptions to describe grain growth and dust destruction efficiencies furthermore depend directly (or indirectly in case of the supernova rate) on the current SFR, which causes this degeneracy between grain growth and dust destruction efficiencies, as long as both processes cancel each other out. To adapt those recipes, we require improved understanding of the grain growth and destruction processes in the interstellar medium. The 2D contour plots suggest that the models are also hampered by degeneracies between the supernova dust survival rate ($f_{\text{survival}}$), the grain growth parameter ($\epsilon$) and the dust destruction efficiency ($M_{\text{cl}}$) in some parts of the 4D parameter space, which results in wide 1D posterior PDFs. 

Our median values of $\log\,\epsilon$=1.5$^{+1.1}_{-0.9}$ -- equivalent to present-day growth timescales $\tau_{\text{growth}}$ of $\gtrsim$\,100\,Myr, with a median of 400\,Myr -- are consistent with the range of values ($\epsilon \in$[10,457]) inferred by \citet{2012MNRAS.423...38M} based on the resolved dust-to-metal gradients observed in a sample of 15 SINGS galaxies, and the accretion timescales ($\tau_{\text{grow}}$=20-200\,Myr, or $\epsilon$=500) that were found adequate to reproduce the dust masses in a sample of high-redshift ($z$>1) submillimetre galaxies \citep{2014MNRAS.441.1040R}. In general, however, our grain growth efficiencies are significantly lower than many other studies. In \citet{2013EP&S...65..213A}, \citet{2014A&A...562A..76Z}, \citet{2015MNRAS.451L..70M} and \citet{2016MNRAS.457.1842S}, fast grain growth timescales of 0.2-2\,Myr have been assumed which causes grain growth to dominate dust production as soon as a critical metallicity threshold has been reached. \citet{2015MNRAS.449.3274F} required similarly short accretion timescales (5\,Myr) to reproduce the dust and metal masses in low-metallicity dwarf galaxies. Also the dust, metal and gas scaling relations for a sample of nearby galaxies were found to be best reproduced by chemical evolution models with $\epsilon$ values of 2500-4000 \citep{2017MNRAS.471.1743D}\footnote{The $\epsilon$ values (5000-8000) from \citet{2017MNRAS.471.1743D} have been corrected to account for their assumed cold gas fraction ($f_{\text{c}}$=0.5) to allow for a direct comparison with our values.}. All of these studies suggest that grain growth dominates dust production for different galaxy populations across a wide range of different redshifts (see also Section \ref{DustSources.sec}). 

Even though recent laboratory studies suggest that SiO$_{\text{x}}$ and more complex silicate-type grains can form without an activation energy barrier under typical molecular cloud ($T_{\text{dust}}$=10-12\,K) conditions \citep{2014ApJ...782...15K,2015arXiv150200388R,2018IAUS..332..312H}, it might be hard for the majority of dust grains in the low-redshift Universe to have formed through accretion of elements onto pre-existing grain seeds given the low accretion rates and the Coulomb barrier that needs to be overcome in diffuse gas clouds, and the efficient formation of ice mantles which prevents efficient grain growth in dense molecular clouds \citep{1978MNRAS.183..417B,2016MNRAS.463L.112F,2018MNRAS.476.1371C}. \citet{2016ApJ...831..147Z} modelled the formation of silicate grains through the accretion of elements in diffuse gas clouds (with gas densities $n_{\text{H}}$ between 5 and 50 cm$^{-3}$) on average timescales of 350\,Myr, while \citet{2018ApJ...857...94Z} suggest that iron grains can grow efficiently in the cold neutral medium on timescales $\lesssim$10\,Myr. Due to the absence of laboratory measurements of diffusion and desorption energies, the latter works assumed that elements sticking to grain surfaces, will have sufficient time to reach a strong active bonding site where these refractory elements can be chemisorbed. Given that the exposure to strong UV radiation in diffuse gas clouds will make these elements prone to photo-desorption processes, and various elements on the grain surface (with differing diffusion energies) might be competing for the same dangling bonds, we argue that a detailed set of laboratory studies, combined with detailed chemical modelling, is needed to verify what kind of grain species can form and what timescales are involved in their formation. We speculate that our longer grain growth timescales (and longer dust lifetimes, see Section \ref{DustDestruction.sec}) might reduce the tension with grain surface chemical models which have so far been incapable of proposing a viable chemical route for grain growth. 

\subsubsection{Dust destruction efficiencies}
\label{DustDestruction.sec}
The dust destruction efficiency has been parameterised through the interstellar mass that is cleared per single supernova event ($M_{\text{cl}}$). In reality, it is unlikely that a single value will apply to all supernova events as $M_{\text{cl}}$ will depend on the ambient density, on the 3D structure of the ambient medium and on the supernova explosion energy. With several models assuming a single value for $M_{\text{cl}}$, we pursue to infer what average values are adequate to reproduce the observed scaling laws in the local Universe. Similar to the grain growth parameter, the 1D posterior PDF for $M_{\text{cl}}$ shows a sharp drop in likelihood beyond $M_{\text{cl}}\gtrsim$10$^{2.4}$\,M$_{\odot}$. Higher values are only allowed in case the dust destruction efficiency is perfectly balanced by the same level of dust production through grain growth. The models are incapable of distinguishing between values of $M_{\text{cl}}$ below this threshold due to degeneracies with the level of supernova dust production and the grain growth parameter.

The peaks in the 1D posterior PDFs occur at low $M_{\text{cl}}$ values, resulting in median values of $M_{\text{cl}}$=10$^{1.4-1.6}$ for galaxies in Bins\,3 and 4, and correspond to long dust lifetimes of 1 to 2\,Gyr. The upper limits in our models for the mass cleared per supernova event ($\lesssim$400\,M$_{\odot}$) are consistent with current dust destruction timescales $\gtrsim$200\,Myr. Our preferred model dust lifetimes of a few Gyr are consistent with the longer dust destruction timescales (2-3\,Gyr) inferred for silicate grains by \citet{2015ApJ...803....7S} by means of supernova remnant models with evolving shock waves. Long dust lifetimes (on the order of a few Gyr) for silicate grains were also suggested by \citet{2011A&A...530A..44J} after accounting for the 3D distribution of interstellar material, while carbonaceous grains are assumed to be processed on short timescales. Our conclusion applies to the ensemble of interstellar grains, and is in agreement with these longer silicate lifetimes. In future work, we hope to make the distinction in our models between the formation and destruction of various grain species such as carbonaceous and silicate dust grains. It is worth noting that our inferred dust destruction timescales are factors of a few longer than the average values reported by other works (e.g., 400-600\,Myr, \citealt{1994ApJ...433..797J,1996ApJ...469..740J}; $<$90\,Myr, \citealt{2014MNRAS.441.1040R}; 20-70\,Myr for dust in the Magellanic clouds \citealt{2015ApJ...799..158T,2015ApJ...799...50L}; 350\,Myr, \citealt{2016ApJ...831..147Z,2018ApJ...857...94Z,2019MNRAS.487.3252H}). 

\subsection{Dominant dust production sources}
\label{DustSources.sec}
\begin{figure}
	\includegraphics[width=8.5cm]{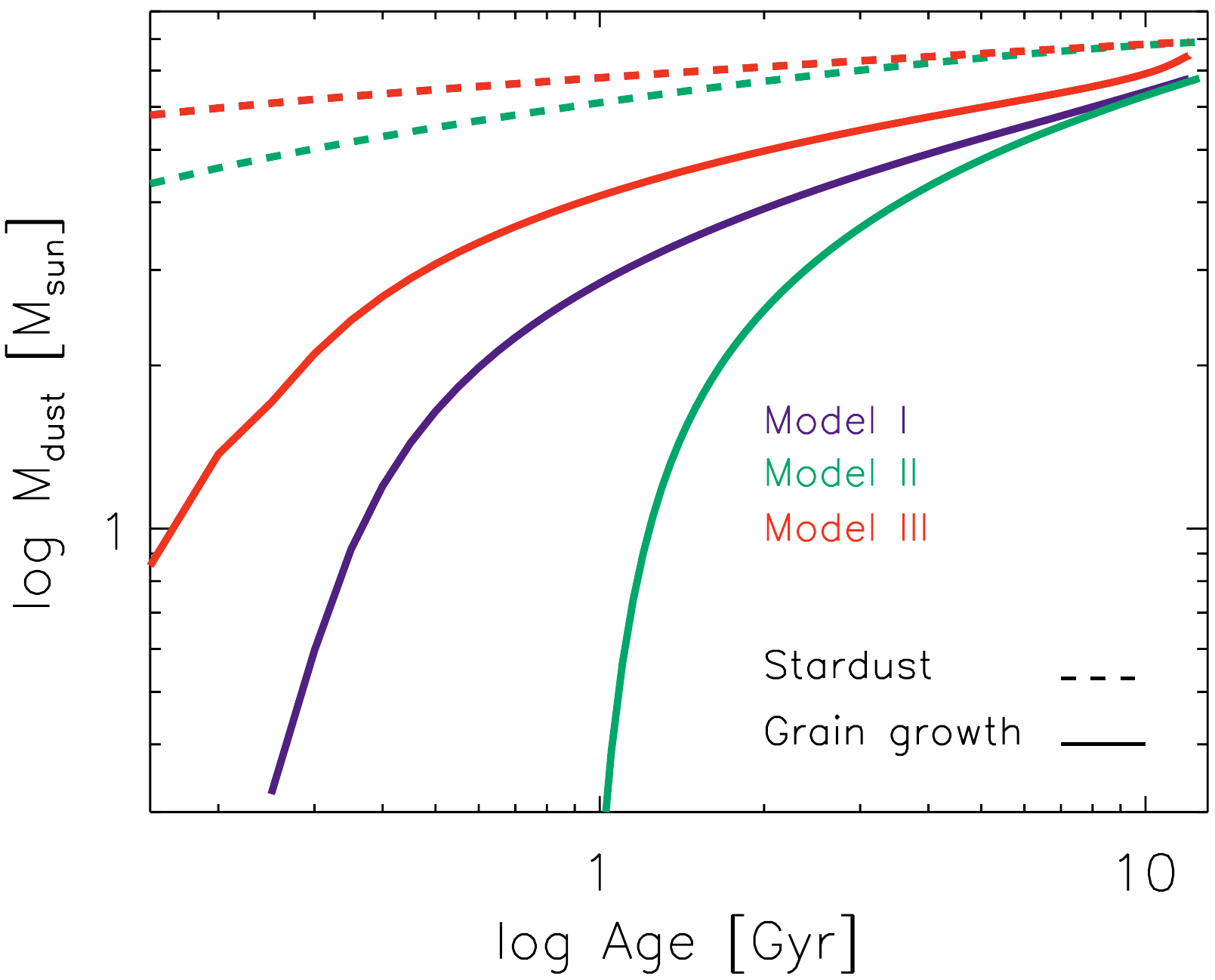} 
   \caption{The amount of dust produced through stellar sources (AGB+SNe, dashed curves) and through interstellar grain growth (solid curves) as a function of galaxy age, as inferred from the median model parameter values for Bin 4. The stardust tracks for Model I and Model III overlap due to the same assumed SFH shape for both models.}
    \label{Scaling_DustFraction}
\end{figure}

Our thorough search of the four-dimensional parameter space, adapted to cover a wide range of different Dust and Element evolUtion modelS, has revealed that local galaxy scaling relations (with the exception of galaxies with low and high specific gas masses) can be reproduced adequately by models with long dust survival rates (on the order of 1-2\,Gyr), low grain growth efficiencies ($\epsilon\sim$30-40) and a predominant contribution of stellar dust production sources to account for the present-day galaxy dust budgets. More specifically, we estimate that most of the dust ($>$90\,$\%$) is produced through stellar sources over a galaxy's lifetime, with a minor contribution from grain growth ($<$10\,$\%$, see Fig. \ref{Scaling_DustFraction}). The contribution of grain growth increases with time for all models, with 50 to 80$\%$ of present-day dust masses resulting from stellar dust, while 20 to 50$\%$ of the dust is suggested to grow through the accretion in interstellar clouds. Models with in- and outflows (Model III) have an increased contribution from grain growth, resulting in more or less equal contributions from grain growth and stellar sources to the dust production over a galaxy lifetime. Given that a fraction of the dust is lost in galactic outflows (i.e., scaled with the dust-to-gas ratio of the galaxy at that point in time), we require more dust production through grain growth to reproduce the observed dust-to-stellar and dust-to-metal mass ratios with Models III. We furthermore note a trend of high relative fractions of stardust for less evolved galaxies (Bins 1 and 2), which is not surprising given the low metal abundances (and hence low grain growth efficiencies) for these galaxies. 

We speculate that through performing a rigorous search of the four-dimensional parameter space, our results provide an alternative for the chemical evolution models with extremely low supernova dust production efficiencies and short grain growth timescales ($\lesssim$a few Myr), which have been invoked to explain the dust, metal and gas scaling laws of local galaxies (e.g. \citealt{2014A&A...562A..76Z,2015MNRAS.449.3274F,2017MNRAS.471.1743D}).  Regardless of our model assumptions on the SFH and gaseous flows, the local dust, H{\sc{i}} gas and metal scaling relations are reproduced well with models that assume long dust lifetimes (1-2\,Gyr), favourable supernova dust injection rates ($f_{\text{survival}}$ of 37-89$\%$) and low grain growth efficiencies ($\epsilon$ of 30-40). These long grain growth timescales could reduce the tension between the high grain growth efficiencies (required to reproduce the large dust masses observed in low-to high-redshift galaxies) and grain surface chemical models, which currently fail to account for efficient grain growth processes in the interstellar medium (e.g., \citealt{1978MNRAS.183..417B,2016MNRAS.463L.112F,2018MNRAS.476.1371C,2019A&A...627A..38J}). 

\subsection{Modelling caveats}
\label{Caveats.sec}
We resorted to making some assumptions in DEUS to avoid introducing various model degeneracies. We briefly discuss the implications of these assumptions.
\begin{itemize}
\item \textbf{Star formation history:} we have assumed a customised SFH for each galaxy bin (i.e., Models I). To test the importance of this model assumption, we also ran models with a delayed star formation history (i.e., Models II). A quick comparison between the inferred model parameters shows that the dependence on the specific shape of the SFH is minimal based on the close resemblance between the Model I and II output parameters (see Table \ref{DEUSresults_bins}). We argue that the minor importance of the specific SFH shape results from the long dust lifetimes, which imply that the current dust reservoir has been built up during the last 1-2\,Gyr, and that variations in the SFH shape on these timescales are less relevant as long as the final produced dust mass remains the same. We should also note that the simplicity of the SFH shapes, and other model assumptions may affect the dependence of the results on the SFH.

\item \textbf{Closed-box vs gaseous flows:} even though the importance of gaseous flows is now well established in the field, the precise nature of these gas-regulated ``bathtub" galaxies still requires further characterisation. In Model III, we have assumed that the infalling gas is pristine (i.e., the gas is not enriched with metals or dust), while the outflowing gas has the same gas-phase metallicity and dust-to-gas ratio as the galaxy at the time of the outflow. This assumption will vary depending on the outflow mechanism and the location of the onset of these gaseous outflows. The outflow rate is often assumed to scale with the SFR, but a time-dependent outflow model with strong outflows at early times has been shown adequate to reproduce the observed gas and stellar metallicities in galaxies \citep{2018MNRAS.474.1143L}. Similar time- or stellar mass-dependent outflows are also consistent with galaxy simulations (e.g., \citealt{2015MNRAS.454.2691M,2017MNRAS.465.1682H}). We suggest that these strong outflows at early times (as implemented in our Model III), result in a slow build-up of a galaxy's metal content, which reduces the efficiency of grain growth processes at early times. Low outflow rates at the present epoch also reduce the need for short grain growth timescales to account for the observed dust masses in galaxies. But, as remarked upon before, different assumptions on the time-dependence of these outflows will affect dust production and destruction efficiencies. To limit these biases, we have assumed closed-boxes for Models I and II to model galaxies which have reached an equilibrium between gaseous infall, outflow and star formation. % while the infall of metal-poor gas plays an important role in sustaining a galaxy's star formation activity (ref), and gas outflows are known to link to the observed relation between the stellar mass, metallicity and star formation rate in galaxies (ref)

\item \textbf{One-zone models:} outflow rates are thought to vary with radial distance from the galaxy centre (e.g., \citealt{2018MNRAS.476.3883L,2019MNRAS.tmp.1173B,2019MNRAS.483.4968V}), which would require resolved galaxy models to take this into consideration. Other than these spatial variations in mass loading factors, the 3D structure and filling factors of various ISM phases that together constitute an entire galaxy, will vary depending on the evolutionary stage and the specific type of galaxies under consideration. Dwarf galaxies in the nearby Universe provide an excellent example of how their low metal and dust content, high degree of porosity and radiation field hardness severely affects their ISM build-up with a highly-ionised, diffuse medium that dominates the ISM volume, and only minor contributions from compact phases (e.g. \citealt{2012A&A...548A..91L,2015A&A...578A..53C,2019IAUS..344..240M,2019A&A...626A..23C}). The detection of highly ionised nebular lines (e.g., \citealt{2014ApJ...784...58S,2016Sci...352.1559I,2017A&A...605A..42C,2017ApJ...837L..21L,2018Natur.557..392H}) suggests that high-redshift galaxies might have an ISM build-up similar to low-metallicity dwarfs in the nearby Universe, and supports the need for spatially resolved chemical evolution modelling to account for radially-dependent gas in/outflows and filling factors of different ISM phases (e.g. \citealt{2017MNRAS.467.4322P}). In the future, we plan to expand DEUS to include a realistic 3D ISM structure to model how the density and temperature distributions of the total ensemble of gas clouds in a galaxy varies with time. %and the residence time of dust grains in specific ISM phases 

\item \textbf{Metal and dust yields:} we had to assume a set of AGB and supernova metal and dust yields, and apply specific prescriptions to describe the efficiency of grain growth and dust destruction processes. We endeavoured to select yields and recipes that correspond to the current state-of-the-art, but these prescriptions remain limited by our current knowledge on how grains are destroyed and whether or not grains can grow either in diffuse or dense clouds of the interstellar medium. If the true yields were to differ significantly from our model assumptions and/or show variations with metallicity (e.g., \citealt{2009MNRAS.397.1661V,2019ApJ...879..109B,2019MNRAS.486.4738D}), this could impact our inferred model parameters. In De Vis in prep., the choice of metal yields is shown to mostly impact the metallicities of galaxies with high specific gas masses. %It is however beyond the scope of this current work to study the effect of altering the assumed metal and dust yields. 

\item \textbf{Time dependence:} we have not accounted for variations in the dust destruction efficiency and grain growth parameter in time, which could be induced if strong variations in the grain size distribution occur throughout a galaxy's lifetime, as the efficiency of grain destruction and grain growth is strongly grain size-dependent (e.g. \citealt{2015MNRAS.447.2937H}). 

\item \textbf{Initial mass function (IMF):}
We have furthermore assumed a fixed \citet{2003PASP..115..763C} IMF. The shape of the IMF has been suggested to vary in different environments (e.g. \citealt{2018MNRAS.474.4169O}), and deviations from this standard IMF will affect the dust and metal yields, and supernova rates in DEUS. 
\end{itemize}
In future work, we aim to explore the effects of varying the IMF and applying different sets of metal and dust stellar yields, to accommodate physically-motivated recipes to describe grain growth and dust destruction processes, and to allow for spatial variations in the efficiencies of these processes with local ISM conditions.

\begin{table*}
%\centering
\caption{The bin number (1st column), $\log$($M_{\text{HI}}$/$M_{\star}$) range (2nd column), sample size (3rd column) and median properties (columns 4-9) inferred for each of the six galaxy bins. These galaxy bins were selected to include less evolved galaxies (Bin\,1) while gradually moving to galaxies with the most advanced stages of evolution (Bin\,6).}
\label{MHIMstar_bins}
\begin{tabular}{lllcccccc} % four columns, alignment for each
\hline
Bin & Range & $N_{\text{gal}}$ & $\log$($M_{\star}$/M$_{\odot}$) & $\log$(sSFR/yr$^{-1}$) & $\log$($M_{\text{HI}}$/$M_{\star}$) & $\log$($M_{\text{dust}}$/$M_{\star}$) & $\log$($M_{\text{dust}}$/$M_{\text{metals}}$) & Metallicity \\ 
\hline
1 & [0.5,-] & 17 & 8.65$\pm$0.79 & -9.36$\pm$0.28 & 0.75$\pm$0.19 & -2.60$\pm$0.65 & -1.12$\pm$0.46 & 8.41$\pm$0.22 \\
2 & [0,0.5[ & 81 & 9.33$\pm$0.44 & -9.62$\pm$0.29 & 0.19$\pm$0.15 & -2.47$\pm$0.29 & -0.86$\pm$0.23 & 8.53$\pm$0.19 \\
3 & [-0.5,0[ & 134 & 9.73$\pm$0.53 & -9.98$\pm$0.35 & -0.26$\pm$0.15 & -2.66$\pm$0.23 & -0.70$\pm$0.19 & 8.67$\pm$0.15 \\
4 & [-1.0,-0.5[ & 132 & 10.19$\pm$0.59 & -10.21$\pm$0.34 & -0.72$\pm$0.14 & -2.86$\pm$0.21 & -0.59$\pm$0.15 & 8.72$\pm$0.11 \\
5 & [-1.5,-1.0[ & 46 & 10.40$\pm$0.38 & -10.57$\pm$0.63 & -1.18$\pm$0.14 & -3.07$\pm$0.30 & -0.41$\pm$0.12 & 8.77$\pm$0.10 \\
6 & [-,-1.5[ & 13 & 10.36$\pm$0.39 & -11.30$\pm$0.91 & -1.72$\pm$0.35 & -3.62$\pm$0.52 & -0.21$\pm$0.15 & 8.79$\pm$0.09 \\
\hline 
\end{tabular}
\end{table*}

\begin{table*}
%\centering
\caption{The median values for the four DEUS parameters as inferred from three different models for each of the six galaxy bins. Models I and II correspond to closed-box models with optimised non-parametric SFHs and with a delayed SFH (with fixed shape), respectively (see Appendix \ref{Sec_SFH}). Model III includes gaseous in- and outflows (see Appendix \ref{Sec_Infall_Outflow}) and the specific set of non-parametric SFHs. The upper and lower limits on the model parameters have been inferred from the posterior PDFs as the 16th and 84th percentiles. In addition to these output model parameters, we calculated the reduced $\chi^{2}_{red}$ statistic by comparing the observed values with the model predictions for the median parameters. We also inferred the fraction of dust produced through stellar sources ($f_{\text{stardust}}$) and through accretion processes in the ISM ($f_{\text{grain growth}}$) throughout the galaxy lifetime, and at the current age of the galaxy (values between square brackets).}% based on DEUS model with the median likelihood values for each of the four parameters The upper and lower limits on these dust fractions result from the uncertainties on the $f_{\text{survival}}$ and $\log\,\epsilon$ parameters (at fixed values for $M_{\text{gas,ini}}$ and $\log\,M_{\text{cl}}$). }
\label{DEUSresults_bins}
\begin{tabular}{lcccccc} % four columns, alignment for each
\hline
Bin & Bin\,1 & Bin\,2 & Bin\,3 & Bin\,4 & Bin\,5 & Bin\,6 \\
\hline
\hline
Model parameters & \multicolumn{6}{c}{Model I. ``Closed-box" model with specific non-parametric SFHs} \\
\hline
\hline
$\log$($M_{\text{gas,ini}}$/M$_{\odot}$) & 9.33$^{+0.11}_{-0.11}$ & 9.77$^{+0.08}_{-0.07}$ & 9.99$^{+0.05}_{-0.04}$ & 10.36$^{+0.03}_{-0.03}$ & 10.51$^{+0.02}_{-0.02}$ & 10.48$^{+0.03}_{-0.03}$ \\
$f_{\text{survival}}$ [$\%$] & 51$^{+33}_{-21}$ & 60$^{+26}_{-27}$ & 63$^{+24}_{-25}$ & 65$^{+22}_{-27}$ & 69$^{+23}_{-35}$ & 61$^{+27}_{-34}$ \\
$\log\epsilon$ & 2.0$^{+1.2}_{-1.3}$ & 1.8$^{+1.2}_{-1.1}$ & 1.6$^{+1.0}_{-1.0}$ & 1.5$^{+0.8}_{-0.9}$ & 2.2$^{+1.0}_{-0.7}$ & 3.0$^{+0.7}_{-0.8}$ \\
$\log$($M_{\text{cl}}$/M$_{\odot}$) & 2.2$^{+1.2}_{-1.4}$ & 1.6$^{+1.1}_{-1.0}$ & 1.6$^{+1.0}_{-1.0}$ & 1.4$^{+1.0}_{-0.9}$ & 1.5$^{+1.7}_{-1.0}$ & 1.8$^{+1.3}_{-1.0}$ \\
\hline
$\chi^{2}_{\text{red}}$ & 9.6 & 5.0 & 1.2 & 1.6 & 11.0 & 15.5 \\
\hline 
$f_{\text{stardust}}$ [$\%$] & 99 [91$^{}_{}$] & 98 [84$^{}_{}$] & 96 [69$^{}_{}$] & 93 [52$^{}_{}$] & 55 [8$^{}_{}$] & 30 [16$^{}_{}$] \\
$f_{\text{grain growth}}$ [$\%$] & 1 [9$^{}_{}$] & 2 [16$^{}_{}$] & 4 [31$^{}_{}$] & 7 [48$^{}_{}$] & 45 [92$^{}_{}$] & 70 [84$^{}_{}$] \\
\hline
\hline
Model parameters & \multicolumn{6}{c}{Model II. ``Closed-box" model with delayed SFHs} \\
\hline
\hline
$\log$($M_{\text{gas,ini}}$/M$_{\odot}$) & 9.33$^{+0.11}_{-0.10}$ & 9.77$^{+0.08}_{-0.07}$ & 9.99$^{+0.05}_{-0.05}$ & 10.35$^{+0.03}_{-0.03}$ & 10.50$^{+0.02}_{-0.02}$ & 10.48$^{+0.03}_{-0.03}$ \\
$f_{\text{survival}}$ [$\%$] & 50$^{+32}_{-27}$ & 59$^{+27}_{-26}$ & 63$^{+24}_{-26}$ & 68$^{+21}_{-27}$ & 66$^{+24}_{-35}$ & 60$^{+28}_{-33}$ \\
$\log\epsilon$ & 2.0$^{+1.3}_{-1.3}$ & 1.7$^{+1.0}_{-1.1}$ & 1.5$^{+1.0}_{-0.9}$ & 1.5$^{+0.8}_{-0.9}$ & 2.1$^{+1.0}_{-0.7}$ & 3.0$^{+0.7}_{-0.7}$ \\
$\log$($M_{\text{cl}}$/M$_{\odot}$) & 2.0$^{+1.3}_{-1.3}$ & 1.7$^{+1.0}_{-1.1}$ & 1.5$^{+1.0}_{-0.9}$ &  1.4$^{+0.9}_{-0.9}$ & 1.5$^{+1.6}_{-1.0}$ & 1.8$^{+1.4}_{-1.1}$  \\
\hline
$\chi^{2}_{\text{red}}$ & 9.4 & 4.9 & 1.2 & 1.6 & 10.9 & 15.3 \\
\hline 
$f_{\text{stardust}}$ [$\%$] & 99 [96$^{}_{}$] & 98 [91$^{}_{}$] & 96 [81$^{}_{}$] & 93 [66$^{}_{}$] & 57 [13$^{}_{}$] & 30 [11$^{}_{}$] \\
$f_{\text{grain growth}}$ [$\%$] & 1 [4$^{}_{}$] & 2 [9$^{}_{}$] & 4 [19$^{}_{}$] & 7 [34$^{}_{}$] & 43 [87$^{}_{}$] & 70 [89$^{}_{}$] \\
\hline
\hline
Model parameters & \multicolumn{6}{c}{Model III. Model with gas in/outflows and specific non-parametric SFHs} \\
\hline
\hline
$\log$($M_{\text{gas,ini}}$/M$_{\odot}$) & 9.43$^{+0.13}_{-0.12}$ & 9.79$^{+0.10}_{-0.10}$ & 9.84$^{+0.04}_{-0.02}$ & 10.27$^{+0.01}_{-0.01}$ & 10.45$^{+0.01}_{-0.01}$ & 10.41$^{+0.01}_{-0.01}$ \\
$f_{\text{survival}}$ [$\%$] & 51$^{+32}_{-31}$ & 66$^{+24}_{-37}$ & 57$^{+31}_{-38}$ & 66$^{+24}_{-38}$ & 56$^{+31}_{-36}$ & 34$^{+32}_{-36}$ \\
$\log\epsilon$ & 2.8$^{+1.8}_{-1.9}$ & 4.0$^{+1.3}_{-2.5}$ & 3.7$^{+0.7}_{-0.8}$ & 2.7$^{+1.0}_{-1.0}$ & 2.9$^{+0.8}_{-0.6}$ & 3.2$^{+0.6}_{-0.6}$ \\
$\log$($M_{\text{cl}}$/M$_{\odot}$) & 2.1$^{+1.5}_{-1.4}$ & 1.9$^{+1.3}_{-1.2}$ & 1.8$^{+1.4}_{-1.2}$ & 1.6$^{+1.5}_{-1.0}$ & 1.7$^{+1.4}_{-1.1}$ & 1.7$^{+1.2}_{-1.1}$ \\
\hline 
$\chi^{2}_{\text{red}}$ & 26.4 & 30.3 & 23.9 & 7.0 & 0.8 & 5.8 \\
\hline 
$f_{\text{stardust}}$ [$\%$] & 99 [95$^{}_{}$] & 71 [33$^{}_{}$] & 54 [12$^{}_{}$] & 73 [10$^{}_{}$] & 57 [8$^{}_{}$] & 49 [17$^{}_{}$] \\
$f_{\text{grain growth}}$ [$\%$] & 1 [5$^{}_{}$] & 29 [67$^{}_{}$] & 46 [88$^{}_{}$] & 27 [90$^{}_{}$] & 43 [92$^{}_{}$] & 51 [83$^{}_{}$] \\
\hline 
\end{tabular}
\end{table*}

\section{Conclusions}
\label{Conclusions.sec}
%Rigorous search of parameter space can be useful!!!
We analysed local dust, H{\sc{i}} gas and metal scaling relations for a diverse sample of 423 nearby galaxies to infer that: 
\begin{itemize}
\item the specific dust and H{\sc{i}} gas masses are tightly linked to a galaxy's specific star formation rate (sSFR), which suggests that the interstellar mass (either traced through H{\sc{i}} gas or dust) plays an important role in setting a galaxy's SFR (through the Kennicutt-Schmidt law).  
\item the H{\sc{i}} gas scaling laws show the largest degree of dispersion, which suggests that variations in $M_{\text{dust}}$/$M_{\star}$, $M_{\text{dust}}$/$M_{\text{metals}}$(gas+dust) and $M_{\text{dust}}$/$M_{\text{HI}}$ ratios are not necessarily influenced by dust production and destruction mechanisms but rather driven by the current H{\sc{i}} gas reservoirs of galaxies. %in dust fractions and dust-to-H{\sc{i}} ratios are predominantly driven by variations in the gas reservoirs available in galaxies. 
\item the strong correlations between $M_{\text{HI}}$/$M_{\star}$, and $M_{\text{dust}}$/$M_{\star}$, $M_{\text{dust}}$/$M_{\text{metals}}$(gas+dust) and $M_{\text{dust}}$/$M_{\text{HI}}$ reinforce the idea that the specific H{\sc{i}} gas mass ($M_{\text{HI}}$/$M_{\star}$) plays an important role in setting the dust and metal content of galaxies.
\item the $M_{\text{dust}}$/$M_{\text{metals}}$(gas+dust) ratio in galaxies is nearly constant ($10^{-0.66\pm0.24}$) across our sample of galaxies for $M_{\star}\ge10^{9}\,M_{\odot}$. Weak (but significant) trends with $M_{\star}$, sSFR and $M_{\text{HI}}$/$M_{\star}$ support a scenario of increasing $M_{\text{dust}}$/$M_{\text{metals}}$(gas+dust) ratios as a galaxy evolves.%, excluding H{\sc{i}}-deficient galaxies with significantly higher DTM ratios ($10^{-0.44\pm0.08}$).
\item the large spread (0.34\,dex) in the $M_{\text{dust}}$/$M_{\text{HI}}$ ratio at a given metallicity should urge caution: total gas masses inferred from dust mass measurements will be uncertain by a factor of $\gtrsim2$ due to variance -- driven by intrinsic galaxy variations, the unknown extent of H{\sc{i}} reservoirs and/or uncertain oxygen abundances -- at a fixed metallicity. %the $M_{\text{dust}}$/$M_{\text{HI}}$ ratio shows a super-linear trend (slope=2.26$\pm$0.07) with the oxygen abundance.  %Part of this spread appears to be driven by intrinsic galaxy variations. Additional uncertainties could arise from the extent of their H{\sc{i}} reservoir. %Any secondary dependencies on the metallicity calibrations could not be accounted for in this study, and will be explored in future work. %, could be responsible for part of the scatter, and will be explored in future work. % which is consistent with previous work quoting a slightly shallower slope of 2.02$\pm$0.28 (\citealt{2014A&A...563A..31R}, due to the contribution of molecular gas predominantly at the high metallicity end)
%\item the high dust and gas fractions observed for JINGLE galaxies demonstrate the richness of their ISM reservoirs, from which many stars will still be able to form in the future. The solar-like metallicities, dust-to-metal ratios ($M_{\text{dust}}$/$M_{\text{metals}}$=10$^{-0.67\pm0.23}$), and dust-to-stellar ratios ($M_{\text{dust}}$/$M_{\star}$=10$^{-2.71\pm0.36}$) characteristic of the entire local galaxy population, suggest that JINGLE galaxies have already been enriched with metals and dust during previous star formation episodes, and makes them less pristine than several low-mass and metal-poor HAPLESS and HiGH galaxies. %compared to the average local galaxy population 
\end{itemize}
 
%We studied variations in H{\sc{i}} and dust fractions, and dust-to-H{\sc{i}} ratios along the star formation main sequence (SFMS) and inferred that:
%\begin{itemize}
%\item Galaxies on and above the SFMS are characterised by high H{\sc{i}} gas fractions, implying that their elevated levels of SF activity are driven by the availability of massive gas reservoirs.
%\item Decreased H{\sc{i}} gas fractions for more massive $\log$($M_{\star}$/M$_{\odot}$)$\geq$9.5 SFMS galaxies can partly be attributed to their higher molecular (H$_{2}$-to-H{\sc{i}}) gas fractions.
%\item A similar decrease in dust fractions (probing the total atomic+molecular gas reservoir) along the SFMS towards higher stellar masses, however, reveal that H$_{2}$-to-H{\sc{i}} variations provide only half of the story, and that true variations in the star formation efficiency occur along the SFMS, resulting in varying dust fractions.
%\item Dust-to-H{\sc{i}} ratios are highest for quenched galaxies below the SFMS which have exhausted most of their H{\sc{i}} gas reservoir. The lowest dust-to-H{\sc{i}} ratios are found in less massive $\log$($M_{\star}$/M$_{\odot}$)$\leq$9.5 SFMS (HAPLESS, HiGH and KINGFISH dwarf) galaxies corresponding to less evolved systems with massive H{\sc{i}} gas reservoirs and low dust-to-metal ratios.
%\end{itemize}
%We constructed a customised SFH for each of these six galaxy bins. 

To model the evolution of the dust and metal budgets in nearby galaxies, we have split up the local sample of galaxies in six ``galaxy evolutionary" bins according to their specific H{\sc{i}} gas masses. The observed $M_{\star}$, metallicity, $M_{\text{dust}}$/$M_{\star}$, $M_{\text{HI}}$/$M_{\star}$ and $M_{\text{dust}}$/$M_{\text{metals}}$(gas+dust) ratios at these six galaxy evolutionary stages were interpreted with a set of Dust and Element evolUtion modelS (DEUS) -- including dust production by asymptotic giant branch stars, supernova remnants, grain growth in the interstellar medium, and dust destruction through astration and processing by supernova shocks. DEUS was coupled to an MCMC method to effectively search a large parameter space and to constrain the relative importance of stellar dust production, grain growth and dust destruction by supernova shocks. We obtained an extensive set of models by varying the initial gas mass ($M_{\text{gas,ini}}$), the survival rate of supernova dust after passage of the reverse shock ($f_{\text{survival}}$), the grain growth parameter ($\epsilon$) and the interstellar mass cleared per supernova event ($M_{\text{cl}}$, which determines the efficiency of dust destruction through supernova shocks). Based on a rigorous search of this four-dimensional parameter space, we conclude that: %and DEUS models including gas in/outflows:
\begin{itemize}
\item the average scaling laws for galaxies with $-1.0\lesssim \log M_{\text{HI}}/M_{\star}\lesssim0$ (which are considered to have reached an equilibrium between gas infall, outflow and star formation) can be reproduced using closed-box models with a high fraction (37-89$\%$) of supernova dust that is able to survive a reverse shock, low grain growth efficiencies ($\epsilon$=30-40), and long dust lifetimes (1-2\,Gyr). 
\item the contribution from stardust ($>$90\,$\%$) outweighs the fraction of dust grown through accretion in the ISM ($<$10\,$\%$) over the entire lifetime of these galaxies, while present-day dust budgets have similar contributions from stellar sources (50-80$\%$) and ISM dust growth (20-50$\%$). %, which reduces the need for (unrealistically) low grain growth timescales.
\item the specific shape of the SFH does not strongly influence the model outcome due to these long dust lifetimes. %, \textbf{and in part may be driven by the relatively simple shape assumed to describe the star formation histories.} % \textbf{In future work, we will explore how more variable SFHs and complex models will effect these dependencies} %due to the gradual build-up of a galaxy's metal and dust content over several 100 Myr.
%\item our models underpredict the metal abundances of galaxies at an early stage of evolution (with $\log M_{\text{HI}}/M_{\star}\ge$0), which suggests that the mass loading factors have been overestimated for these low-mass galaxies, or that these high sSFR galaxies have recently experienced enriched gas inflows that also powered the recent star formation activity in these galaxies.
%\item we require gaseous outflows to explain the low $M_{\text{HI}}$/$M_{\star}$ ratios in more evolved galaxies.
%accounting for gaseous in- and outflows seems to require more dust to be produced through grain growth to account for the removal of dust and metals by these outflows. These DEUS models do underpredict the final metallicities of these galaxies, which suggests that the mass loading factors have been overestimated or that the outflowing gas contains significantly less metals and/or dust.
\end{itemize} %Based on our extensive search of a four-dimensional grid of Dust and Element evolUtion modelS, 
We demonstrate in this paper that local galaxy scaling relations can be accounted for by efficient supernova dust production, low grain growth efficiencies, and long dust lifetimes. We speculate that these models provide an alternative to earlier work that required vigorous dust destruction and efficient grain growth on timescales $\lesssim$ a few Myr (e.g., \citealt{2009ASPC..414..453D,2014A&A...562A..76Z,2015MNRAS.449.3274F,2017MNRAS.471.1743D}) to explain local galaxy scaling relations. These long dust lifetimes and reduced grain growth efficiencies could reduce the tension with grain-surface chemical models (e.g., \citealt{1978MNRAS.183..417B,2016MNRAS.463L.112F,2018MNRAS.476.1371C,2019A&A...627A..38J}) that have not been able to come up with efficient grain growth mechanisms in interstellar clouds. Our model results might furthermore help solving the dust budget problem at high redshifts (e.g., \citealt{2003A&A...406L..55B,2003MNRAS.344L..74P,2015Natur.519..327W}), in case similar dust production and destruction efficiencies would apply to those primordial galaxies.
%our rigorous exploration of a multi-dimensional grid of Dust and Element evolUtion modelS (DEUS) reveals that -to some degree-

We caution that model parameter degeneracies between supernova dust production, grain growth and dust destruction efficiencies can not fully be resolved based on the current set of global galaxy scaling laws presented in this work. In future work, we plan to expand DEUS with radially dependent gaseous flows, to explore alternative recipes to describe grain growth and dust destruction processes, and to include additional observational constraints (e.g., resolved galaxy properties, and depletion factors for various elements).

\section*{Acknowledgements}
We would like to thank the anonymous referee for her/his suggestions which have improved the presentation of the fitting methods and results reported in this paper. IDL gratefully acknowledges the support of the Research Foundation -- Flanders (FWO). LCH was supported by the National Science Foundation of China (11721303) and the National Key R\&D Program of China (2016YFA0400702). MJM acknowledges the support of the National Science Centre, Poland through the SONATA BIS grant 2018/30/E/ST9/00208. \\
The James Clerk Maxwell Telescope is operated by the East Asian Observatory on behalf of The National Astronomical Observatory of Japan, Academia Sinica Institute of Astronomy and Astrophysics, the Korea Astronomy and Space Science Institute, the National Astronomical Observatories of China and the Chinese Academy of Sciences (Grant No. XDB09000000), with additional funding support from the Science and Technology Facilities Council of the United Kingdom and participating universities in the United Kingdom and Canada. Additional funds for the construction of SCUBA-2 were provided by the Canada Foundation for Innovation. This data is being observed under JCMT Project ID: M16AL005. The Starlink software (Currie et al. 2014) used as part of the JINGLE data reduction process is currently supported by the East Asian Observatory. \\
PACS was developed by a consortium of institutes led by MPE (Germany) and including UVIE (Austria); KU Leuven, CSL, IMEC (Belgium); CEA, LAM (France); MPIA (Germany); INAFIFSI/ OAA/OAP/OAT, LENS, SISSA (Italy); IAC (Spain). This development has been sup- ported by the funding agencies BMVIT (Austria), ESA- PRODEX (Belgium), CEA/CNES (France), DLR (Ger- many), ASI/INAF (Italy), and CICYT/ MCYT (Spain). SPIRE was developed by a consortium of institutes led by Cardiff University (UK) and including Univ. Lethbridge (Canada); NAOC (China); CEA, LAM (France); IFSI, Univ. Padua (Italy); IAC (Spain); Stockholm Observatory (Sweden); Imperial College London, RAL, UCL-MSSL, UKATC, Univ. Sussex (UK); and Caltech, JPL, NHSC, Univ. Colorado (USA). This development has been supported by national funding agencies: CSA (Canada); NAOC (China); CEA, CNES, CNRS (France); ASI (Italy); MCINN (Spain); SNSB (Sweden); STFC and UKSA (UK); and NASA (USA).

% List of institutions
\section*{Affiliations}
\label{Affiliations.sec}
$^{1}$Sterrenkundig Observatorium, Ghent University, Krijgslaan 281 - S9, 9000 Gent, Belgium\\
$^{2}$Dept. of Physics \& Astronomy, University College London, Gower Street, London WC1E 6BT, UK\\
$^{3}$Dept. F{\'i}sica Te{\'o}rica y del Cosmos, Universidad de Granada, Spain\\
$^{4}$Instituto Universitario Carlos I de F{\'i}sica Te{\'o}rica y Computacional, Universidad de Granada, 18071, Granada, Spain\\
$^{5}$School of Physics and Astronomy, Cardiff University, Queens Buildings, The Parade, Cardiff CF24 3AA\\
$^{6}$Space Telescope Science Institute, 3700 San Martin Drive, Baltimore, Maryland, 21211, USA \\
$^{7}$Department of Physics \& Astronomy, McMaster University, Hamilton, ON L8S 4M1 Canada\\
$^{8}$Universit\'e Paris-Saclay, CNRS,  Institut d'Astrophysique Spatiale, 91405, Orsay, France\\
$^{9}$Steward Observatory, University of Arizona, 933 N Cherry Avenue, Tucson, AZ  85721-0065, USA\\
$^{10}$George P. and Cynthia Woods Mitchell Institute for Fundamental Physics and Astronomy, Texas A\&M University, College Station, TX 77843-4242, USA\\
$^{11}$Centre for Astrophysics Research, University of Hertfordshire, College Lane, AL10 9AB, UK\\
$^{12}$Sub-department of Astrophysics, University of Oxford, Denys Wilkinson Building, Keble Road, Oxford, OX1 3RH, UK\\
$^{13}$Yonsei Frontier Lab and Department of Astronomy, Yonsei University, 50 Yonsei-ro, Seodaemun-gu, Seoul 03722, Republic of Korea \\
$^{14}$Blackett Laboratory, Physics Department, Imperial College, London, SW7 2AZ, UK \\
$^{15}$Institute of Astronomy \& Astrophysics, Academia Sinica, Taipei, 10617, Taiwan \\
$^{16}$Shanghai Astronomical Observatory, 80 Nandan Road, Xuhui District, Shanghai, China 200030 \\
$^{17}$Purple Mountain Observatory \& Key Lab. of Radio Astronomy, Chinese Academy of Sciences, Nanjing 210034, China \\
$^{18}$Centre for Astronomy, National University of Ireland, Galway, University Road, Galway, Ireland H91 TK33 \\
$^{19}$Kavli Institute for Astronomy and Astrophysics, Peking University, Beijing 100871, China \\
$^{20}$Department of Astronomy, School of Physics, Peking University, Beijing 100871, China \\
$^{21}$Korea Astronomy and Space Science Institute, 776 Daedeokdae-ro, Yuseong-gu, Daejeon 34055, Republic of Korea \\
$^{22}$Astronomical Observatory Institute, Faculty of Physics, Adam Mickiewicz University, ul. S{\l}oneczna 36, 60-286 Pozna{\'n}, Poland \\
$^{23}$Tsinghua Center for Astrophysics and Physics Department, Tsinghua University, Beijing 100084, China \\
$^{24}$Shanghai Astronomical Observatory, Chinese Academy of Sciences, 80 Nandan Road, Shanghai, 200030, PR China \\ 
$^{25}$Max Planck Institute for Astronomy, K{\"o}nigstuhl 17, D-69117 Heidelberg, Germany \\
$^{26}$Astronomy Centre, Department of Physics and Astronomy, University of Sussex, Brighton BN1 9QH, England \\
$^{27}$Department of Physics, Zhejiang University, Hangzhou, Zhejiang 310027, China\\
$^{28}$National Astronomical Observatory of China, 20A Datun Road, Chaoyang District, Beijing, China 100012 \\

\bibliographystyle{mnras}
\bibliography{JINGLE_dustscaling}

%%%%%%%%%%%%%%%%%%%%%%%%%%%%%%%%%%%%%%%%%%%%%%%%%%

%%%%%%%%%%%%%%%%% APPENDICES %%%%%%%%%%%%%%%%%%%%%

\clearpage

\appendix

\section{Dust mass determination}
\label{DustMasses.sec}
\begin{figure*}
	\includegraphics[width=9.8cm]{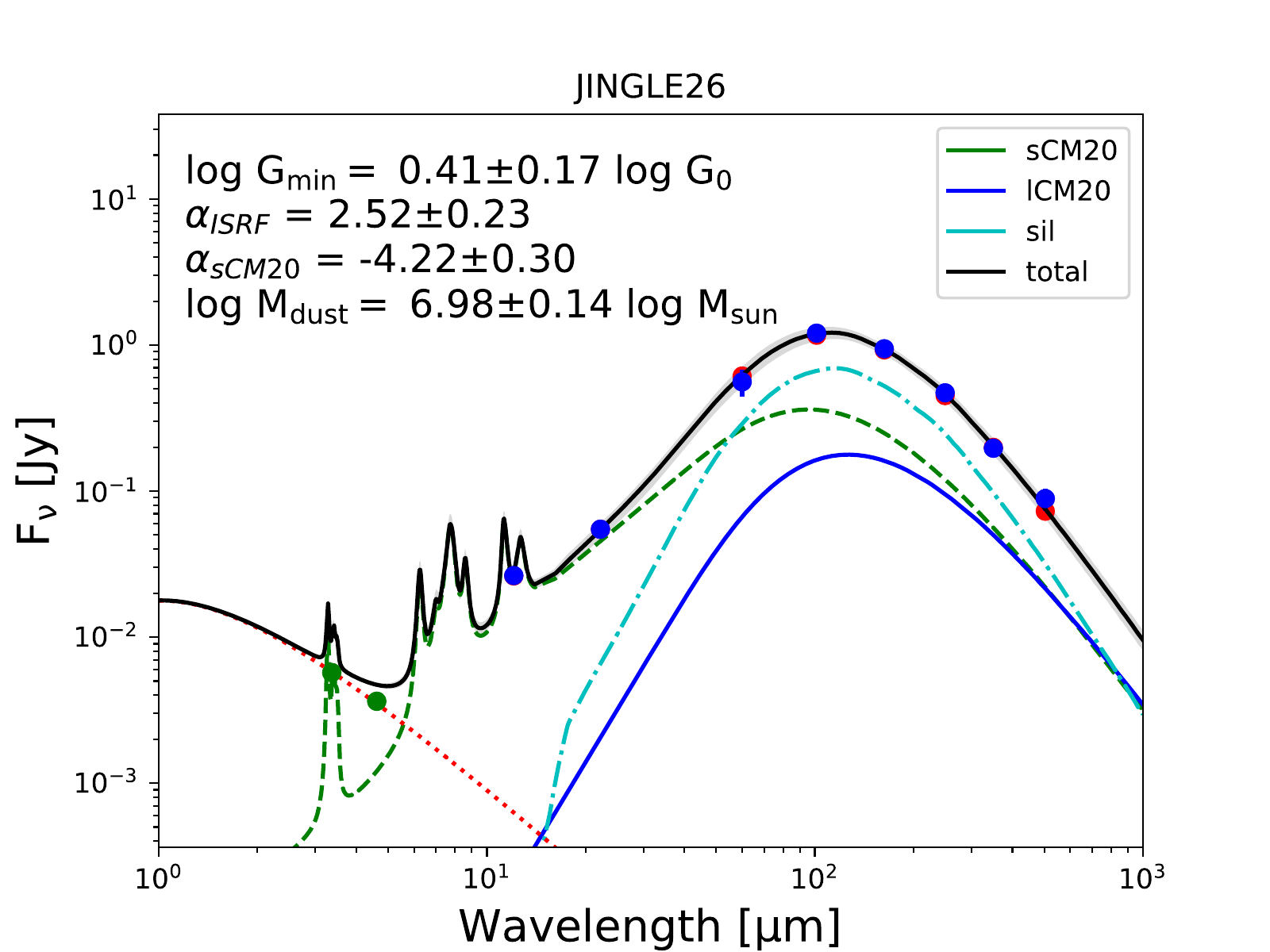}
	\includegraphics[width=7.8cm]{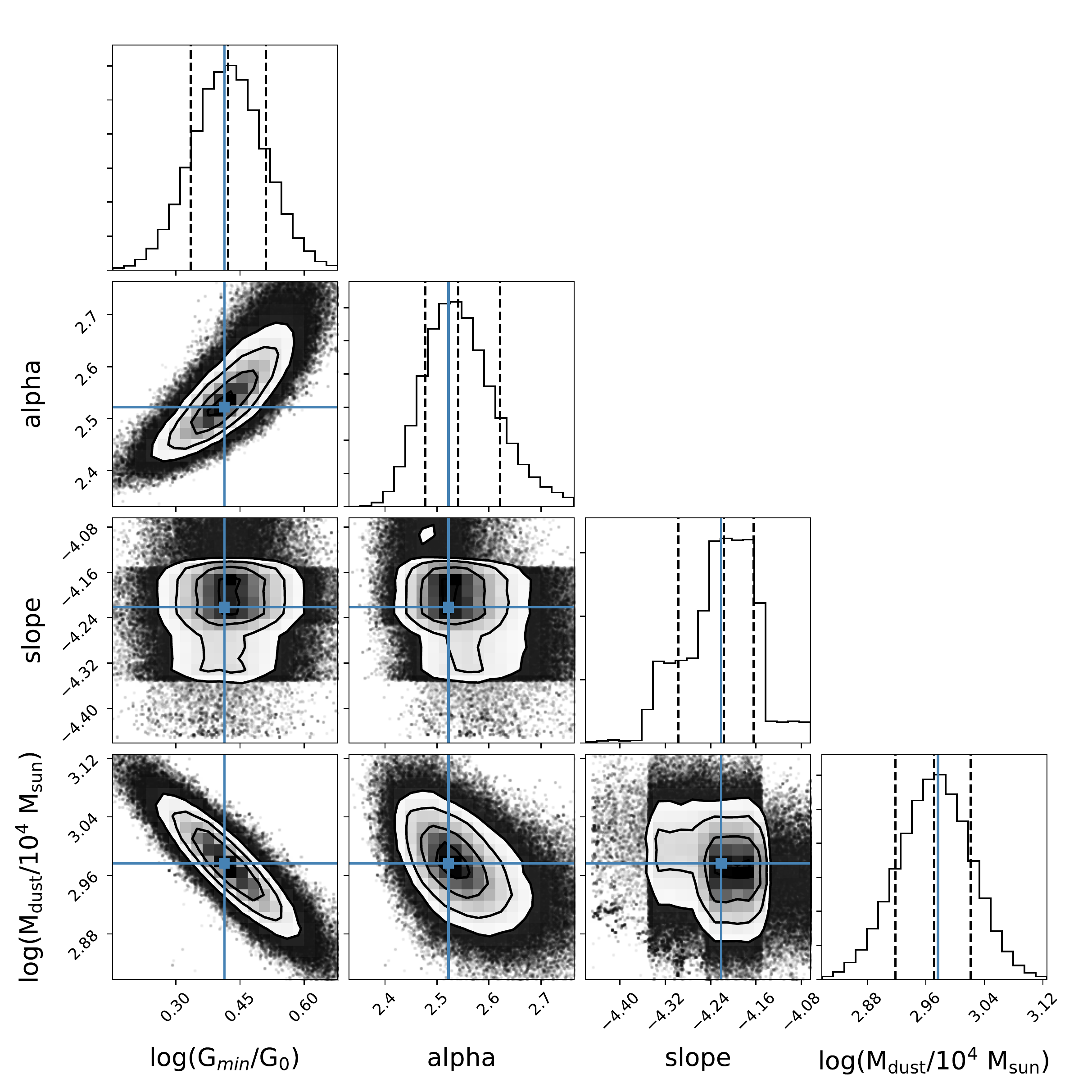}
    \caption{Left panel: a representative model fit for JINGLE\,26 with our Bayesian THEMIS dust SED model. The best-fit SED models for small (sCM20) and large (lCM20) carbonaceous grains and large silicate (sil) grains are indicated with green dashed, blue solid and cyan dash-dotted lines, respectively. The stellar emission at NIR wavelengths is modelled using a blackbody function with temperature $T=5,000K$ (red dotted curve). The total best-fit stellar+dust SED emission is shown in black. The shaded grey region indicate the lower and upper limit uncertainties on the SED models. Right panel: 1D and 2D posterior PDFs which indicate the likelihood of a given output parameter value. The blue line indicates the position of the maximum likelihood (or best-fit model) solution which does not always correspond to the peak of the posterior PDF, while the black dashed lines correspond to the 16th, 50th and 84th percentiles.}   	\label{ExampleSED}   
\end{figure*}

Galaxy dust masses have been inferred from Bayesian dust spectral energy distribution (SED) models fit to the mid-infrared to sub-millimetre emission observed in five samples of nearby galaxies (JINGLE, HRS, KINGFISH, HAPLESS, HiGH). In brief, the Bayesian dust SED models use the THEMIS (The Heterogeneous dust Evolution Model for Interstellar Solids, \citealt{2013A&A...558A..62J,2017A&A...602A..46J}) dust model composition, in addition to two different prescriptions for the radiation field intensity: 1. single interstellar radiation field (ISRF) and 2. multi-component ISRF. In this work, we rely on the dust mass measurements inferred from the second model which relies on the multi-component radiation field prescription from \citet{2001ApJ...549..215D}. More specifically, the starlight intensity is assumed to be distributed between $U_{\text{min}}$ and $U_{\text{max}}$ and the fractional dust mass heated by each ISRF intensity is assumed to be $dM/dU \propto U^{-\alpha_{\text{ISRF}}}$. We fix $U_{\text{max}}$=10$^{7}$ (e.g., \citealt{2019A&A...624A..80N}), and vary the minimum starlight intensity, $U_{\text{min}}$, as well as the slope of the ISRF power-law distribution, $\alpha_{\text{ISRF}}$, in our models. The THEMIS dust mix includes a set of small (sCM20) and large (lCM20) amorphous hydrocarbon grains (a-C(:H)) and large silicates with iron nano-particle inclusions (a-Sil$_{\text{Fe}}$), for which the optical properties were derived from laboratory studies, and the size distribution and abundances of grain species were calibrated to reproduce the extinction and emission observed in the diffuse ISM of the Milky Way \citep{2013A&A...558A..62J,2014A&A...565L...9K}. To model the diversity of dust SEDs observed in the local Universe, we allow for variations in the THEMIS dust mix by varying the slope $\alpha_{\text{sCM20}}$ of the grain size distribution of small hydrocarbons, and the relative dust masses of small hydrocarbons, $M_{\text{sCM20}}$, and large (hydrocarbon and amorphous silicate) dust grains, $M_{\text{lCM20+sil}}$. An example dust SED model fit for the galaxy JINGLE\,26 has been shown in Figure \ref{ExampleSED} (left panel). The lower and upper limit uncertainties on galaxy dust masses have been inferred from the 16th and 84th percentiles in the posterior PDFs (see Fig. \ref{ExampleSED}, right panel) and have been tabulated in Table \ref{DustMasses.tab} for all galaxies.

\newpage
%\begin{center}
\topcaption{Overview of the dust masses inferred for the JINGLE, HRS, HAPLESS, HiGH and KINGFISH galaxies considered in the scaling relations presented in this work. The median dust masses have been inferred from the posterior PDFs while the 16th and 84th percentiles are used to approximate the lower and upper limits on these modelled dust masses. The full galaxy names for JINGLE and HRS galaxies can be retrieved from \citet{2018MNRAS.481.3497S} and \citet{2010PASP..122..261B}.}
\tablefirsthead{\toprule Galaxy & $\log$($M_{\text{dust}}$/$M_{\odot}$) & -dex & +dex\\ \midrule }
\tablehead{\midrule
\multicolumn{4}{l}%
{{Continued from previous column}} \\
\toprule
Galaxy name & $\log$($M_{\text{dust}}$/$M_{\odot}$) & -dex & +dex\\ \midrule}
\tabletail{%
\midrule \multicolumn{4}{r}{{Continued on next column}} \\ \midrule}
\tablelasttail{%
\\\midrule
\multicolumn{4}{r}{{Concluded}} \\ \bottomrule}
\begin{supertabular}{lccc}
\label{DustMasses.tab}
%\multicolumn{4}{c}{JINGLE galaxies} \\
\textbf{JINGLE:} & & & \\
\hline 
JINGLE\,0 & 6.68 & -0.06 & +0.06 \\
JINGLE\,1 & 7.12 & -0.08 & +0.10 \\
JINGLE\,2 & 6.82 & -0.23 & +0.25 \\
JINGLE\,3 & 6.53 & -0.05 & +0.05  \\
JINGLE\,4 & 7.08 & -0.06 & +0.06 \\
JINGLE\,5 & 7.34 & -0.06 & +0.06 \\
JINGLE\,6 & 7.18 & -0.08 & +0.08 \\
JINGLE\,7 & 6.96 & -0.07 & +0.07 \\
JINGLE\,8 & 6.79 & -0.09 & +0.09 \\
JINGLE\,9 & 7.05 & -0.08 & +0.08 \\
JINGLE\,10 & 7.35 & -0.06 & +0.06 \\
JINGLE\,11 & 7.44 & -0.09 & +0.08 \\
JINGLE\,12 & 6.98 & -0.09 & +0.09 \\
JINGLE\,13 & 6.62 & -0.10 & +0.10 \\
%JINGLE\,14 & 6.87 & -0.05 & +0.05 \\
JINGLE\,15 & 7.09 & -0.06 & +0.06 \\
JINGLE\,16 & 7.11 & -0.07 & +0.07 \\
JINGLE\,17 & 6.69 & -0.09 & +0.09 \\
JINGLE\,18 & 6.91 & -0.12 & +0.13 \\
JINGLE\,19 & 7.46 & -0.08 & +0.09 \\
JINGLE\,20 & 7.08 & -0.09 & + 0.09 \\
JINGLE\,21 & 7.23 & -0.13 & +0.15 \\
JINGLE\,22 & 7.73 & -0.06 & +0.05 \\
JINGLE\,23 & 7.34 & -0.05 & +0.05 \\
JINGLE\,24 & 6.93 & -0.14 & +0.16 \\
JINGLE\,25 & 7.24 & -0.05 & +0.05 \\
JINGLE\,26 & 6.97 & -0.05 & +0.05 \\
JINGLE\,27 & 7.33 & -0.08 & +0.09 \\
JINGLE\,28 & 7.43 & -0.07 & +0.08 \\
%JINGLE\,29 & 6.86 & -0.07 & +0.07 \\
JINGLE\,30 & 7.06 & -0.11 & +0.11 \\
JINGLE\,31 & 6.86 & -0.08 & +0.09 \\
JINGLE\,32 & 6.92 & -0.08 & +0.08 \\
JINGLE\,33 & 6.66 & -0.05 & +0.04 \\
JINGLE\,34 & 7.29 & -0.07 & +0.07 \\
JINGLE\,35 & 7.22 & -0.05 & +0.05 \\
JINGLE\,36 & 6.87 & -0.07 & +0.07 \\
%JINGLE\,37  &    7.18    &  -0.06   &   +0.05  \\   
JINGLE\,38  &    6.87   &   -0.06  &    +0.05  \\  
JINGLE\,39  &    7.17   &   -0.08  &    +0.08  \\  
JINGLE\,40  &    7.81  &   -0.05  &    +0.05   \\  
JINGLE\,41  &    7.83  &   -0.04  &    +0.04   \\  
%JINGLE\,42  &    6.99   &   -0.06  &    +0.06  \\  
JINGLE\,43  &    8.02  &   -0.04  &    +0.03   \\  
JINGLE\,44  &    7.81  &   -0.06  &    +0.07   \\  
JINGLE\,45  &    7.95  &   -0.05  &    +0.05   \\ 
JINGLE\,46  &    6.91   &   -0.13  &    +0.14   \\ 
JINGLE\,47  &     7.49  &    -0.05 &     +0.04   \\  
JINGLE\,48  &    7.39   &   -0.04  &    +0.04   \\ 
JINGLE\,49  &     7.55  &    -0.05 &     +0.04   \\  
JINGLE\,50  &    7.36   &   -0.13  &    +0.12   \\
JINGLE\,51  &     7.37  &    -0.04 &     +0.04   \\  
JINGLE\,52  &    7.22   &   -0.09  &    +0.09    \\
JINGLE\,53  &    6.84   &   -0.09  &    +0.08    \\
JINGLE\,54  &    7.76   &   -0.05  &    +0.04    \\
JINGLE\,55  &    7.59   &   -0.04  &    +0.04    \\
JINGLE\,56  &     7.59  &    -0.05 &     +0.05   \\  
JINGLE\,57  &    7.37   &   -0.09  &    +0.09    \\
JINGLE\,58  &     7.10  &    -0.11 &     +0.11   \\ 
%JINGLE\,59  &     7.12  &    -0.06 &     +0.05   \\  
JINGLE\,60  &     7.18  &    -0.10 &     +0.09   \\  
JINGLE\,61  &    7.09   &   -0.07  &    +0.08    \\
%JINGLE\,62 \\
JINGLE\,63  &    6.28   &   -0.13  &    +0.12    \\
JINGLE\,64  &     6.98  &    -0.09 &     +0.08   \\  
%JINGLE\,65  &    7.30   &   -0.06  &    +0.06    \\
JINGLE\,66  &     7.93  &    -0.09 &     +0.05   \\  
%JINGLE\,67  &    7.16   &   -0.05  &    +0.05    \\
%JINGLE\,68  &     7.39  &    -0.06 &     +0.06   \\  
JINGLE\,69  &    6.98   &   -0.06  &    +0.07    \\
JINGLE\,70  &     7.50  &    -0.05 &     +0.06   \\  
JINGLE\,71  &     7.12  &    -0.06 &     +0.06   \\  
JINGLE\,72  &     7.51  &    -0.04 &     +0.04   \\  
JINGLE\,73  &    7.24   &   -0.09  &    +0.09    \\
JINGLE\,74  &     7.24  &    -0.06 &     +0.06   \\  
JINGLE\,75  &    7.48   &   -0.07  &    +0.07    \\
JINGLE\,76  &     7.24  &    -0.04 &     +0.04   \\  
JINGLE\,77  &     7.56  &    -0.04 &     +0.04   \\  
JINGLE\,78  &    7.02   &   -0.10  &    +0.12    \\
JINGLE\,79  &    6.97   &   -0.09  &    +0.09    \\
JINGLE\,80  &    7.08   &   -0.09  &    +0.09    \\
JINGLE\,81  &     7.24  &    -0.05 &     +0.06   \\  
JINGLE\,82  &     7.10  &    -0.09 &     +0.09   \\  
JINGLE\,83  &    7.71   &   -0.06  &    +0.06    \\
JINGLE\,84  &     7.53  &    -0.06 &     +0.06   \\  
JINGLE\,85  &     6.82  &    -0.07 &     +0.07   \\  
JINGLE\,86  &     7.64  &    -0.05 &     +0.05   \\  
JINGLE\,87  &    7.72   &   -0.07  &    +0.06    \\
JINGLE\,88  &    7.30   &   -0.09  &    +0.10    \\
JINGLE\,89  &     7.50  &    -0.05 &     +0.05   \\  
JINGLE\,90  &     7.69  &    -0.07 &     +0.07   \\  
%JINGLE\,91  &    6.98   &   -0.07  &    +0.07    \\
JINGLE\,92  &    7.92   &   -0.06  &    +0.06    \\
JINGLE\,93  &    7.24   &   -0.08  &    +0.08    \\
%JINGLE\,94  &    7.21   &   -0.16  &    +0.19    \\
%JINGLE\,95  &    7.01   &   -0.09  &    +0.11    \\
JINGLE\,96  &    7.09   &   -0.13  &    +0.16    \\
JINGLE\,97  &    7.33   &   -0.09  &    +0.11    \\
JINGLE\,98  &    7.46   &   -0.09  &    +0.08    \\
JINGLE\,99  &     8.11  &    -0.05 &     +0.05   \\  
JINGLE\,100 &      7.35 &     -0.05  &    +0.04  \\ 
JINGLE\,101 &      7.74 &     -0.04  &    +0.04  \\ 
JINGLE\,102 &      7.68 &     -0.06  &    +0.05  \\ 
JINGLE\,103 &     7.38  &    -0.05 &    +0.05    \\
%JINGLE\,104 &     7.11  &    -0.13 &    +0.14    \\
JINGLE\,105 &     7.32  &    -0.07 &    +0.05    \\
JINGLE\,106 &      7.66 &     -0.05  &    +0.06  \\ 
JINGLE\,107 &     7.92  &    -0.10 &    +0.09    \\
JINGLE\,108 &      7.75 &     -0.05  &    +0.05  \\ 
JINGLE\,109 &     7.47  &    -0.08 &     +0.08   \\
JINGLE\,110 &     7.46  &    -0.29 &     +0.27   \\
JINGLE\,111 &      7.48 &     -0.06  &    +0.06  \\ 
JINGLE\,112 &     7.72  &    -0.09 &     +0.09   \\
JINGLE\,113 &     7.40  &    -0.08 &     +0.09   \\
JINGLE\,114 &     7.68  &    -0.06 &     +0.05   \\
JINGLE\,115 &     7.29  &    -0.05 &     +0.05   \\
%JINGLE\,116 &     7.36  &    -0.07 &     +0.06   \\
JINGLE\,117 &     7.58  &    -0.33 &     +0.30   \\
JINGLE\,118 &      8.18 &     -0.04  &    +0.04  \\ 
%JINGLE\,119 &     7.18  &    -0.13 &     +0.14   \\
JINGLE\,120 &     8.04  &    -0.15 &     +0.15   \\
JINGLE\,121 &      7.92 &     -0.06  &    +0.07  \\ 
JINGLE\,122 &      8.00 &     -0.05  &    +0.05  \\  
JINGLE\,123 &      7.57 &     -0.05  &    +0.05  \\ 
%JINGLE\,124 &     7.58  &    -0.11 &     +0.11   \\
JINGLE\,125 &      7.78 &     -0.06  &    +0.06  \\ 
%JINGLE\,126 &     7.84  &    -0.07 &     +0.07   \\
JINGLE\,127 &      7.85 &     -0.06  &    +0.06  \\ 
JINGLE\,128 &     7.87  &    -0.07 &     +0.06   \\
JINGLE\,129 &     7.50  &    -0.10 &     +0.10   \\
%JINGLE\,130 &     7.75  &    -0.10 &     +0.08   \\
JINGLE\,131 &      7.52 &     -0.05  &    +0.05  \\ 
JINGLE\,132 &     8.02  &    -0.12 &     +0.07   \\
JINGLE\,133 &     7.55  &    -0.04 &     +0.04   \\
JINGLE\,134 &     7.71  &    -0.07 &     +0.07   \\
JINGLE\,135 &      7.44 &     -0.06  &    +0.06  \\ 
JINGLE\,136 &      7.75 &     -0.05  &    +0.04  \\ 
JINGLE\,137 &     6.90  &    -0.29 &     +0.51   \\
JINGLE\,138 &      7.22 &     -0.11  &    +0.10  \\ 
JINGLE\,139 &     7.66  &    -0.10 &     +0.09   \\
JINGLE\,140 &     7.53  &    -0.08 &     +0.07   \\
JINGLE\,141 &     7.56  &    -0.07 &     +0.06   \\
%JINGLE\,142 &     7.86  &    -0.12 &     +0.11   \\
JINGLE\,143 &     7.47  &    -0.06 &     +0.06   \\
JINGLE\,144 &      7.56 &     -0.05  &    +0.04  \\ 
JINGLE\,145 &     6.56  &    -0.09 &     +0.10   \\
JINGLE\,146 &      7.49 &     -0.04  &    +0.04  \\ 
JINGLE\,147 &      7.64 &     -0.05  &    +0.04  \\ 
JINGLE\,148 &      7.42 &     -0.05  &    +0.05  \\ 
JINGLE\,149 &      7.31 &     -0.04  &    +0.04  \\
%JINGLE\,150 &      8.06 &     -0.05  &    +0.05  \\ 
JINGLE\,151 &      8.16 &     -0.08  &    +0.06  \\ 
JINGLE\,152 &      7.72 &     -0.05  &    +0.05  \\
%JINGLE\,153 &     7.51  &    -0.07 &     +0.07   \\
JINGLE\,154 &      8.21 &     -0.04  &    +0.03  \\ 
JINGLE\,155 &     7.82  &    -0.08 &     +0.07   \\
JINGLE\,156 &      7.77 &     -0.06  &    +0.05  \\  
%JINGLE\,157 &     7.90  &    -0.06 &     +0.05   \\
%JINGLE\,158 &     7.44  &    -0.06 &     +0.05   \\
%JINGLE\,159 &     7.37  &    -0.06 &     +0.07   \\
JINGLE\,160 &      7.03 &     -0.05  &    +0.05  \\ 
JINGLE\,161 &     7.78  &    -0.12 &     +0.12   \\
JINGLE\,162 &     7.83  &    -0.05 &     +0.06   \\
JINGLE\,163 &      7.59 &     -0.07  &    +0.07  \\ 
JINGLE\,164 &     7.33  &    -0.09 &     +0.09   \\
JINGLE\,165 &      7.60 &     -0.04  &    +0.04  \\ 
JINGLE\,166 &     7.67  &    -0.07 &     +0.07   \\
JINGLE\,167 &     8.16  &    -0.05 &     +0.05   \\
JINGLE\,168 &     8.07  &    -0.08 &     +0.08   \\
JINGLE\,169 &     7.43  &    -0.07 &     +0.07   \\
JINGLE\,170 &     7.74  &    -0.08 &     +0.08   \\
JINGLE\,171 &     7.86  &    -0.11 &     +0.08   \\
JINGLE\,172 &     7.54  &    -0.09 &     +0.09   \\
JINGLE\,173 &      7.35 &     -0.06  &    +0.06  \\  
JINGLE\,174 &     7.97  &    -0.13 &     +0.13   \\
JINGLE\,175 &      7.86 &     -0.08  &    +0.07  \\ 
JINGLE\,176 &     7.77  &    -0.06 &     +0.05   \\
%JINGLE\,177 &      7.47 &     -0.05  &    +0.04  \\ 
JINGLE\,178 &     7.64  &    -0.07 &     +0.07   \\
%JINGLE\,179 &     7.28  &    -0.13 &     +0.13   \\
%JINGLE\,180 &     7.57  &    -0.06 &     +0.07   \\
JINGLE\,181 &     7.71  &    -0.06 &     +0.07   \\
%JINGLE\,182 &     7.65  &    -0.06 &     +0.05   \\
JINGLE\,183 &      7.20 &     -0.05  &    +0.05  \\ 
JINGLE\,184 &      7.92 &     -0.06  &    +0.06  \\
JINGLE\,185 &     7.36  &    -0.09 &     +0.08   \\
%JINGLE\,186 &      7.83 &     -0.03  &    +0.03  \\ 
JINGLE\,187 &     8.05  &    -0.08 &     +0.09   \\
%JINGLE\,188 &     7.75  &    -0.07 &     +0.07   \\
JINGLE\,189 &      7.90 &     -0.05  &    +0.05  \\ 
%JINGLE\,190 &     8.02  &    -0.05 &     +0.06   \\
JINGLE\,191 &      7.55 &     -0.06  &    +0.05  \\  
%JINGLE\,192 &      8.23 &     -0.04  &    +0.04  \\  
\hline \\
\textbf{HRS:} & & & \\
\hline 
%HRS\,1   &   5.84  &    -0.06  &    +0.06 \\     
HRS\,2   &   6.12  &    -0.04  &    +0.04 \\   
HRS\,3 &  5.78 & -0.10 & +0.12 \\      
%HRS\,4   &   7.23   &   -0.04  &    +0.03 \\    
%HRS\,5   &   6.43   &   -0.03  &    +0.04 \\   
%HRS\,6 &  6.67 & -0.13 & +0.13 \\      
%HRS\,7   &   5.89   &   -0.03  &    +0.03 \\  
HRS\,8 &  7.33 & -0.05 & +0.05 \\
HRS\,9   &   6.40   &   -0.05  &    +0.06 \\    
HRS\,10  &    6.13   &   -0.10   &   +0.12 \\   
HRS\,11  &    6.79   &   -0.04   &   +0.04 \\   
HRS\,12  &    5.51   &   -0.06   &   +0.06 \\   
HRS\,13  &    7.41   &   -0.03   &   +0.03 \\  
%HRS\,14 &  5.24 & -0.12 & +0.12 \\ 
HRS\,15  &    7.52   &   -0.05   &   +0.05 \\   
HRS\,16  &    7.10   &   -0.03   &   +0.03 \\   
HRS\,17  &    7.03   &   -0.04   &   +0.04 \\   
%HRS\,18  &    6.44   &   -0.05   &   +0.06 \\   
HRS\,19  &    6.83   &   -0.05   &   +0.05 \\   
HRS\,20  &    6.96   &   -0.03   &   +0.03 \\
%HRS\,21 &  6.37 & -0.10 & +0.10 \\ 
%HRS\,22  &    6.08   &   -0.12   &   +0.13 \\   
HRS\,23  &    7.09   &   -0.03   &   +0.03 \\   
HRS\,24  &    7.37   &   -0.04   &   +0.04 \\   
HRS\,25  &    6.91   &   -0.04   &   +0.04 \\   
HRS\,26  &    6.09   &   -0.08   &   +0.08 \\   
HRS\,27  &    6.35   &   -0.04   &   +0.05 \\   
HRS\,28  &    6.61   &   -0.04   &   +0.04 \\   
HRS\,29  &    6.52   &   -0.05   &   +0.05 \\   
HRS\,30  &    6.51   &   -0.06   &   +0.07 \\   
HRS\,31  &    6.87   &   -0.05   &   +0.05 \\ 
%HRS\,32 &  5.51 & -0.34 & +0.35 \\     
HRS\,33 &  6.87 & -0.04 & +0.04 \\     
HRS\,34 &  7.04 & -0.05 & +0.04 \\     
%HRS\,35 &  5.56 & -0.09 & +0.10 \\
%HRS\,36  &    7.13   &   -0.03   &   +0.03 \\   
HRS\,37  &    6.59   &   -0.04   &   +0.04 \\   
HRS\,38  &    6.62   &   -0.05   &   +0.06 \\
HRS\,39 &  6.60 & -0.07 & +0.09 \\ 
HRS\,40  &    6.51   &   -0.04   &   +0.04 \\   
HRS\,41 &  6.57 & -0.06 & +0.05 \\ 
HRS\,42  &    7.21   &   -0.05   &   +0.05 \\ 
HRS\,44 &  5.98 & -0.06 & +0.06 \\
HRS\,45  &    6.77   &   -0.05   &   +0.05 \\   
HRS\,46 &  6.44 & -0.04 & +0.04 \\     
HRS\,47 &  6.72 & -0.09 & +0.09 \\
HRS\,48  &    7.44   &   -0.04   &   +0.04 \\   
HRS\,50  &    7.04   &   -0.03   &   +0.03 \\   
HRS\,51  &    6.69   &   -0.05   &   +0.05 \\ 
HRS\,52 &  5.94 & -0.07 & +0.08 \\  
HRS\,53  &    6.90   &   -0.04   &   +0.04 \\   
HRS\,54  &    6.60   &   -0.05   &   +0.05 \\   
HRS\,55  &    6.94   &   -0.04   &   +0.04 \\   
HRS\,56  &    7.25   &   -0.03   &   +0.03 \\   
%HRS\,57  &    7.12   &   -0.03   &   +0.03 \\  
%HRS\,58 &  6.18 & -0.06 & +0.05 \\
HRS\,59  &    7.12   &   -0.05   &   +0.06 \\   
HRS\,60  &    6.84   &   -0.06   &   +0.04 \\   
HRS\,61  &    5.97   &   -0.08   &   +0.08 \\
HRS\,62 &  6.90 & -0.10 & +0.11 \\
HRS\,63  &    7.30   &   -0.04   &   +0.04 \\   
HRS\,64  &    6.53   &   -0.08   &   +0.10 \\   
HRS\,65  &    6.43   &   -0.07   &   +0.07 \\
HRS\,66 &  7.13 & -0.04 & +0.03 \\
HRS\,67  &    6.41   &   -0.10   &   +0.10 \\  
HRS\,68 &  5.77 & -0.05 & +0.05 \\
HRS\,69  &    6.99   &   -0.06   &   +0.06 \\   
HRS\,70  &    6.40   &   -0.06   &   +0.06 \\
%HRS\,71 &  7.58 & -0.50 & +0.47 \\
HRS\,72  &    6.07   &   -0.07   &   +0.07 \\   
HRS\,73  &    7.75   &   -0.03   &   +0.03 \\   
HRS\,74  &    6.83   &   -0.05   &   +0.05 \\ 
HRS\,75 &  6.19 & -0.15 & +0.15 \\ 
HRS\,76  &    5.99   &   -0.10   &   +0.11 \\   
HRS\,77  &    7.77   &   -0.04   &   +0.03 \\   
HRS\,78  &    6.50   &   -0.07   &   +0.07 \\   
HRS\,79  &    6.16   &   -0.10   &   +0.10 \\ 
%HRS\,80 &  6.68 & -0.07 & +0.13 \\ 
HRS\,81  &    6.91   &   -0.04   &   +0.04 \\   
HRS\,82  &    5.97   &   -0.05   &   +0.05 \\   
%HRS\,83  &    5.79   &   -0.12   &   +0.13 \\   
HRS\,84  &    6.26   &   -0.04   &   +0.04 \\   
HRS\,85  &    7.22   &   -0.04   &   +0.04 \\   
HRS\,86  &    7.04   &   -0.07   &   +0.07 \\   
%HRS\,87  &    6.06   &   -0.07   &   +0.07 \\   
HRS\,88  &    7.16   &   -0.07   &   +0.06 \\   
HRS\,89  &    7.47   &   -0.06   &   +0.06 \\   
HRS\,91  &    7.72   &   -0.03   &   +0.03 \\   
HRS\,92  &    6.53   &   -0.05   &   +0.05 \\   
HRS\,93  &    6.31   &   -0.06   &   +0.06 \\   
HRS\,94  &    7.13   &   -0.06   &   +0.08 \\   
HRS\,95  &    6.48   &   -0.03   &   +0.03 \\   
HRS\,96  &    7.16   &   -0.03   &   +0.03 \\   
%HRS\,97  &    7.80   &   -0.02   &   +0.02 \\   
HRS\,98   &   6.90   &   -0.04   &   +0.05  \\ 
%HRS\,99 &  5.86 & -0.04 & +0.04 \\ 
%HRS\,100  &    6.90  &   -0.03   &   +0.03  \\ 
HRS\,102  &    7.87  &   -0.03   &   +0.03  \\
%HRS\,103 &  6.53 & -0.05 & +0.05 \\
HRS\,106  &    6.66  &   -0.06   &   +0.07  \\ 
HRS\,107 &  6.30 & -0.06 & +0.07 \\    
%HRS\,108 &  6.12 & -0.09 & +0.09 \\
HRS\,109  &    6.67  &   -0.06   &   +0.06  \\ 
HRS\,110  &    6.74  &   -0.06   &   +0.05  \\ 
HRS\,111  &    7.16  &   -0.03   &   +0.03  \\ 
%HRS\,112  &    6.09  &   -0.07   &   +0.08  \\ 
%HRS\,113  &    7.47  &   -0.03   &   +0.03  \\ 
HRS\,114  &    7.82  &   -0.03   &   +0.03  \\ 
%HRS\,115  &    5.90  &   -0.07   &   +0.07  \\ 
%HRS\,117 &  7.09 & -0.06 & +0.05 \\    
HRS\,118 &  6.07 & -0.17 & +0.18 \\
%HRS\,119  &    6.74  &   -0.04  &   +0.04  \\ 
%HRS\,120  &    6.81  &   -0.04   &   +0.04 \\  
HRS\,121  &    7.11  &   -0.04   &   +0.03 \\  
HRS\,122  &    8.06  &   -0.03   &   +0.03 \\  
HRS\,123  &    6.59  &   -0.05   &   +0.05 \\  
%HRS\,124  &    6.93  &   -0.05   &   +0.05 \\  
%HRS\,127  &    7.02  &   -0.03   &   +0.03 \\  
%HRS\,128  &    6.66  &   -0.06   &   +0.06 \\ 
%HRS\,129 &  5.44 & -0.04 & +0.04 \\
%HRS\,130  &    6.44  &   -0.05   &   +0.06 \\ 
%HRS\,131 &  6.38 & -0.17 & +0.10 \\
HRS\,132  &    6.20  &   -0.07   &   +0.08 \\  
HRS\,133  &    7.06  &   -0.10   &   +0.10 \\  
%HRS\,134  &    6.55  &   -0.05   &   +0.05 \\  
%HRS\,136  &    6.50  &   -0.04   &   +0.04 \\  
HRS\,139 &  6.38 & -0.06 & +0.07 \\
HRS\,140  &    6.96  &   -0.06   &   +0.06 \\  
%HRS\,141  &    7.32  &   -0.04   &   +0.04 \\  
HRS\,142  &    6.54  &   -0.04   &   +0.04 \\  
HRS\,143  &    7.21  &   -0.04   &   +0.05 \\  
%HRS\,144  &    6.96  &   -0.04   &   +0.04 \\  
HRS\,145  &    6.77  &   -0.06   &   +0.07 \\  
HRS\,146  &    6.78  &   -0.05   &   +0.05 \\  
HRS\,147  &    6.84  &   -0.04   &   +0.04 \\  
HRS\,148  &    6.97  &   -0.07   &   +0.06 \\  
%HRS\,149  &    7.22  &   -0.03   &   +0.03 \\  
%HRS\,151  &    6.59  &   -0.05   &   +0.05 \\  
%HRS\,152  &    6.39  &   -0.04   &   +0.04 \\  
%HRS\,153  &    6.61  &   -0.05   &   +0.05 \\  
HRS\,154  &    7.06  &   -0.07   &   +0.08 \\  
%HRS\,156  &    6.89  &   -0.04   &   +0.03 \\  
HRS\,157  &    6.68  &   -0.04   &   +0.03 \\ 
HRS\,158 &  6.89 & -0.21 & +0.21 \\
%HRS\,159  &    6.57  &   -0.04   &   +0.05 \\  
%HRS\,160  &    7.11  &    -0.04  &   +0.04 \\  
%HRS\,161  &    6.08  &    -0.04  &   +0.04 \\  
%HRS\,162  &    6.07  &    -0.06  &   +0.05 \\  
%HRS\,163  &    7.00  &    -0.05  &   +0.04 \\ 
%HRS\,165 &  6.25 & -0.10 & +0.10 \\
%HRS\,167  &    6.67  &    -0.04  &   +0.04 \\  
%HRS\,168  &    6.10  &    -0.07  &   +0.07 \\  
HRS\,169  &    6.59  &    -0.06  &   +0.07 \\  
%HRS\,170  &    7.27  &    -0.07  &   +0.07 \\  
%HRS\,171  &    6.61  &    -0.04  &   +0.04 \\  
%HRS\,172  &    6.57  &    -0.04  &   +0.04 \\  
%HRS\,173  &    6.63  &    -0.04  &   +0.04 \\  
%HRS\,174  &    6.03  &    -0.04  &   +0.03 \\  
%HRS\,176  &    6.58  &    -0.04  &   +0.04 \\  
HRS\,177  &    6.41  &    -0.04  &    +0.04 \\ 
%HRS\,180 &  5.56 & -0.07 & +0.07 \\   
HRS\,182  &    6.69  &    -0.04  &    +0.04 \\  
%HRS\,184  &    5.52  &    -0.07  &    +0.07 \\  
%HRS\,185  &    6.50  &    -0.07  &    +0.07 \\ 
%HRS\,186 &  5.10 & -0.13 & +0.13 \\
HRS\,187  &    7.12  &    -0.05  &    +0.06 \\  
HRS\,188  &    6.91  &    -0.05  &    +0.05 \\  
HRS\,189  &    6.36  &    -0.05  &    +0.05 \\  
%HRS\,190  &    7.94  &    -0.04  &    +0.03 \\ 
%HRS\,191 &  5.73 & -0.09 & +0.11 \\ 
%HRS\,192  &    5.69  &    -0.08  &    +0.07 \\  
%HRS\,193  &    6.57  &    -0.06  &    +0.06 \\  
HRS\,194  &    7.99  &    -0.03  &    +0.03 \\  
HRS\,196  &    6.91  &    -0.06  &    +0.06 \\  
%HRS\,197  &    6.73  &    -0.04  &    +0.04 \\ 
HRS\,198 &  6.75 & -0.07 & +0.08 \\    
%HRS\,199 &  6.03 & -0.09 & +0.09 \\
%HRS\,200  &    6.69  &    -0.03  &    +0.03 \\  
HRS\,201  &    7.69  &    -0.03  &    +0.03 \\  
HRS\,203  &    6.81  &    -0.04  &    +0.04 \\  
HRS\,204  &    7.84  &    -0.03  &    +0.03 \\  
HRS\,205  &    7.53  &    -0.04  &    +0.04 \\  
%HRS\,206  &    6.22  &    -0.05  &    +0.06 \\  
%HRS\,207  &    6.76  &    -0.03  &    +0.03 \\  
%HRS\,208 &  7.60 & -0.04 & +0.04 \\    
%HRS\,209 &  6.23 & -0.20 & +0.21 \\    
%HRS\,210 &  4.80 & -0.15 & +0.07 \\    
HRS\,212  &    6.33  &    -0.07  &    +0.07 \\  
HRS\,213  &    8.27  &    -0.03  &    +0.03 \\  
HRS\,215  &    6.99  &    -0.03  &    +0.03 \\  
HRS\,216  &    7.38  &    -0.03  &    +0.03 \\  
%HRS\,217  &    7.50  &    -0.03  &    +0.03 \\  
%HRS\,220  &    7.61  &    -0.04  &    +0.04 \\  
%HRS\,221  &    6.73  &    -0.03  &    +0.03 \\  
%HRS\,222 &  5.88 & -0.09 & +0.09 \\    
%HRS\,223 &  6.14 & -0.16 & +0.19 \\    
%HRS\,224  &    6.68  &    -0.08  &    +0.09 \\ 
%HRS\,225 &  4.99 & -0.43 & +0.38 \\    
%HRS\,226  &    6.29  &    -0.05  &    +0.05 \\  
HRS\,227  &    6.87  &    -0.07  &    +0.07 \\  
%HRS\,230  &    6.53  &    -0.04  &    +0.05 \\  
%HRS\,231 &  5.47 & -0.04 & +0.04 \\    
%HRS\,232  &    6.22  &    -0.05  &    +0.05 \\  
%HRS\,233  &    6.81  &    -0.04  &    +0.04 \\  
HRS\,237  &    6.62  &    -0.06  &    +0.05 \\  
HRS\,238 &  5.59 & -0.17 & +0.24 \\    
HRS\,239  &    6.85  &    -0.03  &    +0.03 \\  
HRS\,242  &    6.83  &    -0.05  &    +0.05 \\  
%HRS\,243 &  6.94 & -0.06 & +0.07 \\    
HRS\,244  &    7.12  &    -0.03  &    +0.03 \\  
HRS\,246  &    7.27  &    -0.03  &    +0.04 \\  
HRS\,247  &    7.57  &    -0.03  &    +0.03 \\ 
%HRS\,249 &  5.91 & -0.29 & +0.25 \\    
HRS\,251  &    7.87  &    -0.03  &    +0.02 \\  
HRS\,252  &    6.58  &    -0.10  &    +0.11 \\  
%HRS\,253  &    5.37  &    -0.05  &    +0.05 \\  
%HRS\,254  &    7.33  &    -0.04  &    +0.04 \\  
HRS\,255 &  6.88 & -0.11 & +0.11 \\
%HRS\,256  &    6.57  &    -0.03  &    +0.03 \\  
HRS\,257 &  7.27 & -0.05 & +0.05 \\
%HRS\,258  &    5.23  &    -0.08  &    +0.09 \\  
HRS\,259  &    6.78  &    -0.08  &    +0.10 \\  
%HRS\,260  &    6.77  &    -0.03  &    +0.03 \\  
%HRS\,261  &    6.75  &    -0.06  &    +0.07 \\  
HRS\,262  &    6.89  &    -0.04  &    +0.05 \\  
HRS\,263  &    8.19  &    -0.05  &    +0.04 \\  
HRS\,264 &  6.35 & -0.10 & +0.10 \\
%HRS\,265  &    6.13  &    -0.05  &    +0.05 \\ 
HRS\,266 &  7.45 & -0.08 & +0.12 \\
HRS\,267  &    6.77  &    -0.08  &    +0.07 \\  
HRS\,268  &    6.83  &    -0.03  &    +0.03 \\  
%HRS\,270  &    6.99  &    -0.06  &    +0.07 \\  
HRS\,271  &    6.75  &    -0.06  &    +0.07 \\  
HRS\,273  &    7.05  &    -0.04  &    +0.04 \\  
HRS\,274  &    6.79  &    -0.04  &    +0.05 \\  
HRS\,275  &    7.01  &    -0.04  &    +0.04 \\  
HRS\,276  &    6.56  &    -0.06  &    +0.06 \\  
%HRS\,277 &  5.53 & -0.12 & +0.12 \\    
%HRS\,278 &  6.28 & -0.07 & +0.07 \\    
HRS\,279 &  6.93 & -0.07 & +0.08 \\
HRS\,280  &    6.38  &    -0.04  &    +0.04 \\  
HRS\,281  &    5.82  &    -0.07  &    +0.07 \\  
HRS\,283  &    7.04  &    -0.04  &    +0.04 \\  
HRS\,284  &    6.80  &    -0.03  &    +0.03 \\  
%HRS\,285  &    7.06  &    -0.03  &    +0.03 \\  
HRS\,286 &  6.70 & -0.19 & +0.19 \\
HRS\,287  &    6.78  &    -0.04  &    +0.05 \\  
%HRS\,288  &    6.98  &    -0.04  &    +0.18 \\  
HRS\,289  &    7.34  &    -0.03  &    +0.03 \\  
HRS\,290  &    6.04  &    -0.05  &    +0.05 \\  
HRS\,292  &    6.68  &    -0.04  &    +0.04 \\  
HRS\,293  &    6.68  &    -0.04  &    +0.05 \\  
%HRS\,294  &    6.78  &    -0.05  &    +0.05 \\  
HRS\,295  &    7.66  &    -0.03  &    +0.03 \\  
HRS\,296 &  5.49 & -0.08 & +0.09 \\
HRS\,297  &    7.38  &    -0.03  &    +0.04 \\ 
HRS\,298  &    6.52  &    -0.04  &    +0.04 \\  
HRS\,299  &    7.10  &    -0.05  &    +0.06 \\  
%HRS\,300  &    6.08  &    -0.10  &    +0.11 \\  
HRS\,301  &    7.36  &    -0.06  &    +0.08 \\  
HRS\,302 &  6.97 & -0.11 & +0.13 \\ 
%HRS\,303  &    6.29  &    -0.04  &    +0.04 \\  
%HRS\,304  &    7.04  &    -0.03  &    +0.03 \\  
%HRS\,305 &  5.77 & -0.16 & +0.14 \\
%HRS\,306  &    6.52  &    -0.06  &    +0.06 \\  
HRS\,307  &    7.71  &    -0.04  &    +0.04 \\ 
HRS\,308 &  5.15 & -0.08 & +0.09 \\
HRS\,309  &    6.24  &    -0.07  &    +0.07 \\  
%HRS\,310  &    6.91  &    -0.05  &    +0.06 \\  
%HRS\,311 &  7.43 & -0.04 & +0.04 \\    
%HRS\,313 &  7.13 & -0.06 & +0.08 \\  
HRS\,314  &    6.73  &    -0.07  &    +0.07 \\  
HRS\,315 &  6.18 & -0.23 & +0.31 \\    
HRS\,317 &  6.16 & -0.13 & +0.14 \\
HRS\,318  &    6.75  &    -0.05  &    +0.05 \\  
HRS\,319  &    7.10  &    -0.06  &    +0.06 \\  
HRS\,320  &    7.48  &    -0.08  &    +0.09 \\  
HRS\,321  &    6.39  &    -0.04  &    +0.04 \\  
HRS\,322 &  7.28 & -0.10 & +0.12 \\ 
HRS\,323  &    7.06  &    -0.04  &    +0.03 \\  
\hline \\
\textbf{HAPLESS+HiGH:} & & & \\
\hline 
UGC\,06877 &  5.32 & -0.06 & +0.06 \\  % (HAPLESS\,1) 
PGC\,037392 &  5.77 & - 0.26 & + 0.27 \\  % (HAPLESS\,2)
UGC\,09215 &  6.94 & -0.07 & +0.07 \\  %  (HAPLESS3/HIGH23)
UM\,452 &  5.56 & - 0.20 & + 0.21 \\  % (HAPLESS\,4) 
NGC\,4030 &  7.87 & -0.03 & +0.03 \\    % (HAPLESS6/HIGH12)
NGC\,5496 &  7.11 & -0.07 & +0.07 \\    % (HAPLESS7/HIGH21)
UGC\,07000 &  6.36 &  -0.08 & +0.08 \\ % (HAPLESS8/HIGH39)
UGC\,09299 &  6.38 & -0.14 & +0.15 \\   % (HAPLESS9/HIGH27)  
NGC\,5740 &  7.18 & -0.04 & +0.04 \\    % (HAPLESS10/HIGH38)
UGC\,07394 &  7.01 & -0.25 & +0.24 \\   % (HAPLESS11/HIGH18)
PGC\,051719 &  6.19 & -0.08 & +0.09 \\  % (HAPLESS\,12) 
NGC\,5584 &  7.50 & -0.05 & +0.04 \\    % (HAPLESS14/HIGH22)
UGC\,09348 &  6.58 & -0.07 & +0.07 \\  % (HAPLESS\,16)
%NGC\,5733 &  6.15 & -0.09 & +0.09 \\    % (HAPLESS\,18)
UGC\,06780 &  6.78 & -0.34 & +0.28 \\   % (HAPLESS19/HIGH6)   
NGC\,5719 &  7.46 & -0.03 & +0.03 \\    % (HAPLESS20/HIGH35)
NGC\,5746 &  7.99 & -0.03 & +0.03 \\    % (HAPLESS21/HIGH40)
%NGC\,5738 &  5.34 & - 0.14 & + 0.16 \\  % (HAPLESS\,22)   
NGC\,5690 &  7.60 & -0.03 & +0.03 \\    % (HAPLESS23/HIGH29)
NGC\,5750 &  7.12 & -0.05 & +0.06 \\    % (HAPLESS\,25) 
NGC\,5705 &  7.36 & -0.09 & +0.09 \\     % (HAPLESS26/HIGH32)
UGC\,09482 &  5.36 & -0.19 & +0.54 \\   % (HAPLESS27/HIGH36)  
NGC\,5691 &  6.85 & -0.04 & +0.04 \\    % (HAPLESS28/HIGH30)
NGC\,5713 &  7.45 & -0.03 & +0.03 \\    % (HAPLESS29/HIGH34)
UGC\,09470 &  6.05 & -0.12 & +0.14 \\  % (HAPLESS30/HIGH37)
UGC\,06903 &  7.24 & -0.06 & +0.06 \\   % (HAPLESS31/HIGH9)
CGC\,G019-084 &  6.23 & -0.06 & +0.06 \\ % (HAPLESS\,32)
UM\,491 &  5.63 & - 0.30 & + 0.46 \\  % (HAPLESS\,33) 
UGC\,07531 &  5.70 & -0.22 & +0.32 \\   % (HAPLESS34/HIGH19)
UGC\,07396 &  6.59 & - 0.30 & + 0.33 \\  % (HAPLESS\,35)
%UGC\,6879 &  7.26 & - 0.09 & + 0.11 \\    % (HAPLESS\,37)  
UGC\,04684 &  6.66 & -0.19 & +0.18 \\   % (HAPLESS39/HIGH3)
NGC\,5725 &  6.32 & -0.10 & +0.12 \\     % (HAPLESS40/HIGH33) 
UGC\,06578 &  5.23 & -0.26 & +0.25 \\  %  (HAPLESS41/HIGH5) 
UGC\,04673 &  7.41 & -0.37 & +0.32 \\   % (HIGH2) 
UGC\,04996 &  6.95 & -0.11 & +0.13 \\  %  (HIGH4)
UGC\,06970 &  6.45 & -0.36 & +0.30 \\   % (HIGH10)
NGC\,4202 &  7.50 & -0.07 & +0.06 \\     % (HIGH15)
2MASX\,J14265308 &  7.27 & -0.17 & +0.19 \\    % (HIGH24) 2MASX\,J14265308+0057462
IC\,1011 &  7.43 & -0.06 & +0.06 \\        % (HIGH25)
IC\,1010 &  7.96 & -0.12 & +0.11 \\        % (HIGH26)
\hline \\
\textbf{KINGFISH:} & & & \\
\hline 
NGC\,0337 &  7.00 & -0.04 & +0.04 \\
NGC\,0584 &  7.10 & -0.28 & +0.30 \\ 
NGC\,0628 &  7.24 & -0.04 & +0.05 \\   
NGC\,0855 &  5.52 & -0.05 & +0.05 \\   
NGC\,0925 &  7.18 & -0.06 & +0.05 \\   
NGC\,1097 &  7.71 & -0.04 & +0.04 \\   
%NGC\,1266 &  6.55 & -0.03 & +0.03 \\   
NGC\,1291 &  7.12 & -0.04 & +0.04 \\   
%NGC\,1377 &  5.86 & -0.05 & +0.06 \\
%NGC1404 & & & \\
IC\,0342 &  7.45 & -0.05 & +0.05 \\   
NGC\,1482 &  7.14 & -0.04 & +0.04 \\   
NGC\,1512 &  7.20 & -0.06 & +0.06 \\   
NGC\,2146 &  7.42 & -0.04 & +0.04 \\
HoII &  4.40 & -0.04 & +0.04 \\
DDO\,053 &  3.53 & -0.06 & +0.06 \\   
NGC\,2798 &  6.82 & -0.05 & +0.04 \\   
NGC\,2841 &  7.65 & -0.04 & +0.04 \\   
NGC\,2915 &  4.62 & -0.05 & +0.05 \\
HoI &  4.77 & -0.17 & +0.21 \\   
NGC\,2976 &  6.15 & -0.05 & +0.05 \\   
NGC\,3049 &  6.54 & -0.05 & +0.06 \\   
NGC\,3077 &  5.83 & -0.04 & +0.04 \\
M81dwB &  4.07 & -0.10 & +0.08 \\ 
NGC\,3184 &  7.38 & -0.04 & +0.04 \\   
NGC\,3190 &  7.03 & -0.04 & +0.04 \\   
NGC\,3198 &  7.44 & -0.05 & +0.05 \\
IC\,2574 &  5.55 & -0.05 & +0.05 \\  
NGC\,3265 &  6.03 & -0.05 & +0.05 \\   
NGC\,3351 &  7.07 & -0.04 & +0.04 \\   
NGC\,3521 &  7.76 & -0.04 & +0.04 \\   
NGC\,3621 &  7.15 & -0.04 & +0.04 \\   
NGC\,3627 &  7.39 & -0.04 & +0.04 \\   
NGC\,3773 &  5.66 & -0.04 & +0.04 \\   
NGC\,3938 &  7.50 & -0.04 & +0.04 \\
NGC\,4236 &  6.21 & -0.05 & +0.05 \\  
NGC\,4254 &  7.66 & -0.04 & +0.04 \\   
NGC\,4321 &  7.73 & -0.04 & +0.04 \\   
NGC\,4536 &  7.27 & -0.05 & +0.04 \\   
NGC\,4559 &  6.81 & -0.05 & +0.05 \\   
NGC\,4569 &  6.94 & -0.03 & +0.04 \\   
NGC\,4579 &  7.40 & -0.03 & +0.03 \\   
NGC\,4594 &  7.16 & -0.04 & +0.04 \\   
NGC\,4625 &  6.14 & -0.04 & +0.04 \\   
NGC\,4631 &  7.38 & -0.05 & +0.04 \\
%DDO154 & & & \\
NGC\,4826 &  6.43 & -0.03 & +0.03 \\
%DDO165 & & & \\
NGC\,5055 &  7.65 & -0.03 & +0.04 \\   
NGC\,5398 &  5.61 & -0.06 & +0.05 \\
NGC\,5408 &  4.43 & -0.05 & +0.05 \\   
NGC\,5457 &  7.69 & -0.05 & +0.05 \\   
NGC\,5474 &  6.14 & -0.05 & +0.05 \\   
NGC\,5713 &  7.22 & -0.04 & +0.04 \\   
NGC\,5866 &  6.61 & -0.03 & +0.04 \\   
NGC\,6946 &  7.64 & -0.04 & +0.04 \\   
NGC\,7331 &  7.90 & -0.04 & +0.04 \\   
NGC\,7793 &  6.65 & -0.05 & +0.05 
\end{supertabular}

%free parameters (show example fit!)

\section{Galaxy specific properties}
\label{OtherSamples.sec}
To compare the dust properties and dust scaling trends of JINGLE galaxies to other local galaxy samples, we assembled data for several well-studied nearby galaxy samples (HRS, KINGFISH, HiGH and HAPLESS). We repeat the dust SED modelling procedure for each of these galaxy samples to allow for an unbiased comparison with JINGLE. The same set of IR/submm filters has been used (where possible) to infer the contribution from stellar emission (WISE\,3.4 and 4.6\,$\mu$m), and to fit the dust emission (WISE\,12 and 22\,$\mu$m, IRAS\,60\,$\mu$m, PACS\,100 and 160\,$\mu$m, and SPIRE 250, 350 and 500\,$\mu$m flux). %Additional photometric measurements in the 3-1000\,$\mu$m wavelength range are sometimes available for these nearby galaxy samples, but to avoid biasing the SED fitting algorithm, we opt to use a similar set of IR/submm filters for all galaxy samples. %\footnote{We used the 2MASS\,$K_{\text{s}}$ fluxes to infer the stellar emission for HRS galaxies rather than the two shortest wavelength WISE bands, due to completeness in the 2MASS\,$K_{\text{s}}$ catalog. We verified that the use of the WISE fluxes renders similar results.} %\footnote{We preferred to rely on the PACS\,70\,$\mu$m fluxes rather than the IRAS\,60\,$\mu$m constraints to avoid any contamination from neighbouring galaxies within the large IRAS beam.}, with the exception of the 2MASS\,$K_{\text{s}}$ measurements for HRS galaxies and the PACS\,70\,$\mu$m fluxes for KINGFISH galaxies. 
For each of these galaxy samples, galaxy properties (i.e., metallicities, star formation rates and stellar masses etc.) have been derived in a consistent way (where possible). A short description and details about the data assembly have been presented below.

\subsection{JINGLE}
We have adopted the median $M_{\star}$ and SFR parameters (and uncertainties) inferred from a panchromatic SED fitting procedure using \texttt{MagPhys} \citep{2010MNRAS.403.1894D}. Gas-phase metallicities have been calculated using the O3N2 calibration of \citet{2004MNRAS.348L..59P} based on optical strong emission lines observed in the SDSS spectra of JINGLE galaxies \citep{2018MNRAS.481.3497S}. Due to the lack of uncertainty measurements for some galaxy samples, we have assumed a fixed 0.05\,dex uncertainty on the oxygen abundances. The H{\sc{i}} masses and uncertainties were extracted from the ALFALFA catalog \citep{2018ApJ...861...49H}. For JINGLE galaxies not covered or detected by ALFALFA, we have completed our own JINGLE H{\sc{i}} observing campaign with Arecibo (PI: M.W.L.\,Smith) and have taken the H{\sc{i}} masses and uncertainties inferred from these recent observations. Combining both datasets, we have H{\sc{i}} masses available for 161 JINGLE galaxies. With distances ranging from 56 to 223\,Mpc, it is not always easy to assign a Hubble type to each of the JINGLE galaxies. The distinction for JINGLE galaxies is therefore made only between early-type and late-type galaxies. Due the selection criteria, the JINGLE sample is dominated by late-type galaxies (186 galaxies) with a minority of 7 early-type galaxies (JINGLE\,61, 63, 76, 85, 95, 104, 125). % (see Appendix ???). 

\subsection{Herschel Reference Survey}
The \textit{Herschel} Reference Survey (HRS, \citealt{2010PASP..122..261B}) is a volume-limited sample of 322\footnote{We removed HRS\,228 from the sample as it was identified as a background object, rather than a nearby galaxy \citep{2012A&A...543A.161C}.} nearby galaxies with distances ranging from 15\,Mpc to 25\,Mpc, selected based on near-infrared magnitude limits of K$<$12\,mag for late-type galaxies (Sa-Sd-Im-BCD) and K$<$8.7\,mag for early-type galaxies (E, S0 and S0a). Due to the K-band selection criteria, the HRS is a stellar mass-selected sample dominated by more evolved sources which have already converted most of their gas into stars. It is thus not surprising that the HRS sample contains a large fraction of early-type galaxies (62), in addition to 261 late-type galaxies. Even though early-type galaxies are considered to be red and dead, nearly one quarter of the HRS ellipticals and up to 62$\%$ of HRS lenticular galaxies (S0s) have been observed to contain a non-negligible amount of dust \citep{2012ApJ...748..123S}. %Previous studies of HRS galaxies have shown that the  %some results, e.g. Cortese

For the HRS flux measurements and errors, we relied on the aperture-matched photometry measurements from CAAPR (Comprehensive \& Adaptable Aperture Photometry Routine, \citealt{2018A&A...609A..37C}). The PACS and SPIRE observations were presented in \citet{2014MNRAS.440..942C} and \citet{2012A&A...543A.161C}. Metal abundances (using the O3N2 calibration) have been taken from \citet{2013A&A...550A.115H}. We use the stellar masses and star formation rates and corresponding uncertainties inferred from \texttt{MagPhys}\footnote{\texttt{MagPhys} results are missing for four HRS galaxies (HRS\,138, 150, 183, 241) due to possible contamination from an AGN, hot X-ray halo and/or synchrotron component \citep{2017MNRAS.465.3125E}.} (as presented in \citealt{2017MNRAS.464.4680D}). The H{\sc{i}} and H$_{2}$ masses and uncertainties were taken from \citet{2014A&A...564A..65B}, with H{\sc{i}} mass measurements available for 315 HRS galaxies (of which 52 are upper limits). 

\subsection{KINGFISH}
The KINGFISH sample is composed of 61 nearby (D$\leq$30\,Mpc) galaxies with a wide range of morphological classifications \citep{2011PASP..123.1347K}. The sample is not complete as such, but with KINGFISH galaxies covering more than four orders of magnitude in stellar mass and star formation activity, the KINGFISH galaxies stretch across most of the parameter space occupied by local galaxies.
Photometric measurements and errors were taken from \citet{2017ApJ...837...90D}, who presented an updated set of multi-wavelength photometry for all KINGFISH and SINGS galaxies. The H{\sc{i}} and H$_{2}$ masses and uncertainties were taken from \citet{2014A&A...563A..31R} (including three galaxies with H{\sc{i}} upper limits and three galaxies without H{\sc{i}} masses). Stellar masses and SFRs (and uncertainties) have been adopted from the \texttt{MagPhys} fitting results presented in \citet{2019A&A...621A..51H}. Oxygen abundances (based on the O3N2 calibration) have been extracted from \citet{2019A&A...623A...5D}, with metallicities available for 46 out of 61 KINGFISH galaxies. For the missing 15 galaxies, oxygen abundances (relying on the \citealt{2004ApJ...617..240K} metallicity calibration) have been taken from \citet{2011PASP..123.1347K}, and were converted to the 03N2 calibration from \citet{2004MNRAS.348L..59P} following the conversion formula from \citet{2008ApJ...681.1183K}. For two metal-poor KINGFISH galaxies (DDO\,053 and DDO\,165), the metal abundances reported in \citet{2011PASP..123.1347K} were used due to the absence of a reliable conversion at these low metallicities \citep{2008ApJ...681.1183K}. To avoid any contamination from an AGN, we have removed three KINGFISH galaxies (NGC\,1316, NGC\,4725, NGC\,4736; \citealt{2015A&A...582A.121R}) from our sample. %, which leaves us with 58 KINGFISH galaxies.%We have assumed a 0.15\,dex uncertainty for these converted metal abundances to be conservative. 

\subsection{HAPLESS and HiGH samples}
The HAPLESS and HiGH galaxy samples have been selected from the \textit{Herschel} Astrophysical Terahertz Large Area Survey (H-ATLAS, \citealt{2010PASP..122..499E}). The HAPLESS sample was selected based on bright 250\,$\mu$m emission, resulting in a local sample of 42 dusty galaxies at distances between 15 and 46\,Mpc. The selection criteria and galaxy properties have been outlined in \citet{2015MNRAS.452..397C}; they show that HAPLESS galaxies are predominantly blue, star-forming galaxies in an early stage of evolution. HiGH galaxies were selected based on their H{\sc{i}} detections. The 40 H{\sc{i}}-rich galaxies span distances from 11.3 to 158.9\,Mpc, and typically have low stellar masses ($\leq$10$^{9}$\,M$_{\odot}$) and relatively low dust masses for their stellar mass content, indicating that these galaxies are also at an early stage of evolution. Due to similarities in the selection criteria, it is not surprising that the dust-selected HAPLESS and H{\sc{i}}-selected HiGH galaxy samples have 22 galaxies in common. 

Photometry measurements and errors for HAPLESS and HiGH samples have been derived using the CAAPR software from \citet{2018A&A...609A..37C}, and the \texttt{MagPhys} fitting results (for stellar masses and SFRs) and H{\sc{i}} masses (and uncertainties) were taken from \citet{2015MNRAS.452..397C} and \citet{2017MNRAS.464.4680D}, respectively. H{\sc{i}} mass measurements were available for all HiGH galaxies, and 38/42 HAPLESS galaxies. Metal abundances were taken from \citet{2017MNRAS.471.1743D}.

\section{Estimating star formation histories}
\label{Sec_SFH}
A reasonable estimation of a galaxy's SFH is vital to constrain stellar dust production and supernova rates. To avoid biasing the inferred SFHs by assuming a particular SFH shape, we resort to non-parametric star formation histories, where we determine the average SFR in three ``look-back time" bins: 0-10\,Myr, 10-100\,Myr and 100\,Myr-12\,Gyr, and where we have assumed that galaxies started forming stars roughly 12\,Gyr ago. To determine the average SFR during those three distinct periods, we assembled H$\alpha$, WISE\,22\,$\mu$m, FUV, and TIR luminosities, and inferred two different measurements of the recent star formation activity: SFR(H$\alpha$+WISE\,22\,$\mu$m) and SFR(FUV+TIR)\footnote{We have used the prescriptions from \citet{2012ARA&A..50..531K}, and have assumed that the WISE\,22\,$\mu$m filter is equivalent to the MIPS\,24\,$\mu$m or IRAS\,25\,$\mu$m filter.} (see Table \ref{MHIMstar_bins_SFRs} for the estimated SFRs). With H$\alpha$ and WISE\,22\,$\mu$m emission being mostly sensitive to young stellar populations with ages $\lesssim$10\,Myr, while FUV and TIR probe star formation on recent timescales of $\sim$100\,Myr (e.g., \citealt{2012ARA&A..50..531K}), the ratio of these SFR estimates, SFR(H$\alpha$+WISE\,22\,$\mu$m)/SFR(FUV+TIR), provides an indication of how much star formation occurred during the last 10\,Myr as opposed to a longer 100\,Myr look-back time period. For each sample of galaxies at a similar evolutionary stage, as indicated by the ratio of their H{\sc{i}} gas mass versus stellar mass, $M_{\text{HI}}$/$M_{\star}$, we have calculated the average ratio of these SFRs. These ratios, in comparison to the average stellar mass and SFR (which corresponds to a constant SFR over the last 100 Myr as inferred by \texttt{MagPhys}), allow us to constrain the average SFR during the last 10 and 100\,Myr. The average (constant) SFR during the 100\,Myr-12\,Gyr period is then constrained based on the current ``average" stellar mass for each galaxy bin.

\begin{figure*}
	\includegraphics[width=5.8cm]{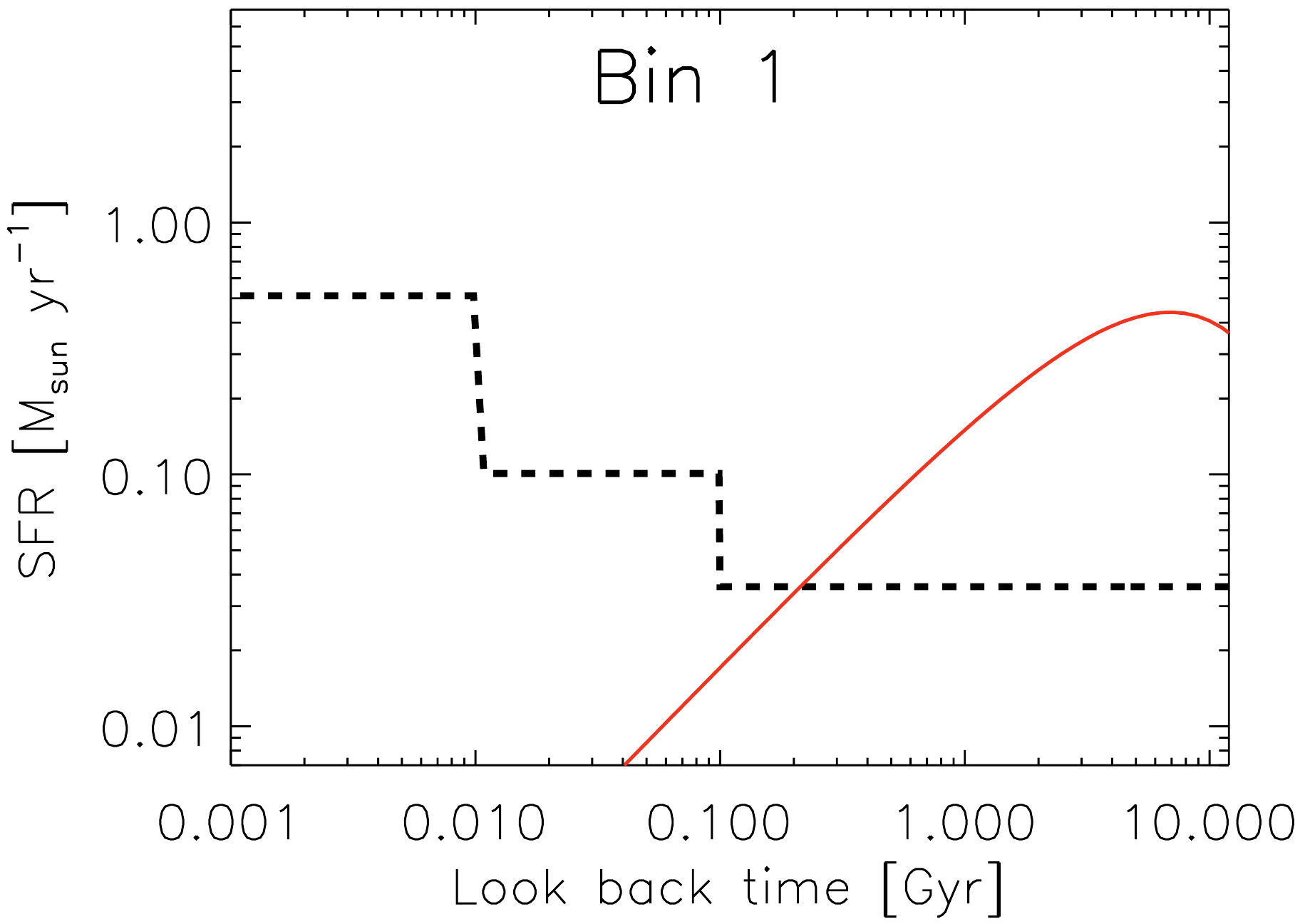}
	\includegraphics[width=5.5cm]{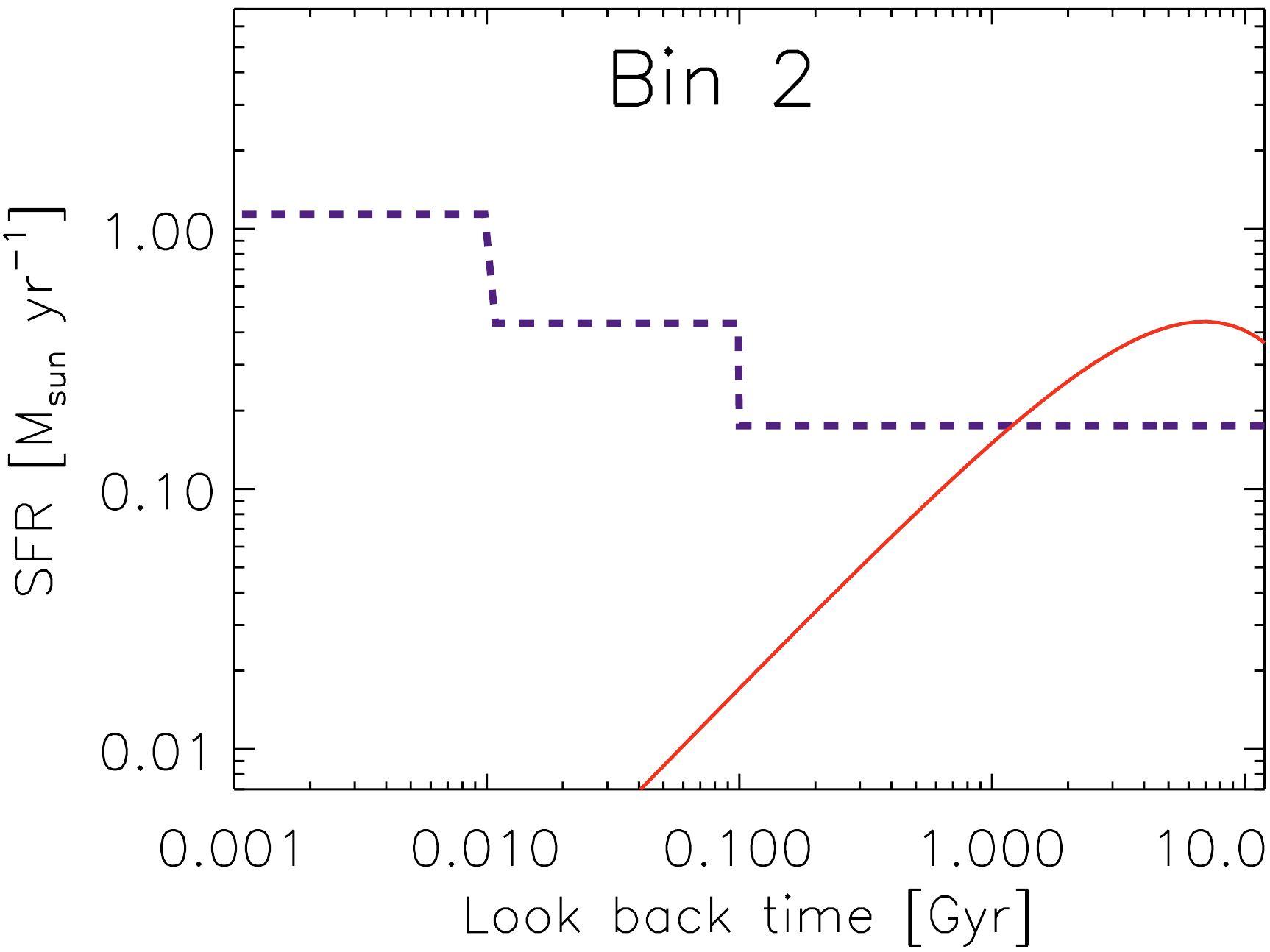}
	\includegraphics[width=5.5cm]{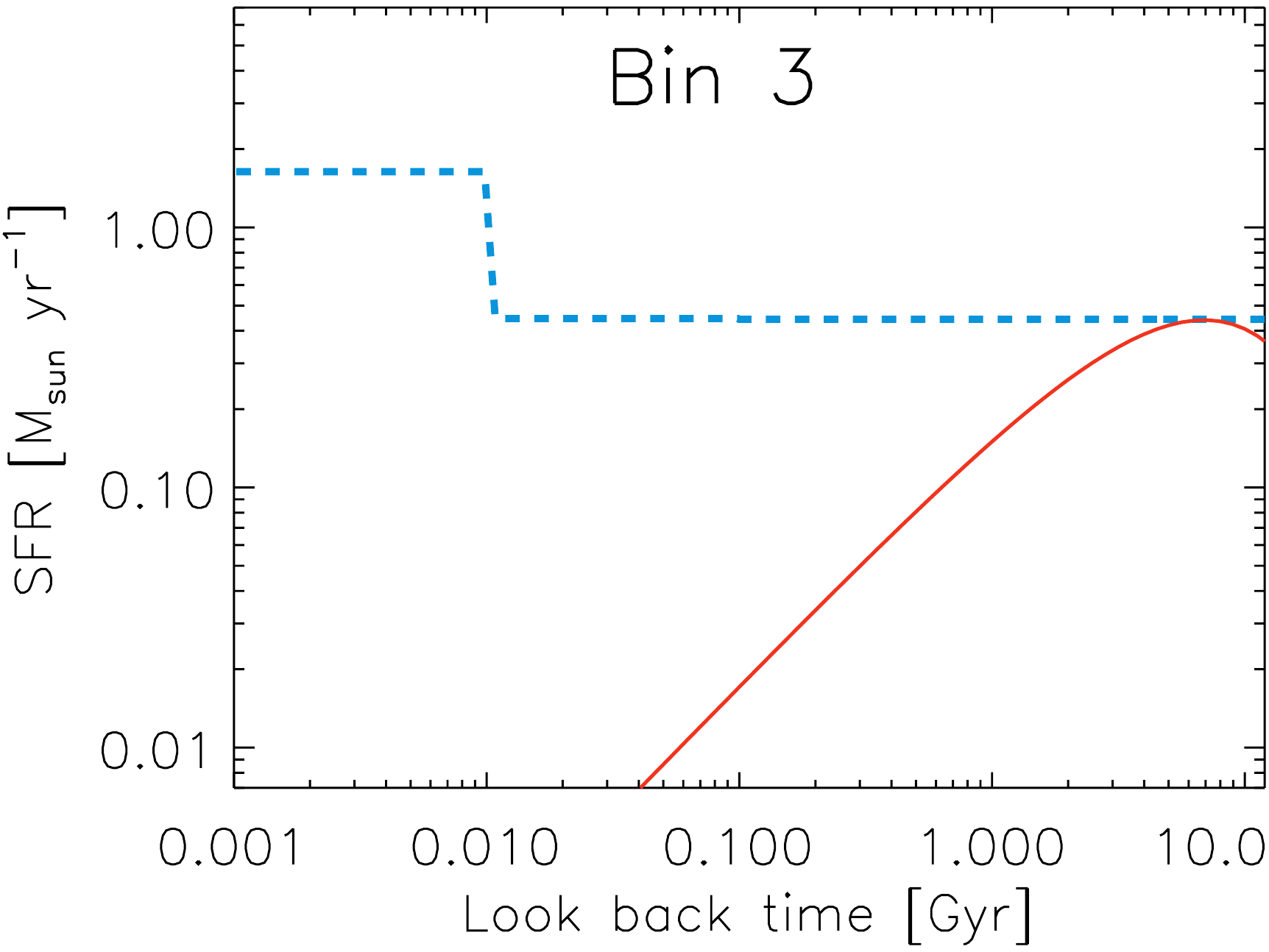}
	\includegraphics[width=5.5cm]{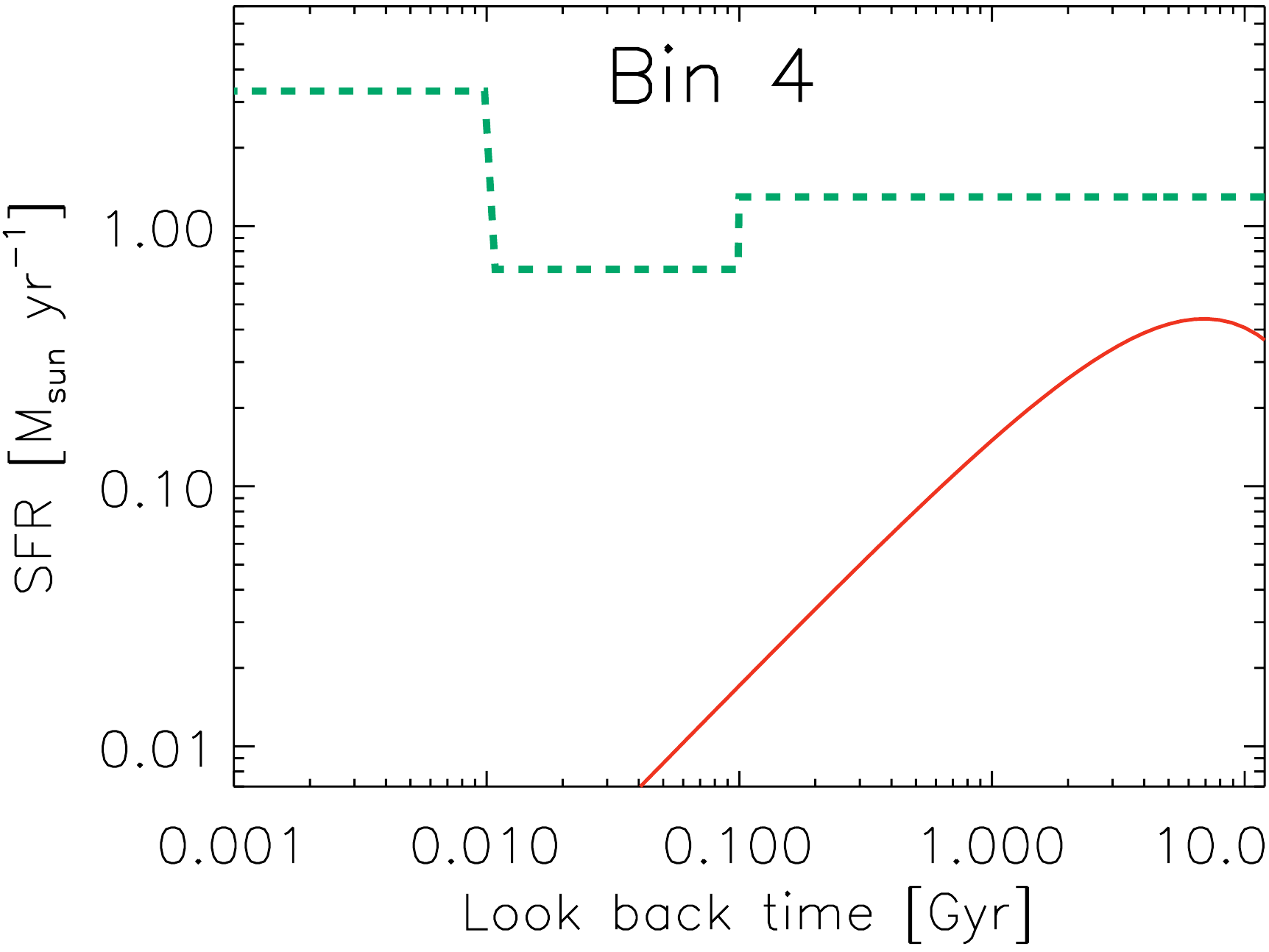}
	\includegraphics[width=5.5cm]{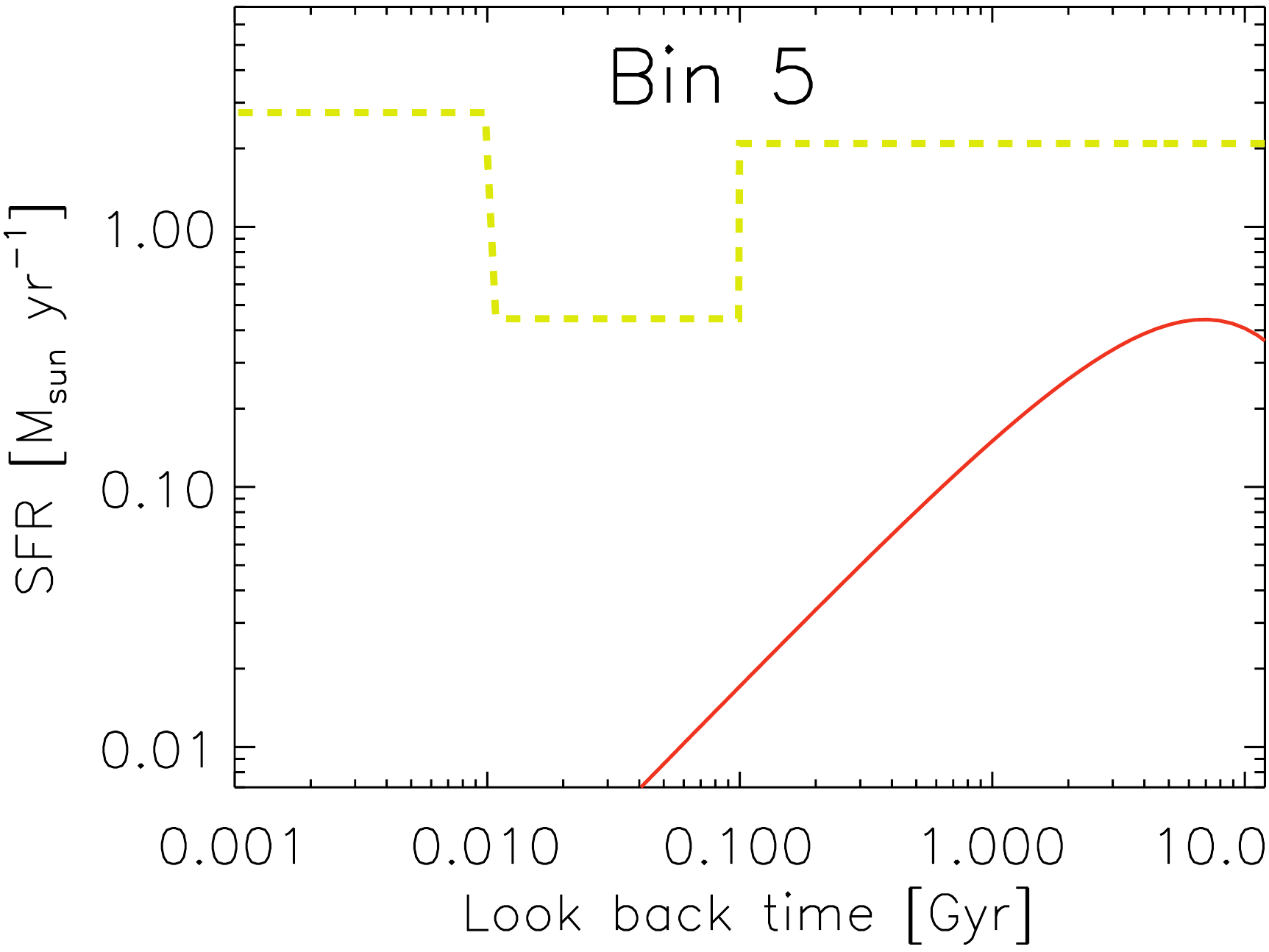}
	\includegraphics[width=5.5cm]{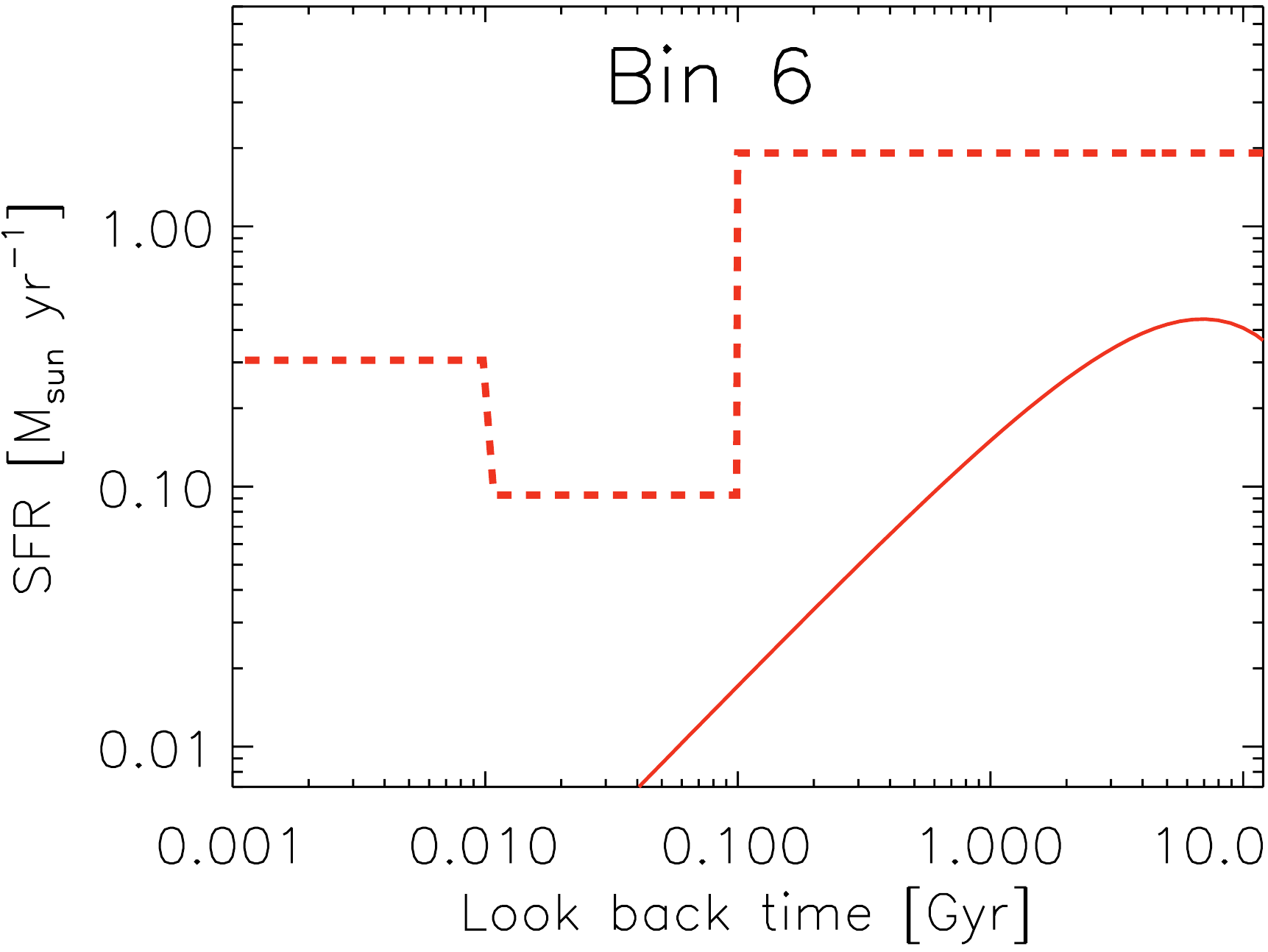}
    \caption{The customised star formation histories that have been inferred for each of the galaxy subsamples following the method presented in Appendix \ref{Sec_SFH}, and were used in our Model I. Galaxies evolve from Bin 1 (corresponding to an early stage of evolution as inferred from the high $M_{\text{HI}}$/$M_{\star}$ ratios) through to Bin 6 (characteristic of evolved galaxies with low $M_{\text{HI}}$/$M_{\star}$ ratios). The customised SFHs are compared to the delayed star formation history (red curve) applied by \citet{2017MNRAS.471.1743D}, which has been used to predict galaxy's dust, H{\sc{i}} and metal content in our Model II.}    \label{Image_SFH} 
\end{figure*}

Figure \ref{Image_SFH} shows the SFHs that have been inferred in this way for the six $M_{\text{HI}}$/$M_{\star}$ bins. For comparison, the delayed SFH model used by \citet{2017MNRAS.471.1743D} is indicated with a red, solid line and was used in Models II to test the sensitivity of our model results to the assumed SFH. Galaxies at an early stage of evolution (Bin\,1) did experience some star formation in the past, but have converted gas into stars at an increased rate during the last 100\,Myr, with a further increment in their star formation activity during the last 10\,Myr. Similar SFHs were inferred for the two subsequent bins (Bins 2 and 3), but show less pronounced differences in their average SFRs during the three subsequent epochs. More evolved galaxies (Bins 4-5) show a dip in their star formation activity over 100\,Myr timescales, followed by an increased star formation activity during the last 10\,Myr. The most evolved galaxies (Bin\,6) show an overall decrease in their star formation activity during the last 100\,Myr, which can be expected if most gas has been consumed in earlier star formation episodes (or has been partly removed from the galaxy).

\begin{figure}
\includegraphics[width=8.5cm]{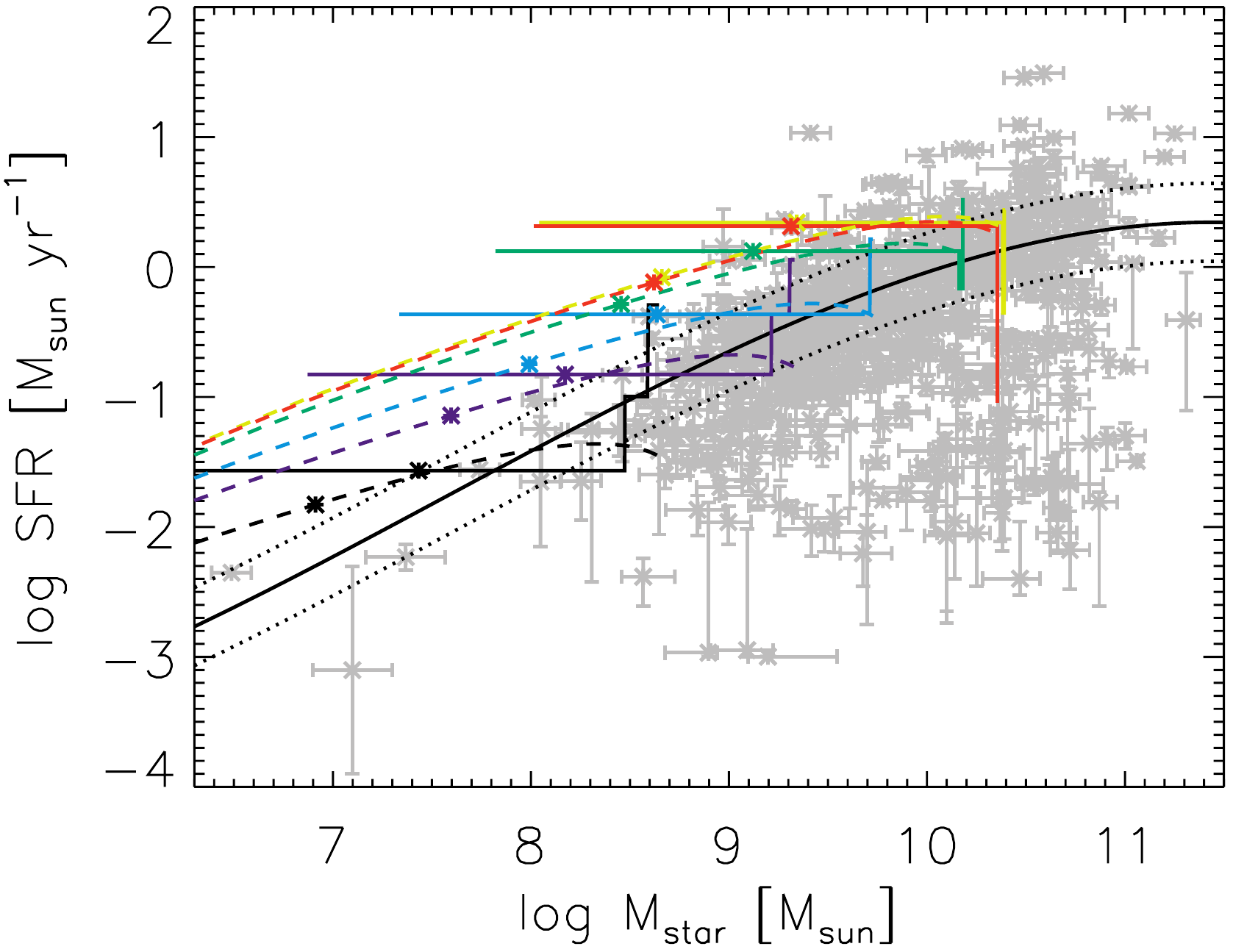} 
\caption{The star formation main sequence for JINGLE, HRS, KINGFISH, HAPLESS and HiGH galaxies (grey symbols), with the average trend found by \citet{2016MNRAS.462.1749S} overlaid as a solid black line. An offset of 0.3\,dex on either side of this trend (similar to the spread observed by \citealt{2007ApJ...660L..43N,2009ApJ...698L.116P,2010A&A...518L..25R}) has been indicated by dashed black lines, and encloses galaxies located on the SF main sequence. The evolutionary path for the customised and delayed SFHs have been overlaid with solid and dashed curves, respectively, for galaxy Bins 1 through to 6 in black, purple, blue, green, yellow and red. The location of the asterisk corresponds to a lookback time of 2\,Gyr.} \label{Scaling_SFmain_SFH}
\end{figure}

We realise that these SFHs will not be representative for all galaxies presented in this paper. However, the star formation activity during the last 10 and 100\,Myr, and how this relates to any earlier star formation activity in the galaxy, will be important to assess how much dust has formed over a galaxy lifetime. %, and in particular the last evolutionary stages that have shaped the content of its current interstellar reservoir. %, but will rather provide a first guidance of how the different SFH shapes will lead to the current dust, metal and gas mass content
The non-parametric SFHs for the six galaxy bins have been overlaid on the star formation main sequence (see Figure \ref{Scaling_SFmain_SFH}). All model SFH tracks display a horizontal trend in the $M_{\star}$-SFR plane, due to the fixed SFR at lookback times older than 100\,Myr, and reside on the SF main sequence during most of a galaxy's lifetime. Only for lookback times $>$11\,Gyr, the model SFH tracks are positioned above the SF main sequence inferred for the local Universe, which is consistent with the expected shift of the main sequence at earlier times \citep{2017ApJ...847...76S}.

%Table including the SFR values
\begin{table*}
%\centering
\caption{We present the $\log\,M_{\text{HI}}$/$M_{\star}$ range and sample size in each galaxy bin, along with the subsample size with available H$\alpha$, FUV, WISE\,22 and TIR measurements which were used to estimate the past ($>$100\,Myr), recent (10-100\,Myr) and present (0-10\,Myr) SFRs inferred for each of the six galaxy bins.}
\label{MHIMstar_bins_SFRs}
\begin{tabular}{lllcccc} % four columns, alignment for each
\hline
Bin & $\log\,M_{\text{HI}}$/$M_{\star}$ range & $N_{\text{galaxies}}$ & $N_{\text{SFR}}$ & $SFR_{\text{>100Myr}}$ [M$_{\odot}$ yr$^{-1}$] & $SFR_{\text{10-100Myr}}$ [M$_{\odot}$ yr$^{-1}$] & $SFR_{\text{<10Myr}}$ [M$_{\odot}$ yr$^{-1}$] \\ 
\hline
1 & [0.5,-] & 17 & 3 & 0.04 & 0.10 & 0.51 \\
2 & [0,0.5[ & 81 & 31 & 0.18 & 0.43 & 1.14  \\
3 & [-0.5,0[ & 134 & 47 & 0.44 & 0.45 & 1.64 \\
4 & [-1.0,-0.5[ & 132 & 62 & 1.30 & 0.68 & 3.30 \\
5 & [-1.5,-1.0[ & 46 & 20 & 2.09 & 0.44 & 2.75 \\
6 & [-,-1.5[ & 13 & 2 & 1.91 & 0.09 & 0.31 \\
\hline 
\end{tabular}
\end{table*}

\section{Dust and Element evolUtion modelS: DEUS}
\label{DEUS.sec}
We here describe the evolution of stars, metals and dust, as it is implemented in DEUS. We have adopted the same notation as commonly used throughout the literature (e.g., \citealt{2014MNRAS.441.1040R,2017MNRAS.471.1743D}). 
\subsection{Model implementation}
\label{Sec_Implement}
\subsubsection{Stellar populations}
The stellar mass of a galaxy, $M_{\star}$, evolves with time according to:
\begin{equation}
\frac{dM_{\star}}{dt}~=~\psi(t) - e(t) %\,\times\,dt
\end{equation}
where $\psi(t)$ is the SFR and $e(t)$ is the ejected mass lost from stars throughout their lives. We assume the mass loss occurs at the end of stellar evolution:
\begin{equation}
e(t) = \int_{m_{\tau_{\text{m}}}}^{m_{\text{max}}}[m-m_{\text{R}}(m)] \psi(t-\tau_{\text{m}}) \phi(m) dm,    
\end{equation}
where the integral runs over stars with masses ($m_{\tau_{\text{m}}}$) with a lifetime $\tau_{\text{m}}$, which formed at a time $t-\tau_{\text{m}}$. The ejected masses are calculated as the difference between the initial stellar mass, $m$, and the remnant mass, $m_{\text{R}}(m)$ for a star with a given initial mass $m$. The stellar lifetimes are adopted from \citet{1992A&AS...96..269S}, while the remnant masses are taken from \citet{1993ApJ...403..630P}. We assume a \citet{2003PASP..115..763C} initial mass function $\phi(m)$, maximum stellar mass $m_{\text{max}}$ of 100\,M$_{\odot}$, with the masses of stars ranging from 0.1 to 100\,M$_{\odot}$ in stellar mass bins of 0.1\,M$_{\odot}$.

\subsubsection{Gaseous reservoirs}
\label{Gas.sec}
The total (interstellar) gas content of a galaxy, $M_{\text{gas}}$, evolves as:
\begin{equation}
\frac{dM_{\text{gas}}}{dt}~=~-\psi(t)~+~e(t)~+~I(t)~-~O(t), 
\end{equation}
where the first term accounts for the gas lost through ``astration", as the gas is being consumed to form new stars at a rate equal to the SFR, $\psi$(t). The second term accounts for the mass gain through mass loss during late stellar evolutionary phases, while the third and fourth term represent the infall and outflow of gas. For Models I and II, we have assumed closed-box models with no infalling or outflowing gas (i.e., $I(t)=0$ and $O(t)=0$). In Model III, we abandoned this closed-box assumption (see Section \ref{Sec_Infall_Outflow}) and have explored how the DEUS parameters are affected by the in- and outflow of gas. Our current picture of galaxy evolution suggests that relatively pristine gas is being funneled through cold gas streams and is slowly accreted from the halo onto the galaxy disk during its lifetime (e.g., \citealt{2009Natur.457..451D}). Neglecting this process, which works to dilute metal fractions in the interstellar medium, will accelerate the buildup of a galaxy's metal content over time. Due to poor constraints on the initial gas mass and metallicity of the halo from which a galaxy forms, it is quite tedious to constrain the rates of infalling and outflowing gas, and their mass loading factors (as they will be influenced by the initial gas and metal abundances) based on the observations at hand. To work around this problem, we leave the initial gas mass of the halo as a free parameter in DEUS and infer what gas mass is needed to reproduce the observed present-day specific H{\sc{i}} gas masses ($M_{\text{gas}}$/$
M_{\star}$) and oxygen abundances. In this manner, the initial gas mass will be adapted to account for any gas mass accreted/lost throughout a galaxy's lifetime. Naturally, this approach will only work for galaxies that have not experienced any recent gas in- or outflows, as those would affect their dust and gas masses on timescales significantly shorter than the time needed to replenish a galaxy's dust and gas content. In this paper, we therefore refrain from modelling the sub-population of H{\sc{i}}-deficient HRS galaxies which have experienced massive gas outflows that have shut down star formation on galaxy-wide scales in these systems over the last $\lesssim$1.5\,Gyr (e.g., \citealt{2006ApJ...651..811B,2010A&A...514A..33P,2016A&A...585A..43C}).   %The fact that we can reproduce them both means that we are OK!?

Rather than total gas masses, we only have H{\sc{i}} gas mass measurements available for most galaxies in our local galaxy sample. The determination of molecular gas masses for JINGLE galaxies has been deferred to JINGLE Paper III for a subsample of JINGLE galaxies, and to future work, in anticipation of the completion of our JCMT CO(2-1) survey. Molecular gas mass estimates exist for a few nearby galaxy samples (e.g., HRS, KINGFISH), but these measurements are either hampered by the extrapolation of localized CO observations to galaxy-wide scales, or by the uncertainties involved in the conversion from CO-to-H$_{2}$ masses, which varies non-linearly with the metallicity and ISM phase filling factors in galaxies (e.g., \citealt{2012AJ....143..138S,2016A&A...588A..23A,2017MNRAS.470.4750A}, Madden et al.\,in prep.). 

To compare the total gas masses in our model to the observed H{\sc{i}} content, which are related through:
\begin{equation}
M_{\text{gas}}~=~\xi~(M_{\text{HI}}+M_{\text{H}_{2}}),    
\end{equation} 
we need to estimate the correction factor ($\xi$) to take into account the gas fraction of elements heavier than hydrogen. Rather than a canonical correction factor of 1.36, often used in the literature, we follow \citet{2016MNRAS.459.1646C} to define the correction factor $\xi$ as:
\begin{equation}
\label{Eq_heavyelcorr}
\xi~=~\frac{1}{1-\left(f_{\text{He}_{\text{p}}}~+~f_{\text{Z}}\left[\frac{\Delta f_{\text{He}}}{\Delta f_{\text{Z}}} \right] \right) - f_{\text{Z}}}
\end{equation}
which depends on the primordial helium mass fraction ($f_{\text{He}_{\text{p}}}$=0.2485$\pm$0.0002, \citealt{2013JCAP...11..017A}), the metal mass fraction $f_{\text{Z}}$ of the galaxy at that point in time, and the evolution of the helium mass fraction with metallicity, $\left[\frac{\Delta f_{\text{He}}}{\Delta f_{\text{Z}}} \right]$, which is assumed to be equal to $\left[\frac{\Delta f_{\text{He}}}{\Delta f_{\text{Z}}} \right]$=1.41$\pm$0.62 \citep{2006AJ....132.2326B}.  The metal mass fraction is defined as $f_{\text{Z}}=f_{\text{Z}_{\odot}} Z_{\text{O}}$ with the Solar metal mass fraction assumed to be $f_{\text{Z}_{\odot}}$=0.0134 \citep{2009ARA&A..47..481A}. The oxygen-based metallicity $Z_{\text{O}}$ is defined as:
\begin{equation}
Z_{\text{O}} = \delta_{\text{O}} \left[\frac{O}{H}\right]/\left[\frac{O}{H}\right]_{\odot}    
\end{equation}
with the Solar oxygen abundance [12+$\log\frac{O}{H}$]=8.69$\pm$0.05 \citep{2009ARA&A..47..481A}, and a correction factor $\delta_{\text{O}}$ (=1.32$\pm$0.09, \citealt{2009MNRAS.395..855M}) to account for the depletion of oxygen (mostly in H{\sc{ii}} regions). 

In addition to this $\xi$ correction factor, we need to correct our model gas masses for a contribution from molecular gas, to allow a direct comparison with the observed H{\sc{i}} masses. Hereto, we rely on the scaling relation of the H$_{2}$-to-H{\sc{i}} mass ratio as a function of the stellar mass inferred by \citet{2014MNRAS.442.2398P} for disc-dominated galaxies using metallicity-based H$_{2}$ formation recipes in their semi-analytic models of galaxy evolution. Their modelled scaling laws agree well with various literature sets of observed data from \citet{2008AJ....136.2782L}, \citet{2011MNRAS.415...32S} and \citet{2014A&A...564A..66B}, and are applicable up to redshifts of z$\sim$2.

\subsubsection{Metal budgets}
\label{Metals.sec}
The mass of metals in the ISM ($M_{\text{metals}}$(gas)) evolves with time according to:
\begin{equation}
\label{Eq_evol_metalmass}
\frac{dM_{\text{metals}}(gas)}{dt}~=~-Z_{\text{M}}(t)\psi(t)~+~e_{\text{Z}}(t)~+~Z_{\text{I}}I(t)~-~Z_{\text{O}}O(t)    
\end{equation}
where $Z_{\text{M}}(t)$ represents the mass fraction of heavy elements in the gas phase (i.e., $Z_{\text{M}}=M_{\text{metals}}\text{(gas)}/M_{\text{gas}}$). $e_{\text{Z}}$ accounts for the ejected metals through stellar mass loss during late stellar evolutionary stages, and can be calculated as:
\begin{equation}
\begin{split}
e_{\text{Z}}(t)~=~\int_{m_{\tau_{\text{m}}}}^{m_{\text{max}}} \left(\left[m-m_{\text{R}(m)}\right]Z(t-\tau_{\text{m}})+mp_{\text{Z}} \right) \\
\times\psi(t-\tau_{\text{m}})\phi(m)dm    
\end{split}
\end{equation}
where the first term between parentheses accounts for the metals ejected through stellar mass loss (based on the metallicity of the gas from which the star formed t~-~$\tau_{\text{m}}$ ago), and the second term ($mp_{\text{Z}}(m)$) represents the heavy elements produced by a star with initial mass $m$ and metallicity $Z$ (where $p_{\text{Z}}(m)$ represents the stellar yields). The metal yields for low- and intermediate mass stars (LIMS, with progenitor masses $M_{\text{prog}}\leq$8\,M$_{\odot}$) and for massive stars (8\,M$_{\odot} <M_{\text{prog}}\leq$40M$_{\odot}$) were taken from \citet{1997A&AS..123..305V} and \citet{1995ApJS..101..181W}, respectively. The third and fourth term of Eq. \ref{Eq_evol_metalmass}, which are related to gas inflows and outflows, are not considered in our closed-box Models I and II. %Figure \ref{} presents the assumed yields for + assumptions made (e.g., between 8 and 11 Msun etc)

Rather than tracking metal fractions, our observations constrain the oxygen abundances in galaxies. To convert these oxygen abundances into metal mass fractions, we rely on the conversion from Solar oxygen abundance (12+$\log$(O/H)$_{\odot}$=8.69) to the Solar metal fraction ($f_{Z_{\odot}}$=0.134), which results in the following relation\footnote{This conversion factor translates into an oxygen mass fraction, compared to the total mass of metals, of 44$\%$, if we first convert the oxygen abundance into oxygen mass fractions through $X_{\text{O}}=12\frac{O}{H}$. The latter value corresponds to the lower limit of predictions between 45-60$\%$ from \citet{2002ApJ...581.1019G}.}: $f_{\text{Z}}=27.36\times10^{(12+\log(O/H))-12}$. Note that this relation has been inferred based on the conversion from total metal fraction to oxygen abundance at solar metallicity, but that the exact conversion at other metallicities likely deviates from this relation due to oxygen abundance variations. At the onset of our simulation, we have assumed a metallicity $Z_{\text{ini}}$=$M_{Z}$/$M_{\text{gas}}$=0.0001\footnote{We verified that assuming an absence of any metals at the start of our simulation renders similar results.}, and an oxygen mass $M_{\text{O}}$=0.44$\times M_{Z}$. %The initial gas mass was left to vary in our models to reproduce the current specific H{\sc{i}} gas masses and metallicities observed in local galaxies.

\subsubsection{Dust reservoirs}
\label{Dust.sec}
The dust mass evolves with time according to:
\begin{equation}
\begin{split}
\frac{dM_{\text{dust}}}{dt}=\int_{m_{\tau_{\text{m}}}}^{m_{\text{max}}}\left([m-m_{\text{R}}(m)]Z(t-\tau_{\text{m}})\delta_{\text{LIMS}}+mp_{\text{Z}}\delta_{\text{dust}}\right)\\  \times\psi(t-\tau_{\text{m}})\phi(m)dm - (M_{\text{dust}}/M_{\text{gas}})\psi(t) \\ - M_{\text{dust}}\delta_{\text{destr}}(t) + M_{\text{dust}}\delta_{\text{grow}}(t) \\ + (M_{\text{dust}}/M_{\text{gas}})_{\text{I}}I(t) - (M_{\text{dust}}/M_{\text{gas}})_{\text{O}}O(t)
\end{split}
\end{equation}
where the first term accounts for dust production by LIMS (during late stellar evolutionary stages) and massive stars (after their explosion as core-collapse supernovae). The second and third term account for dust destruction through astration (i.e., dust that is incorporated into new stars) and dust destruction (mostly) through supernova shocks in the interstellar medium. The fourth term accounts for dust grown through accretion of elements onto pre-existing grain seeds in interstellar clouds. The last two terms have been neglected for closed-box Models I and II, while dust outflows (no dust inflows) have been accounted for in Model III (see App. \ref{Sec_Infall_Outflow}).

We rely on the model dust yields, $\delta_{\text{dust}}$, from \citet{2006A&A...447..553F} for LIMS and from \citet{2019MNRAS.484.2587M} (for their non-rotating CE models\footnote{The ``CE" models from \citet{2019MNRAS.484.2587M} correspond to a set of models for which the properties of the explosions have been calibrated to reproduce empirically inferred values of the $^{56}$Ni mass, as opposed to their ``FE" models for which a fixed explosion energy is assumed.}) for supernovae to model stellar dust production\footnote{We have interpolated between the yields for evolved stars and supernovae to infer dust yields for progenitor masses between 7 and 13\,M$_{\odot}$.}. The supernova dust yields from \citet{2019MNRAS.484.2587M} are at the high end of values reported in the literature, and have not yet been corrected for the destruction of freshly condensed supernova dust by a reverse shock. There is a large uncertainty inherent to the efficiency of reverse shock dust destruction, due to its dependence on the grain size and composition, the clumpiness of the dust distribution in the supernova ejecta, and the reverse shock velocity (which is set by the ambient circum- and interstellar densities). Values have been quoted ranging from a few percent to 100$\%$ dust destruction efficiencies (e.g., \citealt{2007MNRAS.378..973B,2010ApJ...715.1575S,2016A&A...587A.157B,2019MNRAS.489.4465K}). Rather than fixing the survival fraction of supernova dust for destruction by the reverse shock, we implement the reverse shock dust destruction efficiency as a free parameter ($f_{\text{survival}}$) of our model. We furthermore assume that stars more massive than 40\,M$_{\odot}$ will end up as black holes, and do not contribute to the dust enrichment of the ISM. 

To model the dust destruction efficiency, $\delta_{\text{destr}}$, we infer an estimate for the average dust destruction timescale $\tau_{destr}$, which describes the destruction of dust with time through $M_{\text{dust}}(t)=M_{\text{dust}}(0)\exp(-t/\tau_{\text{destr}})$ and which is assumed to be the inverse of the dust destruction efficiency, i.e., $\delta_{\text{destr}}$=$\tau_{destr}^{-1}$. We define the dust destruction timescale as:
\begin{equation}
\label{Eq_dustdestruction}
\tau_{\text{destr}} = \frac{M_{\text{gas}}}{\delta_{\text{SN}} R_{\text{SN}} M_{\text{cl}}}    
\end{equation}
where $R_{\text{SN}}$ is the supernova rate per gas mass, $M_{\text{gas}}$, and per 20\,Myr (i.e., the time interval in our chemical evolution model), which can be inferred from the star formation history and the number of massive stars with progenitor masses between 8 and 40\,M$_{\odot}$ that have reached the end of their lives during the last 20\,Myr period. The average gas mass cleared by each supernova, $M_{\text{cl}}$, (which is a free parameter in our models) depends on the supernova energy, and the structure, clumpiness and density of the ambient interstellar medium \citep{2015ApJ...803....7S,2019MNRAS.487.3252H}. The gas mass cleared of dust, $M_{\text{cl}}$, is thought to vary from 1000\,M$_{\odot}$ under average warm neutral medium conditions ($n_{\text{H}}\sim$0.1\,cm$^{3}$) to 700\,M$_{\odot}$ for the cold neutral medium ($n_{\text{H}}\sim$10\,cm$^{3}$) up to $\lesssim$350\,M$_{\odot}$ in molecular clouds with $n_{\text{H}}>$100\,cm$^{3}$ \citep{2019MNRAS.487.3252H}. The correction factor $\delta_{\text{SN}}$ accounts for ``delayed" or ``clustered" supernova explosions occurring above the galactic plane and within superbubbles, where they will neglect to destroy a significant mass of pre-existing dust. We fix $\delta_{\text{SN}}$ to 0.4, estimated from simulations with a random positional occurrence of supernova \citep{2019MNRAS.487.3252H}, and consistent with earlier estimates (0.36, \citealt{1989IAUS..135..431M}). %We leave the gas mass cleared of dust as a free parameter in our model.

The efficiency of grain growth processes $\delta_{\text{grow}}$(=$\tau_{\text{grow}}^{-1}$) is estimated from the average grain growth timescale:
\begin{equation}
\label{Eq_graingrowth}
\tau_{\text{grow}} = \frac{M_{\text{gas}}}{\epsilon Z_{\text{M}} \psi} \left(1-\frac{\eta_{\text{dust}}}{Z_{\text{M}}}\right)^{-1} 
\end{equation}
following \citet{2012MNRAS.423...26M,2012MNRAS.423...38M} (see Section \ref{GrainGrowth.sec} for a quick summary of their derivation of the grain growth parameter), where $Z_{\text{M}}$ (=$M_{\text{metals}}$(gas)/$M_{\text{gas}}$) represents the metallicity (in the gas phase), $\psi$ corresponds to the star formation rate and $\eta_{\text{dust}}$ is the dust-to-gas ratio. The $\epsilon$ parameter determines the grain growth parameter, and is set as a free parameter in our model.

\subsection{Grain growth parameter}
\label{GrainGrowth.sec}

In this paragraph, we reiterate the derivation of the grain growth prescription presented by \citet{2012MNRAS.423...26M} to illustrate the assumptions that have gone into the derivation of Eq. \ref{Eq_graingrowth} and to give a physical interpretation of the grain growth parameter $\epsilon$. 

The rate (per unit volume) at which the number of atoms, $N_{\text{A}}$, increases in dust grains through accretion of metals on to these dust grains can be expressed as:
\begin{equation}
\frac{\text{d}N_{\text{A}}}{\text{d}t}~=~f_{\text{s}}\,\pi a^{\text{2}}\,n_{\text{Z}}\,n_{\text{gr}}\,\langle v_{\text{gas}}\rangle  
\end{equation}
where $f_{\text{s}}$ is the sticking coefficient (which gives the probability that an atom will stick to a grain), $a$ is the typical grain radius, and $n_{\text{Z}}$ and $n_{\text{gr}}$ are the average number densities for atomic metals and dust grains, respectively, while $\langle v_{\text{gas}}\rangle$ is the mean thermal speed of the gas particles. Rather than number densities, this expression can be rewritten in terms of dust surface densities ($\Sigma_{\text{dust}}$) of molecular gas clouds, i.e.:
\begin{equation}
\frac{\text{d} \Sigma_{\text{dust}}}{\text{d}t}~=~\frac{f_{\text{s}}\,\pi a^{\text{2}}\,\widetilde{\Sigma}_{\text{Z}}\,\Sigma_{dust}\,\langle v_{\text{gas}}\rangle}{\langle m_{\text{gr}}\rangle\,d_{\text{c}}}
\end{equation}
where $\widetilde{\Sigma}_{\text{Z}}$ is the surface density of free (atomic) metals, $\langle m_{\text{gr}}\rangle$ is the mean mass of interstellar dust grains, and $d_{\text{c}}$ is the size of the molecular cloud.
The inferred grain growth timescale can then be written as:
\begin{equation}
\tau_{\text{gr}}~=~\tau_{\text{0}}\left(1-\frac{\eta_{\text{dust}}}{Z_{\text{M}}} \right)^{-1}
\end{equation}
with 
\begin{equation}
\label{Eq_tau0}
\tau_{\text{0}}~=~\frac{\langle m_{\text{gr}}\rangle\,d_{\text{c}}}{f_{\text{s}}\,\pi a^{\text{2}}\,\widetilde{\Sigma}_{\text{Z}}\,\langle v_{\text{gas}}\rangle} \approx \frac{\langle m_{\text{gr}}\rangle\,d_{\text{c}}}{f_{\text{s}}\,\pi a^{\text{2}}\,Z_{\text{M}}\,\Sigma_{\text{mol}}\,\langle v_{\text{gas}}\rangle} 
\end{equation}
where $\Sigma_{\text{mol}}$ is the molecular gas surface density and $Z_{\text{M}}$ the metallicity.
It is assumed that $\Sigma_{\text{mol}}\approx\Sigma_{\text{H}_{\text{2}}}$, and that the star formation rate surface density scales with $\Sigma_{\text{H}_{\text{2}}}$ (where $\alpha$ is a constant):
\begin{equation}
\Sigma_{\text{SFR}}~=~\alpha\, \Sigma_{\text{H}_{2}}%~=~\frac{1}{\alpha}\frac{d \Sigma_{\text{s}}}{\text{d}t},
\end{equation}
through the Kennicutt-Schmidt relation \citep{1959ApJ...129..243S,1998ApJ...498..541K} for molecular (rather than total) gas mass surface densities (e.g., \citealt{2008AJ....136.2846B}). The mean thermal speed $\langle v_{\text{gas}}\rangle$ is furthermore assumed to be roughly constant, and the typical grain radius $a$ is assumed not to vary much, which allows to reduce Eq.\,\ref{Eq_tau0} to:
\begin{equation}
\label{Eq_surfacedensities}
\tau^{-1}_{\text{0}}~=~\frac{\epsilon\,Z_{\text{M}}\,\Sigma_{\text{SFR}}}{\Sigma_{\text{gas}}}
\end{equation}
which only depends on the metallicity $Z_{\text{M}}$, the gas ($\Sigma_{\text{gas}}$) and star formation rate ($\Sigma_{\text{SFR}}$) surface densities, with a constant (dimensionless) factor $\epsilon$ that is left as a free parameter. This factor $\epsilon$ will set the grain growth efficiency and will be sensitive to the average grain size, the mean thermal speed of gas particles, and other assumptions that have gone into this derivation. As we do not consider the resolved nature of galaxies in this work, we convert the surface densities from Eq. \ref{Eq_surfacedensities} to total gas mass and SFR measurements in Eq. \ref{Eq_graingrowth}. 

The derivation of this grain growth prescription from \citet{2012MNRAS.423...26M} is based on the assumption that the accretion rate of gas-phase elements onto grain surfaces scales with the molecular cloud surface density, and does not account for any barriers which could reduce the grain growth efficiency (e.g., ice mantle formation, or Coulomb barriers). Due to the molecular gas mass not always being readily available from observations or simulations, the star formation rate (assumed to scale with the molecular gas content) is used to parameterise the efficiency of grain growth processes. %We note that this grain growth prescription has its limitations, but is  

\subsection{Including gas infall and outflows}
\label{Sec_Infall_Outflow}
It is commonly believed that the infall from primordial gas along dense filaments from the cosmic web plays an important role in fueling and sustaining star formation in galaxies (e.g. \citealt{2009Natur.457..451D,2012RAA....12..917S}). This gas accretion of metal-poor gas is also required to explain the fundamental relation between the stellar mass, metallicity and star formation rate of galaxies (e.g., \citealt{2014A&ARv..22...71S}). For this work, we assume that gas infall scales directly with the star formation rate, and that the gas is pristine and dust-free (i.e., the infalling gas does not contribute to the overall metal and dust budget in galaxies).

Multi-phase galactic outflows, on the other hand, have been shown to play an important role in the quenching of star formation activity, but the main driving force of these outflows (supernovae, stellar wind, accreting black holes, cosmic rays) has yet to be identified (see \citealt{2017ARA&A..55...59N} for a recent review). Other than simulations, the copious number of detections of massive galactic outflows (e.g., \citealt{2014A&A...562A..21C,2017ApJ...835..265W,2019MNRAS.483.4586F}) during recent years has reinforced the importance of these outflows in regulating galaxy evolution. Galactic outflows are thought to be most powerful in galaxies at high redshift which undergo strong bursts of star formation; while low redshift galaxies with $M_{\star}$ $\gtrsim$ 10$^{10}$\,M$_{\odot}$ tend to have a suppressed galactic outflows due to the lower gas fraction and turbulent velocity dispersion in these galaxies, which makes it tenuous to drive outflows with high mass loading factors \citep{2017MNRAS.465.1682H}. The mass loading factor $\eta$ (=$\dot{M}_{\text{out}}$/$\dot{M}_{\star}$) is defined as the ratio of the mass outflow rate to the star formation rate. To quantify how the mass loading factor varies across a galaxy's lifetime, we rely on the prescription from \citet{2017MNRAS.465.1682H} (see their Eq. 44) which relates $\eta$ to the stellar mass $M_{\star}$ and gas fraction $f_{\text{gas}}$ of galaxies through:
\begin{equation}
\eta~=~14\left(\frac{f_{\text{gas}}M_{\star}}{10^{10}M_{\odot}}\right)^{-0.23}\exp\left(\frac{-0.75}{f_{\text{gas}}}\right)
\end{equation}
and accounts for the decreased mass loading factors in the local Universe. The prescriptions from \citet{2017MNRAS.465.1682H} furthermore agree well with the mass-loading factors inferred from the FIRE simulations \citep{2015MNRAS.454.2691M}.

\section{Tables}
\label{Tables.sec}
%\begin{sidewaystable}[ph!]
\begin{landscape}
\vspace*{-1.5\baselineskip}
\captionsetup{singlelinecheck=false}
\captionof{table}{Overview of the results obtained from performing Mann-Whitney U-tests to verify whether two samples have the same median of distribution. The output nearly-normal test statistic Z (top value) and probability level p (bottom value) of this test are presented for each pair of galaxy samples. Probabilities p$>$0.05 (shown in blue) indicate that the hypothesis that both samples have the same median of distribution could not be rejected, and that both samples are therefore not significantly different.}
\label{Table_MannWhitneyU}
\begin{tabularx}{\linewidth}{llcccccccccc} % four columns, alignment for each
\hline
Properties & & JINGLE & JINGLE & JINGLE & JINGLE & HRS & HRS & HRS & KINGFISH & KINGFISH & HAPLESS \\ 
 & & vs. & vs. & vs. & vs. & vs. & vs. & vs. & vs. & vs. & vs.\\
 & & HRS & KINGFISH & HAPLESS & HiGH & KINGFISH & HAPLESS & HiGH & HAPLESS & HiGH & HiGH \\ 
\hline
$\log$\,$M_{\star}$ & Z~= & -6.1 & -2.7 & -4.1 & -3.2 & 0.3 & -3.0 & -1.0 & -1.9 & -0.4 & 1.8 \\
 & p~= & $<10^{-6}$ & 3.4$\times$10$^{-3}$ & $1.8\times10^{-5}$ & $7.3\times10^{-4}$ & \textcolor{blue}{$4.0\times10^{-1}$} & $1.5\times10^{-3}$ & \textcolor{blue}{$1.7\times10^{-1}$} & $2.7\times10^{-2}$ & \textcolor{blue}{$3.5\times10^{-1}$} & $3.7\times10^{-2}$ \\
$\log$\,SFR & Z~= & -12.7 & -5.3 & -5.4 & -2.8 & 1.8 & -1.5 & 4.2 & -1.9 & 1.8 & 4.0 \\
 & p~= & $<10^{-6}$ & $<10^{-6}$ & $<10^{-6}$ & $2.5\times10^{-3}$ & $3.7\times10^{-2}$ & \textcolor{blue}{$6.3\times10^{-2}$} & $1.2\times10^{-5}$ & $3.1\times10^{-2}$ & $3.8\times10^{-2}$ & $3.8\times10^{-5}$ \\
$\log$\,sSFR & Z~= & -5.9 & -0.7 & -0.1 & 2.2 & 2.7 & 1.6 & 4.5 & 0.2 & 2.1 & 1.3 \\
 & p~= & $<10^{-6}$ & \textcolor{blue}{$2.4\times10^{-1}$} & \textcolor{blue}{$4.6\times10^{-1}$} & $1.6\times10^{-2}$ & $3.0\times10^{-3}$ & \textcolor{blue}{$5.8\times10^{-2}$} & $<10^{-6}$ & \textcolor{blue}{$4.2\times10^{-1}$} & $1.9\times10^{-2}$ & \textcolor{blue}{$1.0\times10^{-1}$} \\
 $\log$\,$M_{\text{HI}}$ & Z~= & -8.8 & -3.2 & -3.6 & 1.3 & 2.2 & -1.6 & 6.3 & -2.4 & 3.8 & 4.4 \\
 & p~= & $<10^{-6}$ & $6.1\times10^{-4}$ & $1.7\times10^{-4}$ & \textcolor{blue}{$9.7\times10^{-2}$} & $1.3\times10^{-2}$ & \textcolor{blue}{$5.1\times10^{-2}$} & 0.0 & $9.3\times10^{-3}$ & $7.9\times10^{-5}$ & $5.8\times10^{-6}$ \\
 12+$\log$(O/H) & Z~= & -13.6 & -2.4 & -4.5 & -5.7 & 6.9 & 0.5 & 2.0 & -3.0 & -3.0 & 1.0 \\
  & p~= & $<10^{-6}$ & $9.3\times10^{-3}$ & $3.3\times10^{-6}$ & 0.0 & 0.0 & \textcolor{blue}{$3.0\times10^{-1}$} & $2.2\times10^{-2}$ & $1.5\times10^{-3}$ & $1.3\times10^{-3}$ & \textcolor{blue}{$1.6\times10^{-1}$} \\
$\log$\,$M_{\text{dust}}$/$M_{\star}$ & Z~= & -7.5 & -3.6 & -2.1 & -0.6 & 0.9 & 0.2 & 2.9 & -0.3 & 2.0 & 1.5 \\
& p~= & $<10^{-6}$ & $1.5\times10^{-4}$ & $1.9\times10^{-2}$ & \textcolor{blue}{$2.6\times10^{-1}$} & \textcolor{blue}{$1.9\times10^{-1}$} & \textcolor{blue}{$4.1\times10^{-1}$} & $2.1\times10^{-3}$ & \textcolor{blue}{$4.0\times10^{-1}$} & $2.2\times10^{-2}$ & \textcolor{blue}{$6.8\times10^{-2}$} \\
$\log$\,$M_{\text{HI}}$/$M_{\star}$ & Z~= & -3.1 & 0.9 & 0.2 & 4.5 & 2.7 & 0.6 & 5.8 & -0.5 & 3.0 & 2.4 \\
& p~= & $10^{-3}$ & \textcolor{blue}{$1.8\times10^{-1}$} & \textcolor{blue}{$4.4\times10^{-1}$} & $2.7\times10^{-6}$ & $3.7\times10^{-3}$ & \textcolor{blue}{$2.7\times10^{-1}$} & 0.0 & \textcolor{blue}{$3.2\times10^{-1}$} & $1.5\times10^{-3}$ & $9.4\times10^{-3}$ \\
$\log$\,$M_{\text{dust}}$/$M_{\text{metals}}$ & Z~= & 5.4 & -1.1 & -0.3 & -2.6 & -5.9 & -2.4 & -4.9 & 0.2 & -1.5 & -1.3 \\
& p~= & 0.0 & \textcolor{blue}{$1.4\times10^{-1}$} & \textcolor{blue}{$3.8\times10^{-1}$} & $4.9\times10^{-3}$ & 0.0 & $7.6\times10^{-3}$ & $<10^{-6}$ & \textcolor{blue}{$4.2\times10^{-1}$} & \textcolor{blue}{$6.4\times10^{-2}$} & \textcolor{blue}{$9.9\times10^{-2}$} \\
$\log$\,$M_{\text{dust}}$/$M_{\text{HI}}$ & Z~= & -0.4 & -2.4 & -1.9 & -5.7 & -2.2 & -1.3 & -5.6 & -0.2 & -2.9 & -1.3 \\
& p~= & \textcolor{blue}{$3.5\times10^{-1}$} & $9.3\times10^{-3}$ & $2.7\times10^{-2}$ & 0.0 & $1.4\times10^{-2}$ & \textcolor{blue}{$9.2\times10^{-2}$} & 0.0 & \textcolor{blue}{$4.4\times10^{-1}$} & $2.1\times10^{-3}$ & \textcolor{blue}{$9.4\times10^{-2}$} \\
\hline 
\end{tabularx}
%\end{sidewaystable}
\end{landscape}

\section{List of acronyms and symbols}
\label{Acronyms.sec}
\begin{table*}
%\centering
\caption{We collected the acronyms used throughout this paper in this summary table for quick reference.}
\label{Acronyms.tab}
\begin{tabular}{ll} % four columns, alignment for each
\hline
Symbol & Explanation \\
\hline 
AGB & asymptotic giant branch \\
DEUS & Dust and Element evolUtion modelS \\
FUV & far-ultraviolet \\
HAPLESS & \textit{Herschel}-ATLAS Phase-1 Limited-Extent Spatial Survey\\
H{\sc{i}} def & H{\sc{i}} deficiency \\
HIGH & H{\sc{i}}-selected Galaxies in \textit{Herschel}-ATLAS\\
HRS & \textit{Herschel} Reference Survey \\
IMF & initial mass function \\
IR & infrared \\
ISM & interstellar medium \\
JINGLE & JCMT dust and gas In Nearby Galaxies Legacy Exploration \\
KINGFISH & Key Insights on Nearby Galaxies: A Far-Infrared Survey with \textit{Herschel} \\
LIMS & low-and intermediate mass stars \\
MCMC & Markov Chain Monte Carlo \\
PDF & probability density function \\
PP04 & metallicity calibration from \citet{2004MNRAS.348L..59P} \\
SED & spectral energy distribution \\
SFH & star formation history \\
SFR & star formation rate \\
sSFR & specific star formation rate \\
THEMIS & The Heterogeneous dust Evolution Model for Interstellar Solids\\
TIR & total infrared \\
UV & ultraviolet \\
\hline 
\end{tabular}
\end{table*}

\begin{table*}
%\centering
\caption{We collected the symbols used throughout this paper in this summary table for quick reference.}
\label{Symbols.tab}
\begin{tabular}{ll} % four columns, alignment for each
\hline
Symbol & Explanation \\
\hline 
12+$\log$(O/H) & oxygen abundance (as proxy of metallicity) \\
D & distance \\
DTM & dust-to-metal ratio \\
$\delta_{\text{destr}}$ & dust destruction efficiency \\
$\delta_{\text{grow}}$ & grain growth efficiency \\
$\delta_{\text{LIMS}}$ & dust yields for low- and intermediate mass stars (LIMS)\\
$\delta_{\text{dust}}$ & dust yields for supernovae\\
$\delta_{\text{SN}}$ & correction factor to account for ``clustered" or ``delayed" supernova explosions (which will not clear interstellar material) \\
$e$($t$) & mass loss (at time $t$) during late stellar evolutionary stages \\
$\epsilon$ & grain growth parameter -- DEUS parameter \\
$f_{\text{c}}$ & cold gas fraction \\
$f_{\text{gas}}$ & gas fraction (i.e., $M_{\text{gas}}$/($M_{\text{gas}}$+$M_{\star}$)) \\
$f_{\text{survival}}$ & fraction [in $\%$] of freshly condensed supernova dust capable of surviving the reverse shock -- DEUS parameter\\
$f_{\text{Z}}$ & metal mass fraction \\
$\eta$ & mass loading factor (= ratio of the mass outflow rate to the star formation rate)\\
$\eta_{\text{dust}}$ & dust-to-gas ratio \\
$I$($t$) & infalling gas at time $t$\\
$lcm20$ & large hydrocarbon grains (in the THEMIS dust model) \\
$m_{\text{R}}$ & remnant mass for a star with initial mass $m$\\
M$_{\sun}$ & solar mass \\
$M_{\text{gas,ini}}$ & initial gas mass (in M$_{\sun}$) -- DEUS parameter \\
$M_{\text{cl}}$ & interstellar cleared mass (in M$_{\sun}$) per single supernova event -- DEUS parameter \\
$M_{\text{gas}}$ & total (interstellar) gas mass, i.e. $\xi$($M_{\text{HI}}$+$M_{\text{H}_{2}}$)\\
$M_{\text{HI}}$ & atomic gas mass \\
$M_{\text{H}_{2}}$ & molecular gas mass \\
$M_{\text{metals}}$(gas) & metal mass in the gas phase \\
$M_{\text{metals}}$(gas+dust) & metal mass in the gas phase + locked in dust grains\\
$M_{\text{prog}}$ & progenitor mass (where progenitor refers to the star prior to the supernova event) \\
$M_{\star}$ & stellar mass \\
$M_{\text{HI}}$/$M_{\star}$ & specific H{\sc{i}} gas mass \\
$M_{\text{dust}}$/$M_{\star}$ & specific dust mass \\
$n_{\text{H}}$ & hydrogen density \\
$N_{\text{burn}}$ & MCMC steps in the warm-up phase\\
$N_{\text{chain}}$ & the length of the MCMC chain \\
$N_{\text{eff}}$ & effective sample size, defined as $N_{\text{chain}}$/$\tau_{\text{int}}$\\
$\xi$ & correction factor to account for gas fraction heavier than hydrogen\\
$O$($t$) & outflowing gas at time $t$\\
$p_{\text{Z}}$($m$) & stellar yields for a star with initial mass $m$ and metallicity $Z$\\
$p$ & probability (significance) level (both for Spearman rank correlation and Mann-Whitney U-tests) \\
$\rho$ & Spearman rank correlation coefficient \\
$sCM20$ & small hydrocarbon grains (in the THEMIS dust model) \\
$\sigma$ & standard deviation \\
$\Sigma_{\text{dust}}$ & dust mass surface density \\
$\Sigma_{\text{gas}}$ & gas mass surface density \\
$\Sigma_{\text{SFR}}$ & SFR surface density \\
$sil$ & silicates (in the THEMIS dust model) \\
$R_{\text{SN}}$ & supernova rate per gas mass, $M_{\text{gas}}$ \\
$T_{\text{dust}}$ & dust temperature \\
$\tau_{\text{destr}}$ & dust destruction timescale (= $\delta_{\text{destr}}^{-1}$)\\
$\tau_{\text{int}}$ & integrated auto-correlation time of the MCMC chain\\
$\tau_{\text{grow}}$ & grain growth timescale (= $\delta_{\text{grow}}^{-1}$)\\
$\tau_{\text{m}}$ & lifetime of a star with initial mass $m$\\
$\phi_{\text{m}}$ & initial mass function (IMF) \\
$\chi^{2}$ & chi-squared statistic\\
$\chi^{2}_{\text{red}}$ & reduced chi-squared statistic\\
$\psi$($t$) & star formation rate (SFR) at time $t$ \\
$X_{\text{CO}}$ & CO-to-$H_{\text{2}}$ conversion factor \\
$Z_{\text{I}}$ & metallicity of infalling gas\\
$Z_{\text{M}}$ & metallicity (=$M_{\text{metals}}$(gas)/$M_{\text{gas}}$)\\
$Z_{\text{O}}$ & metallicity of outflowing gas\\
\hline 
\end{tabular}
\end{table*}

\section{Figures}
\label{Figures.sec}
%Bin 1
\begin{figure*}
	\includegraphics[width=17.5cm]{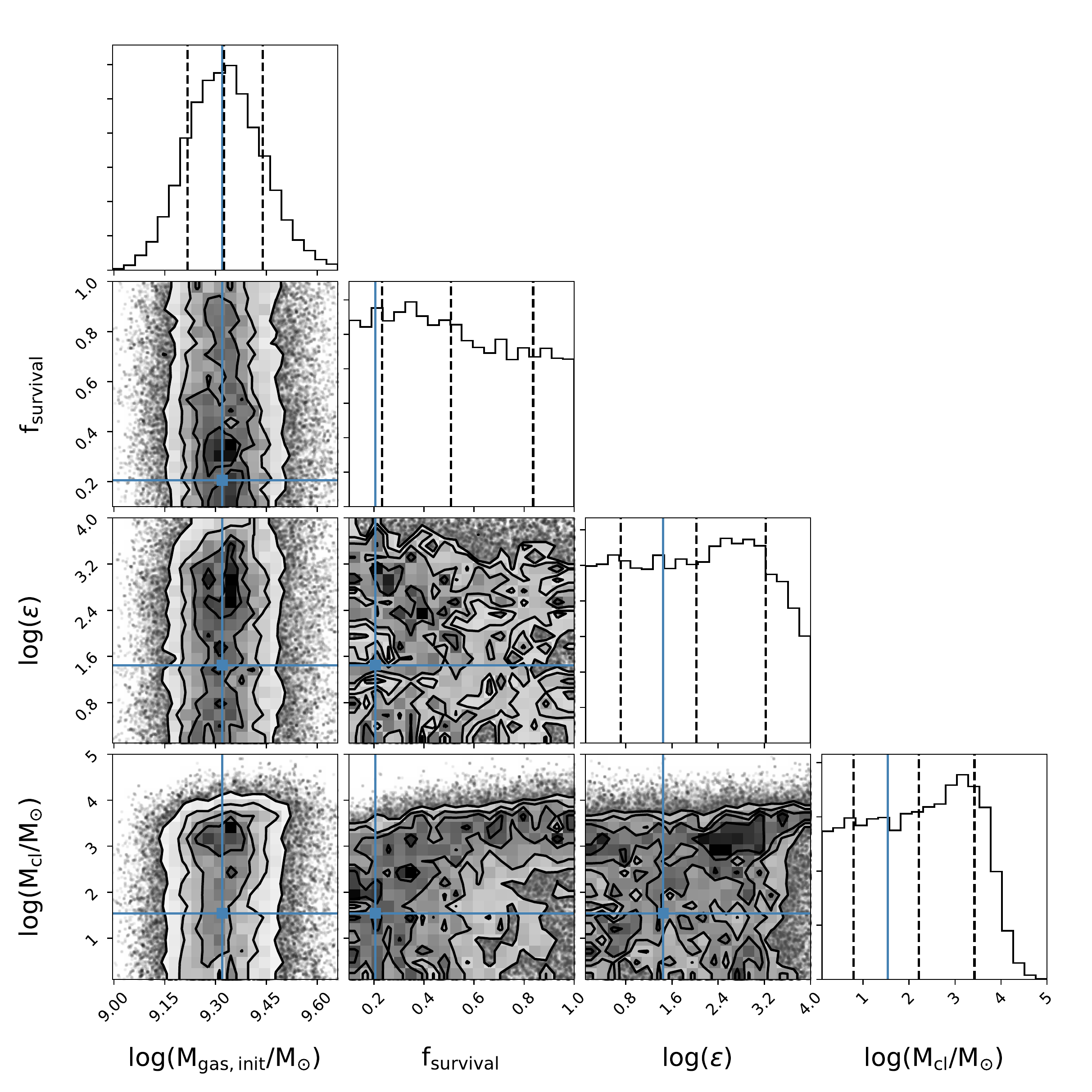}	
    \caption{Corner plot for Model I (i.e., closed-box models with a customised SFH) for galaxy bin 1. The contour plots correspond to 2D posterior PDFs indicating the probability of two parameters in a 2D plane, where contours represent the 0.5$\sigma$, 1.0$\sigma$, 1.5$\sigma$ and 2.0$\sigma$ likelihoods. The histograms correspond to 1D marginalised posterior PDFs showing the likelihood that a certain value will be assigned to a given parameter (by marginalising over the other parameters). The maximum likelihood (blue solid curve) corresponds to the best-fit solution. The black dashed lines correspond to the 16th, 50th and 84th percentiles of the 1D posterior PDFs to reflect the uncertainties on these median model parameter values. }     \label{Image_cornerplot_ModelI_bin1} 
\end{figure*}

\begin{figure*}
	\includegraphics[width=17.5cm]{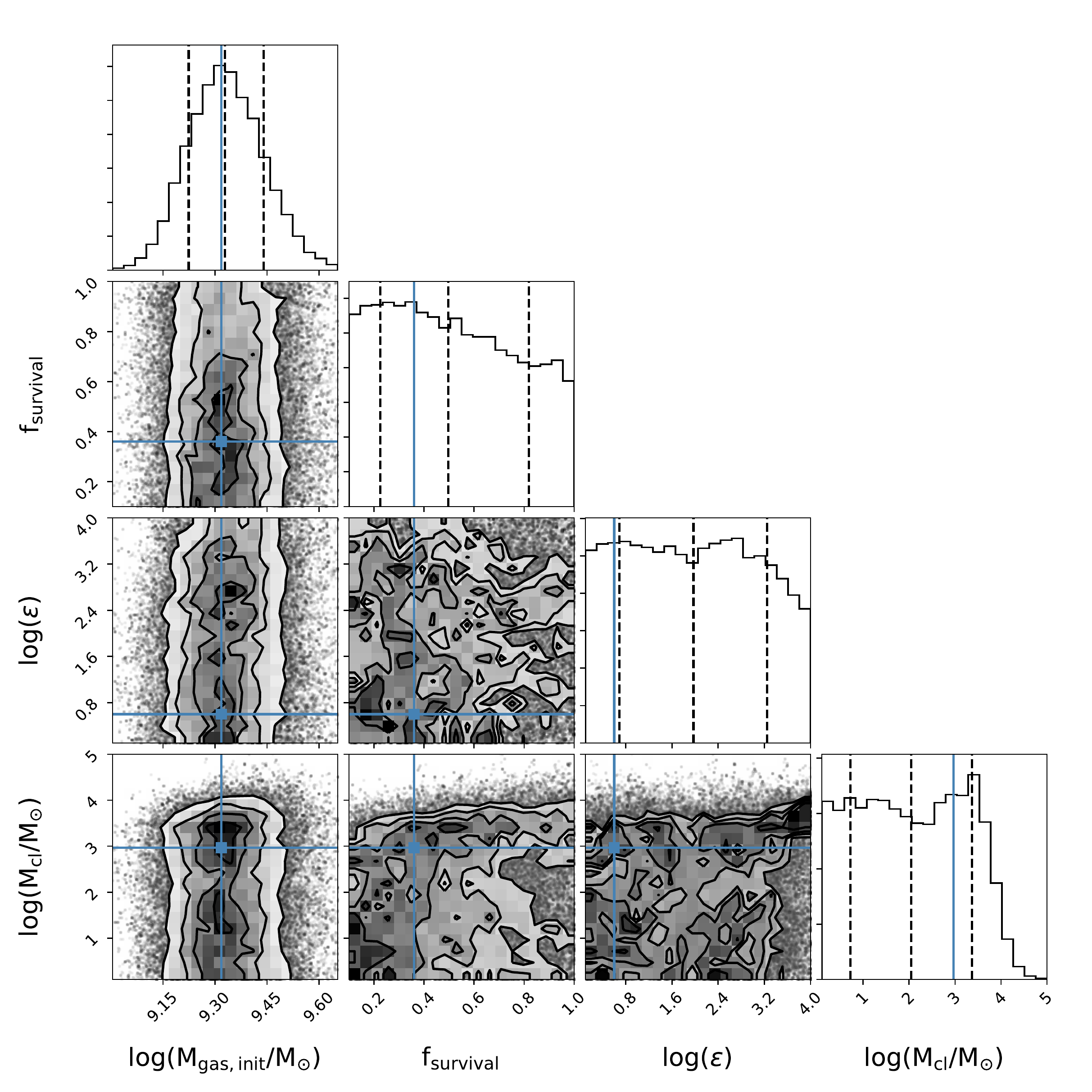}	
    \caption{Corner plot for Model II (i.e., closed-box models with a delayed SFH) for galaxy bin 1. See caption of Fig. \ref{Image_cornerplot_ModelI_bin1} for more information.}     \label{Image_cornerplot_ModelII_bin1} 
\end{figure*}

\begin{figure*}
	\includegraphics[width=17.5cm]{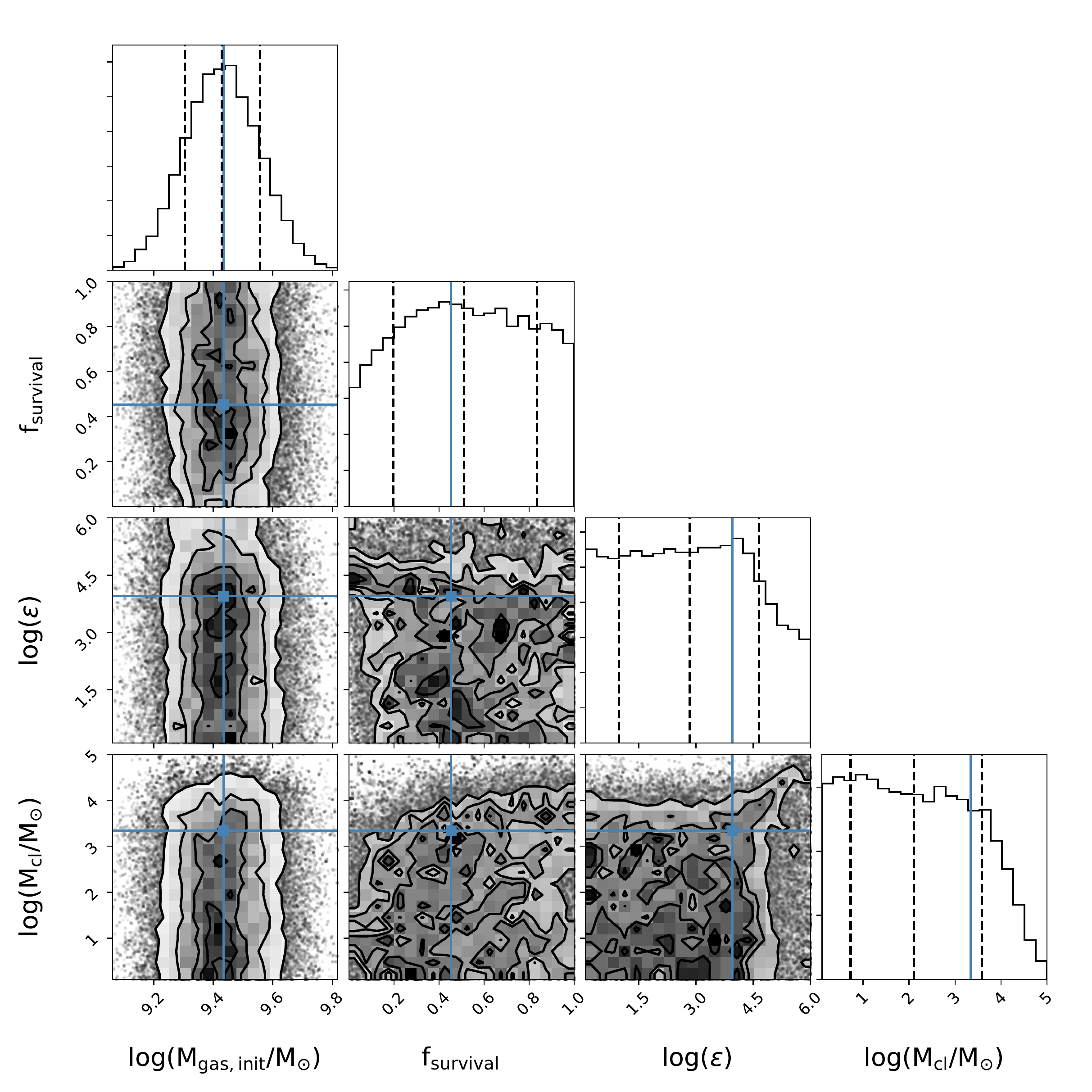}	
    \caption{Corner plot for Model III (i.e., models with gaseous flows and a customised SFH) for galaxy bin 1. See caption of Fig. \ref{Image_cornerplot_ModelI_bin1} for more information.}     \label{Image_cornerplot_ModelIII_bin1} 
\end{figure*}

%Bin 2
\begin{figure*}
	\includegraphics[width=17.5cm]{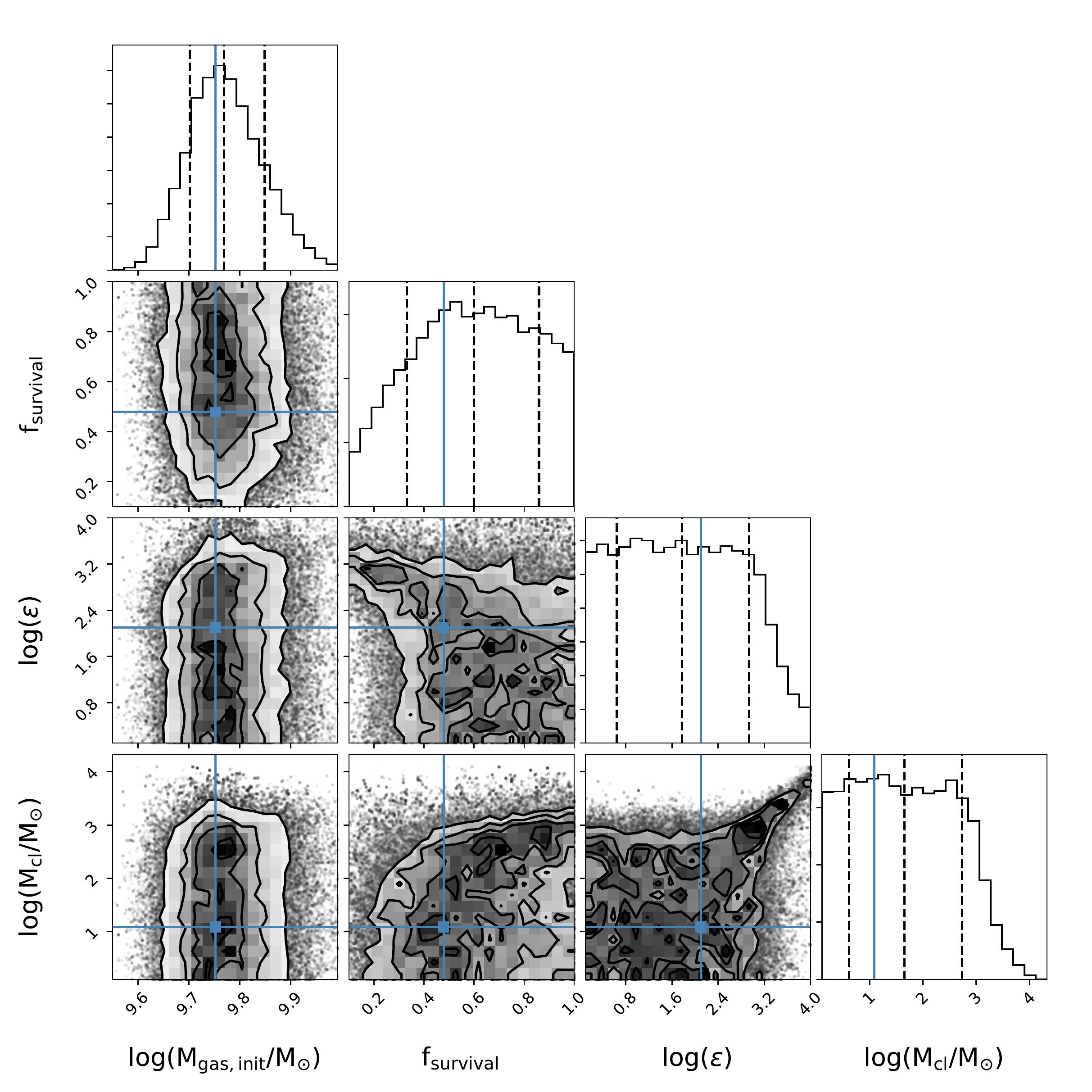}	
    \caption{Corner plot for Model I (i.e., closed-box models with a customised SFH) for galaxy bin 2. See caption of Fig. \ref{Image_cornerplot_ModelI_bin1} for more information.}     \label{Image_cornerplot_ModelI_bin2} 
\end{figure*}

\begin{figure*}
	\includegraphics[width=17.5cm]{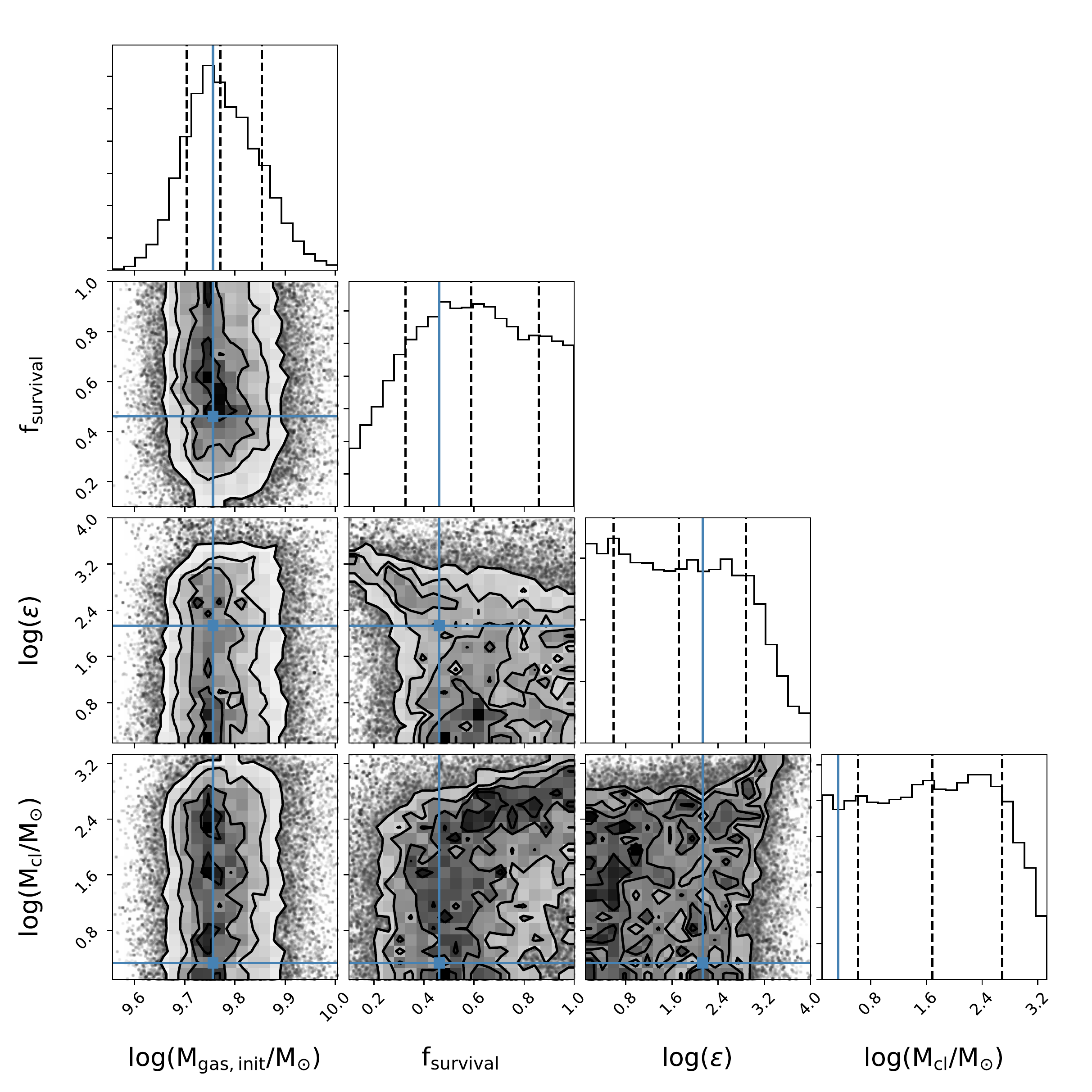}	
    \caption{Corner plot for Model II (i.e., closed-box models with a delayed SFH) for galaxy bin 2. See caption of Fig. \ref{Image_cornerplot_ModelI_bin1} for more information.}     \label{Image_cornerplot_ModelII_bin2} 
\end{figure*}

\begin{figure*}
	\includegraphics[width=17.5cm]{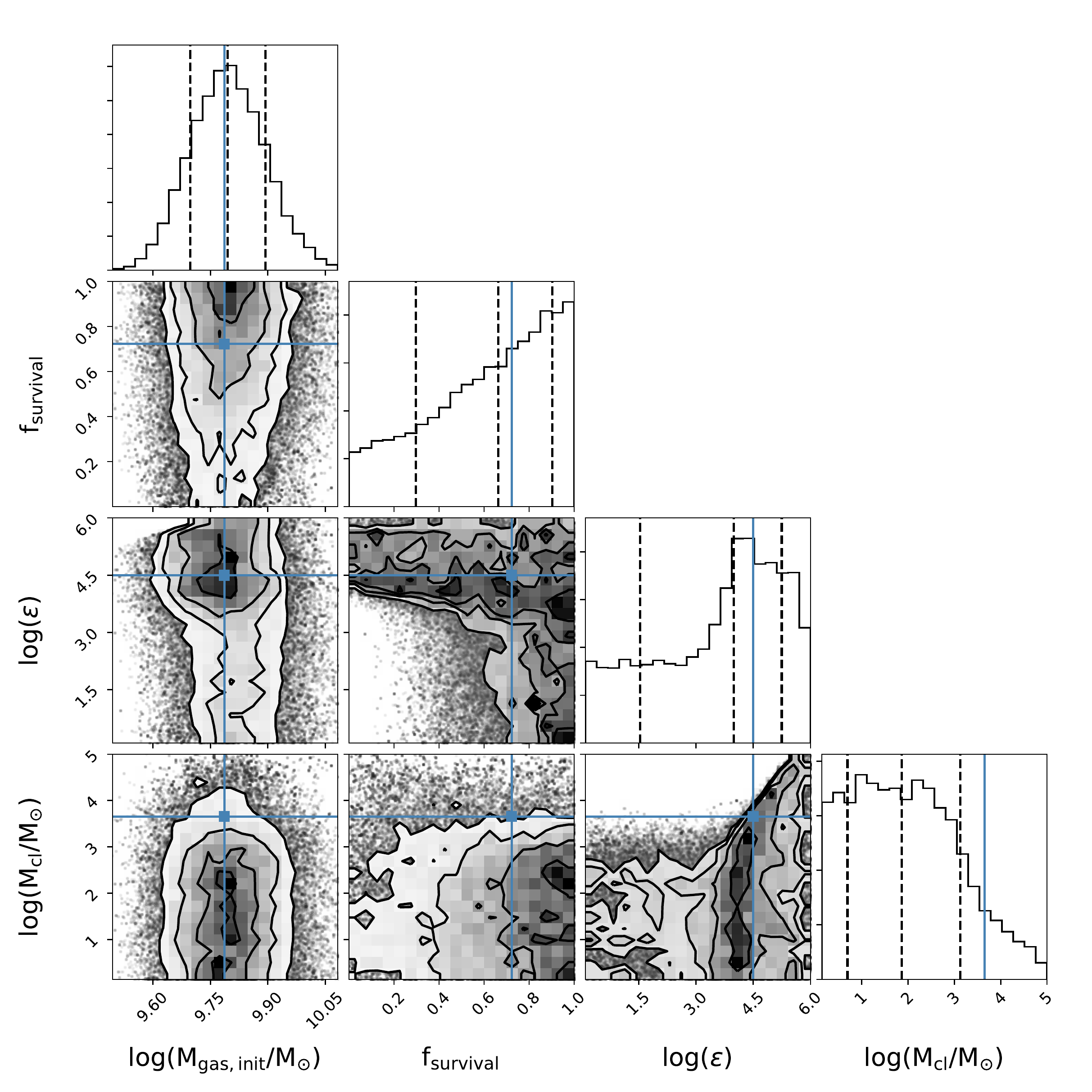}	
    \caption{Corner plot for Model III (i.e., models with gaseous flows and a customised SFH) for galaxy bin 2. See caption of Fig. \ref{Image_cornerplot_ModelI_bin1} for more information.}     \label{Image_cornerplot_ModelIII_bin2} 
\end{figure*}

%Bin 3
\begin{figure*}
	\includegraphics[width=17.5cm]{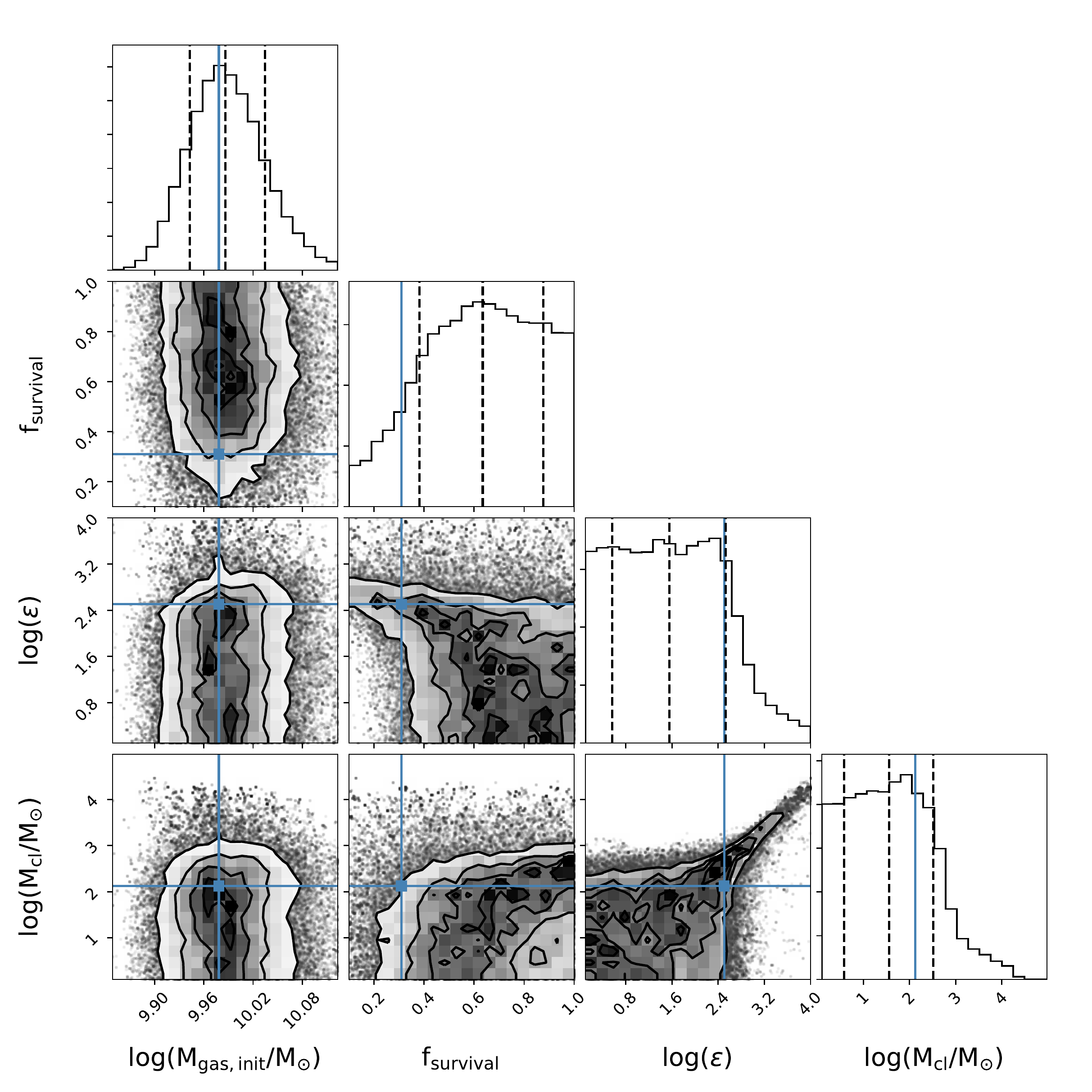}	
    \caption{Corner plot for Model I (i.e., closed-box models with a customised SFH) for galaxy bin 3. See caption of Fig. \ref{Image_cornerplot_ModelI_bin1} for more information.}     \label{Image_cornerplot_ModelI_bin3} 
\end{figure*}

\begin{figure*}
	\includegraphics[width=17.5cm]{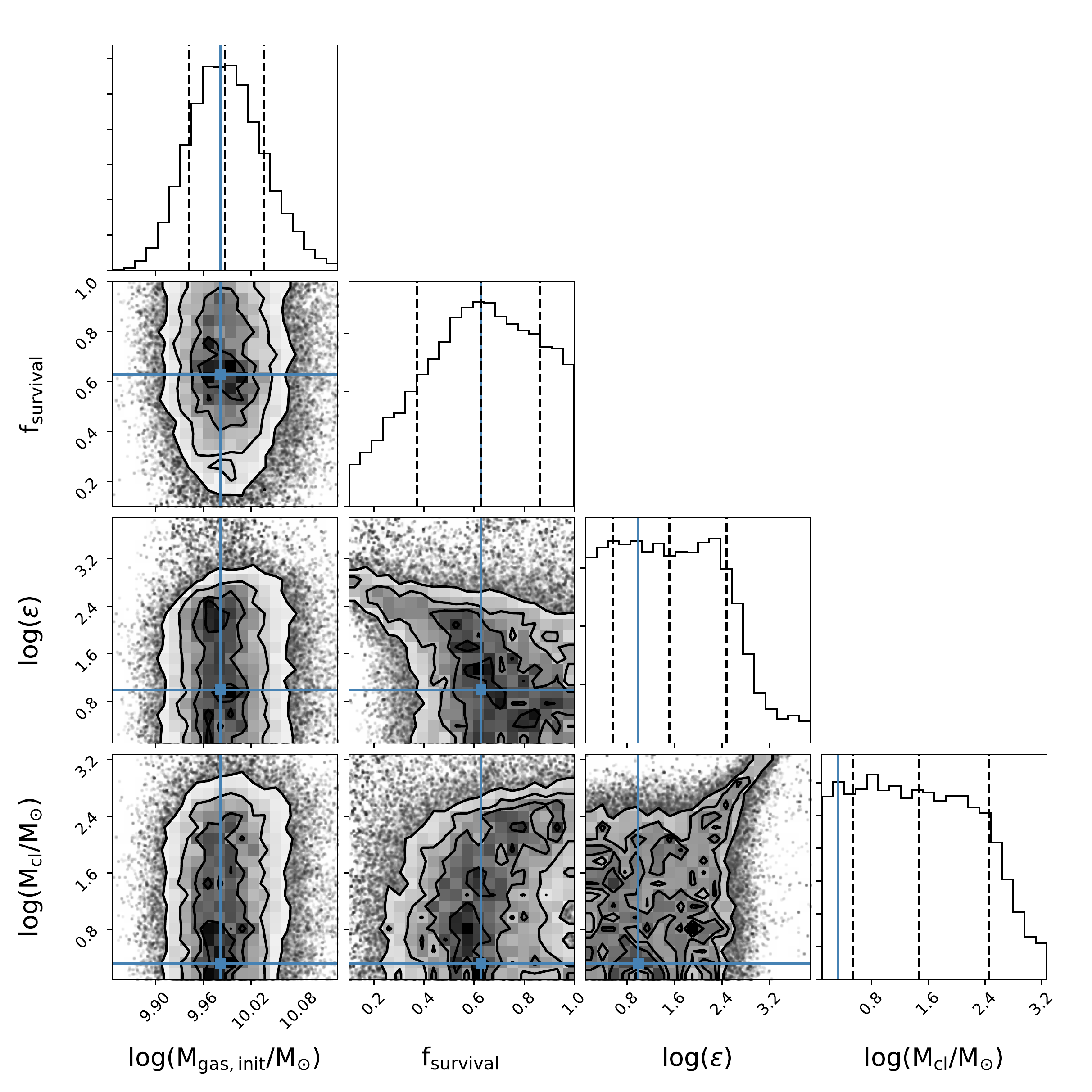}	
    \caption{Corner plot for Model II (i.e., closed-box models with a delayed SFH) for galaxy bin 3. See caption of Fig. \ref{Image_cornerplot_ModelI_bin1} for more information.}     \label{Image_cornerplot_ModelII_bin3} 
\end{figure*}

\begin{figure*}
	\includegraphics[width=17.5cm]{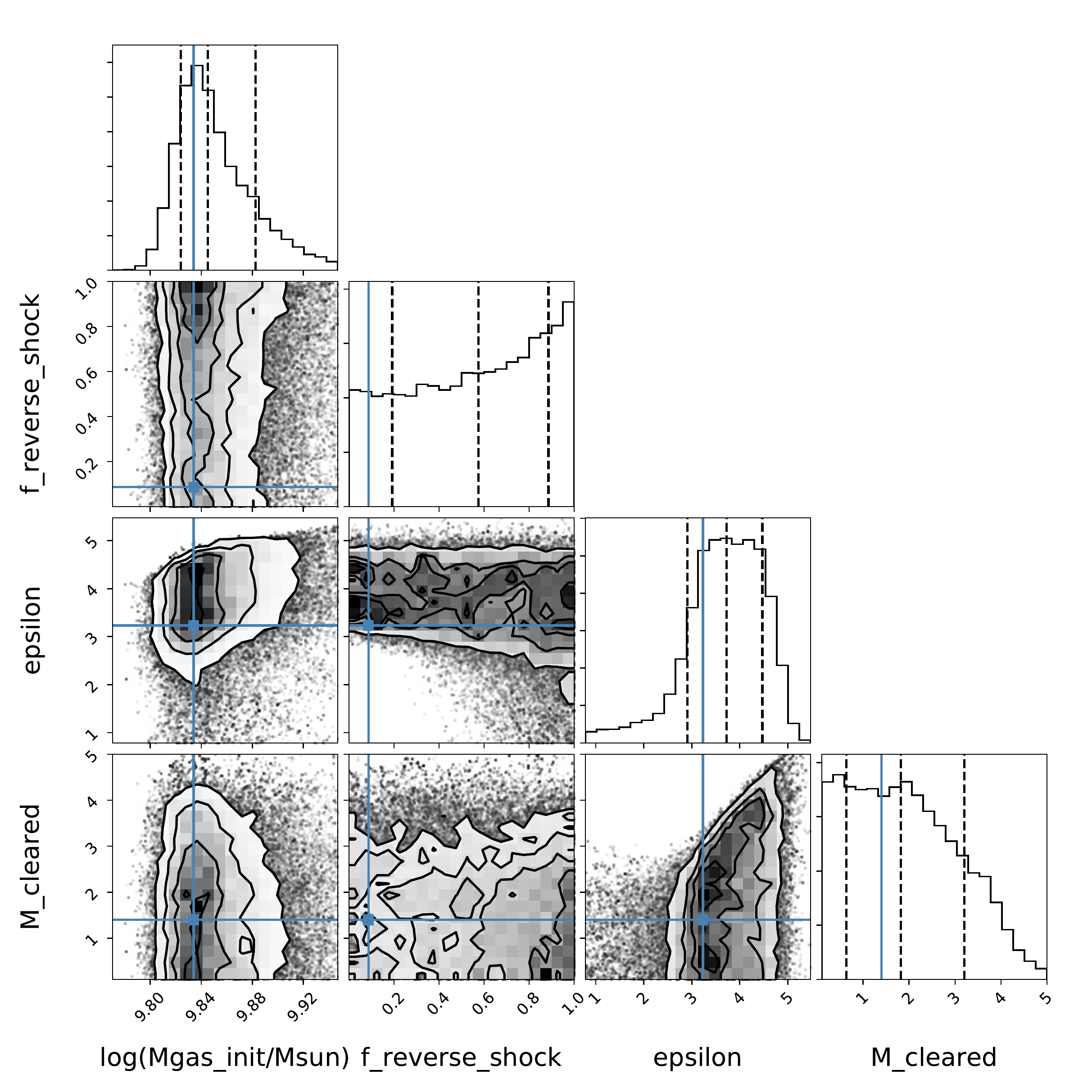}	
    \caption{Corner plot for Model III (i.e., models with gaseous flows and a customised SFH) for galaxy bin 3. See caption of Fig. \ref{Image_cornerplot_ModelI_bin1} for more information.}     \label{Image_cornerplot_ModelIII_bin3} 
\end{figure*}

%Bin 4
\begin{figure*}
	\includegraphics[width=17.5cm]{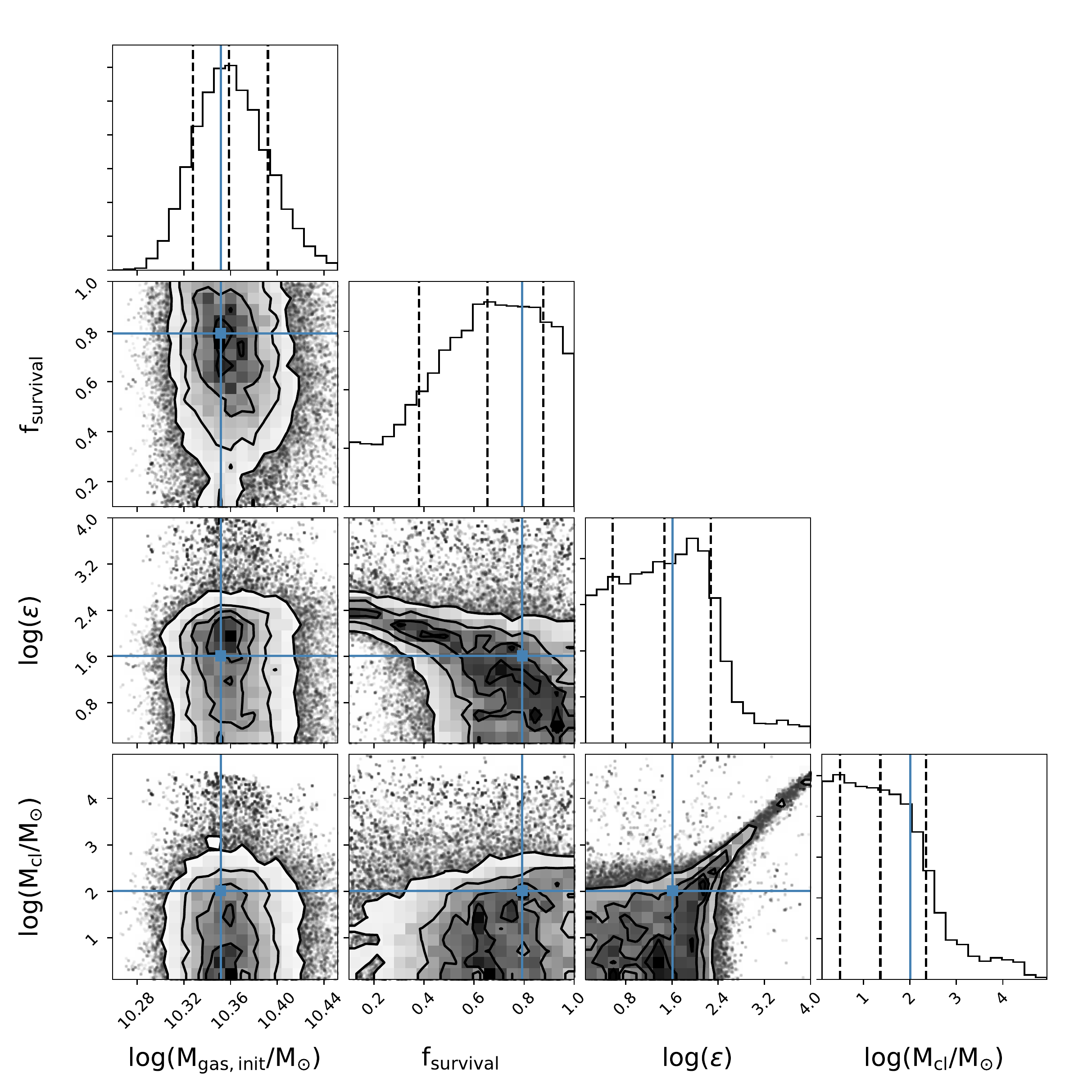}	
    \caption{Corner plot for Model I (i.e., closed-box models with a customised SFH) for galaxy bin 4. See caption of Fig. \ref{Image_cornerplot_ModelI_bin1} for more information.}     \label{Image_cornerplot_ModelI} 
\end{figure*}

\begin{figure*}
	\includegraphics[width=17.5cm]{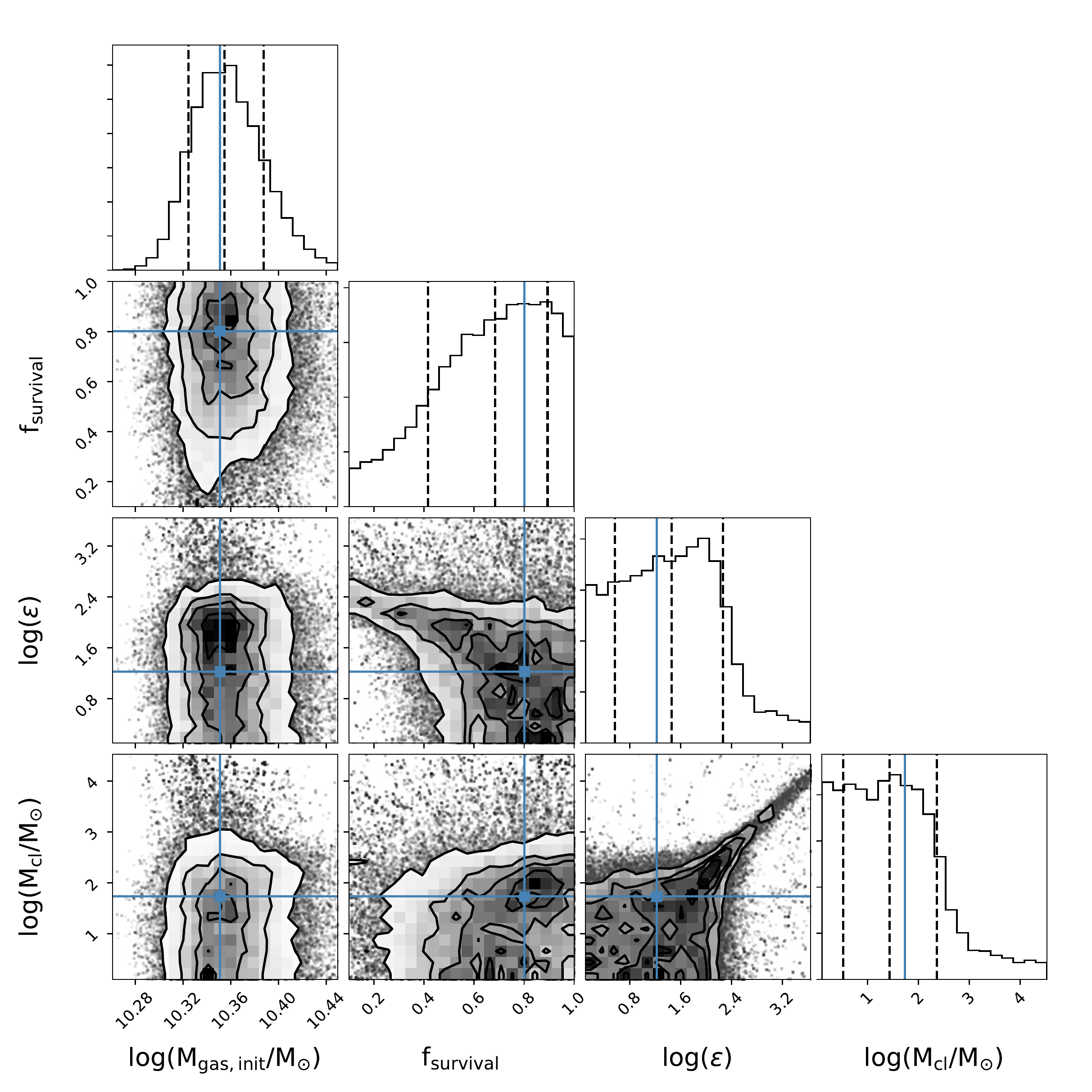}	
    \caption{Corner plot for Model II (i.e., closed-box models with a delayed SFH) for galaxy bin 4. See caption of Fig. \ref{Image_cornerplot_ModelI_bin1} for more information.}     \label{Image_cornerplot_ModelII} 
\end{figure*}

\begin{figure*}
	\includegraphics[width=17.5cm]{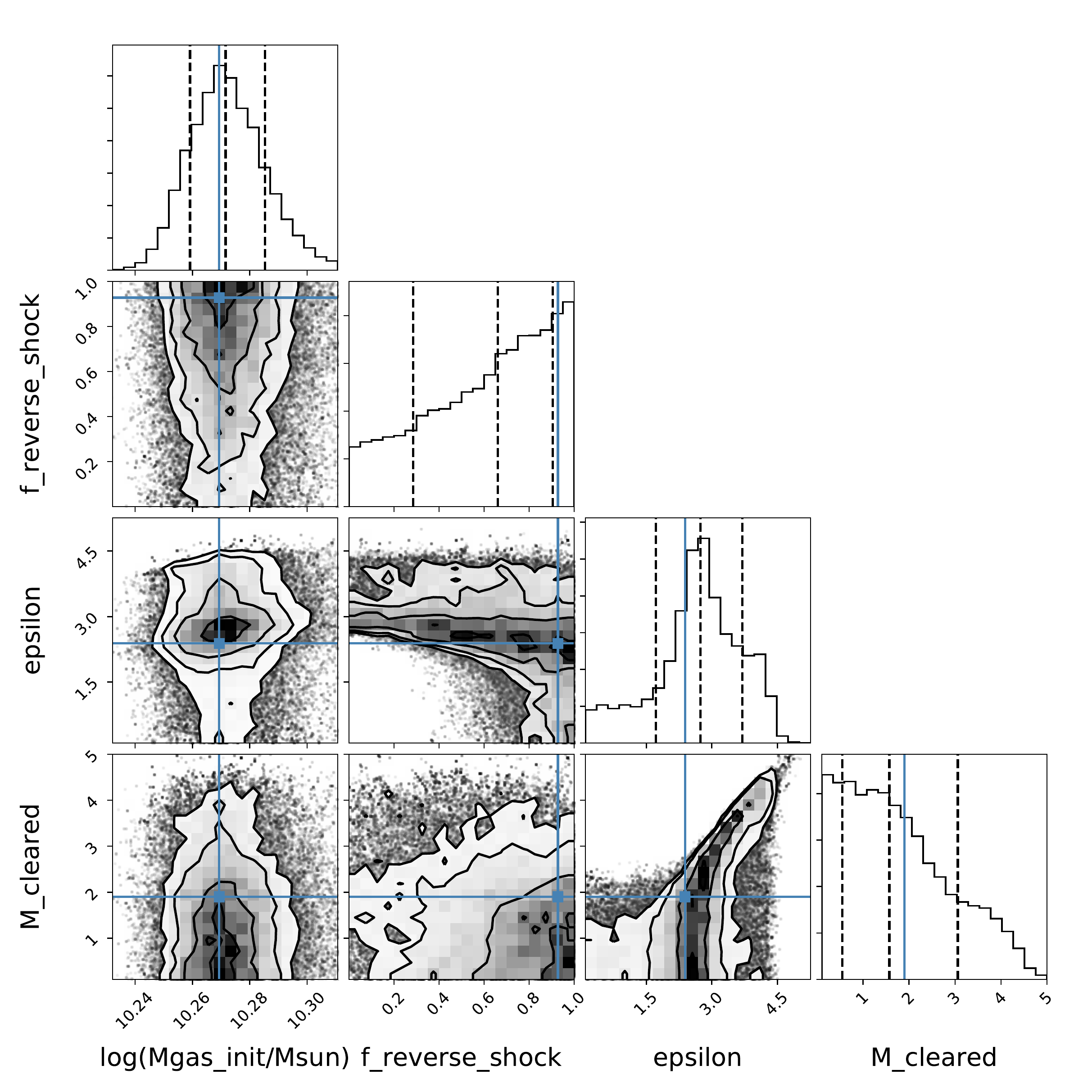}	
    \caption{Corner plot for Model III (i.e., models with gaseous flows and a customised SFH) for galaxy bin 4. See caption of Fig. \ref{Image_cornerplot_ModelI_bin1} for more information.}     \label{Image_cornerplot_ModelIII} 
\end{figure*}

%Bin 5
\begin{figure*}
	\includegraphics[width=17.5cm]{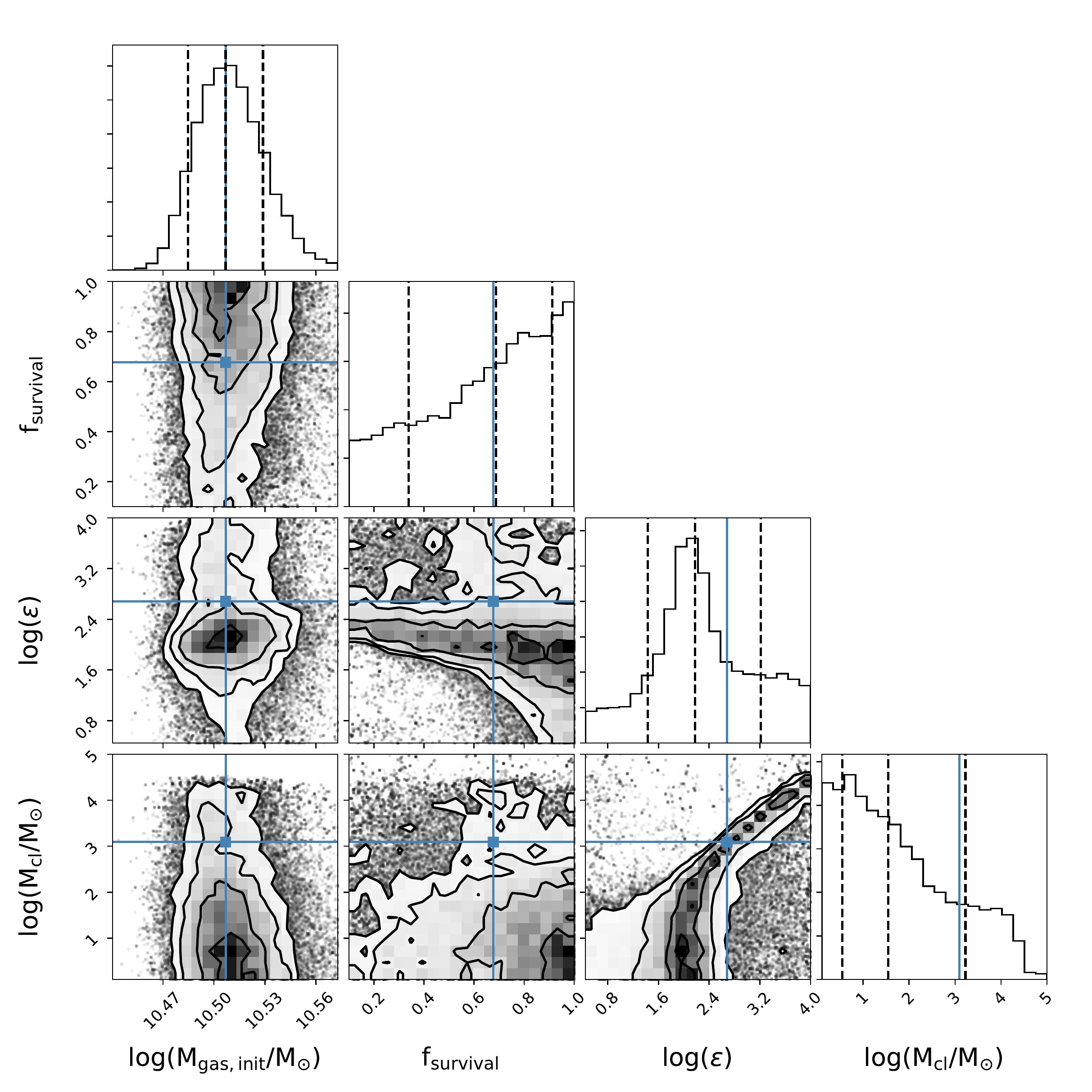}	
    \caption{Corner plot for Model I (i.e., closed-box models with a customised SFH) for galaxy bin 5. See caption of Fig. \ref{Image_cornerplot_ModelI_bin1} for more information.}     \label{Image_cornerplot_ModelI_bin5} 
\end{figure*}

\begin{figure*}
	\includegraphics[width=17.5cm]{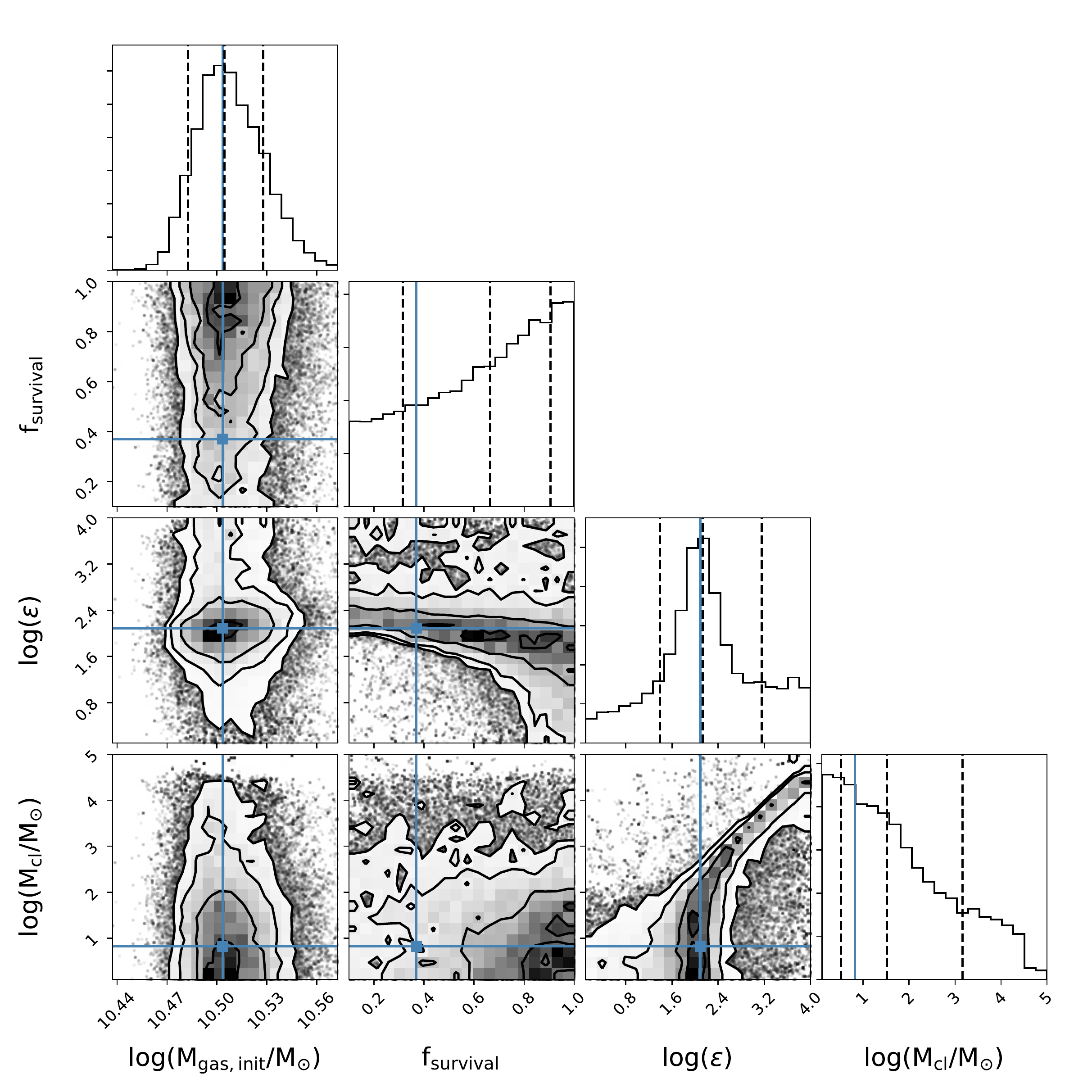}	
    \caption{Corner plot for Model II (i.e., closed-box models with a delayed SFH) for galaxy bin 5. See caption of Fig. \ref{Image_cornerplot_ModelI_bin1} for more information.}     \label{Image_cornerplot_ModelII_bin5} 
\end{figure*}

\begin{figure*}
	\includegraphics[width=17.5cm]{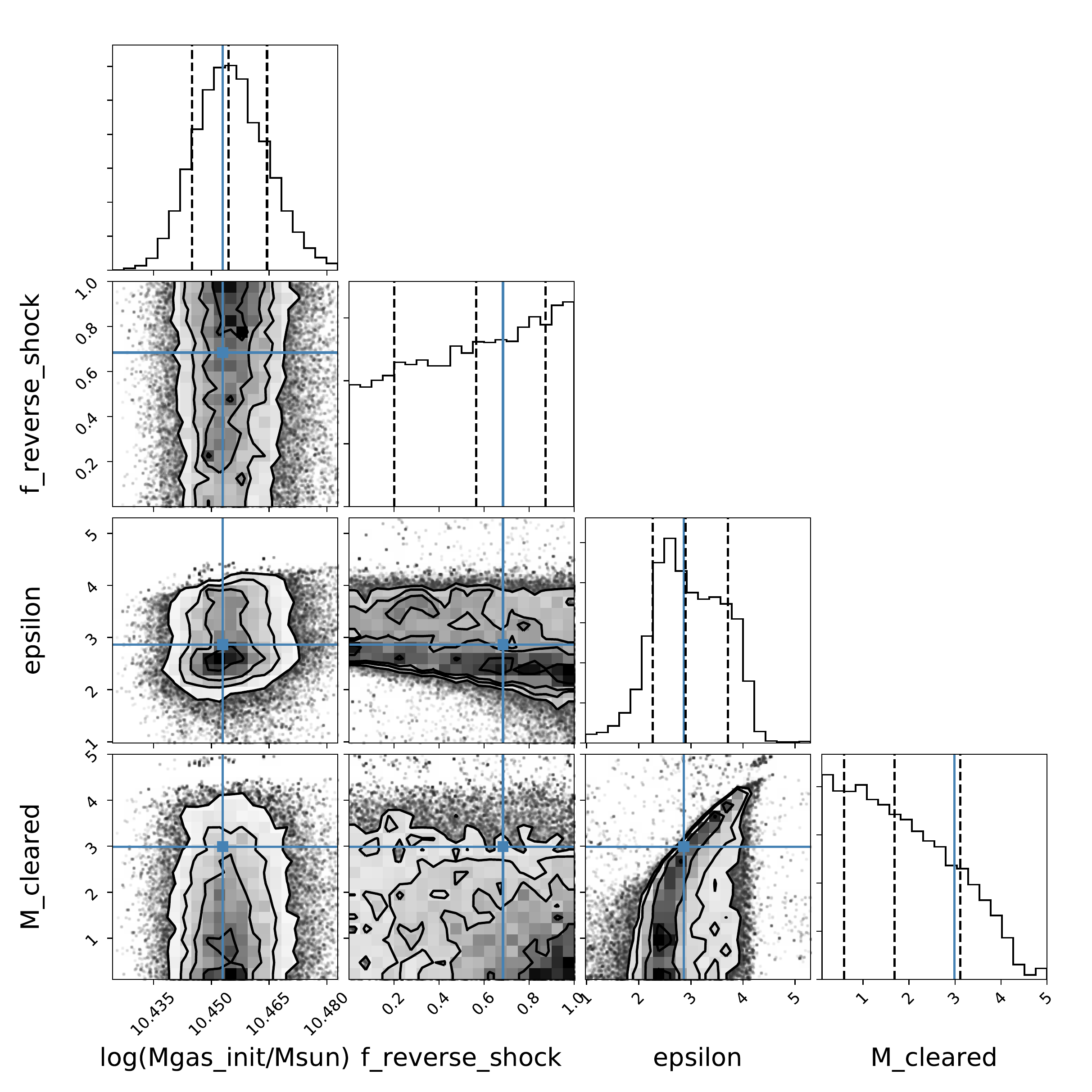}	
    \caption{Corner plot for Model III (i.e., models with gaseous flows and a customised SFH) for galaxy bin 5. See caption of Fig. \ref{Image_cornerplot_ModelI_bin1} for more information.}     \label{Image_cornerplot_ModelIII_bin5} 
\end{figure*}

%Bin 6
\begin{figure*}
	\includegraphics[width=17.5cm]{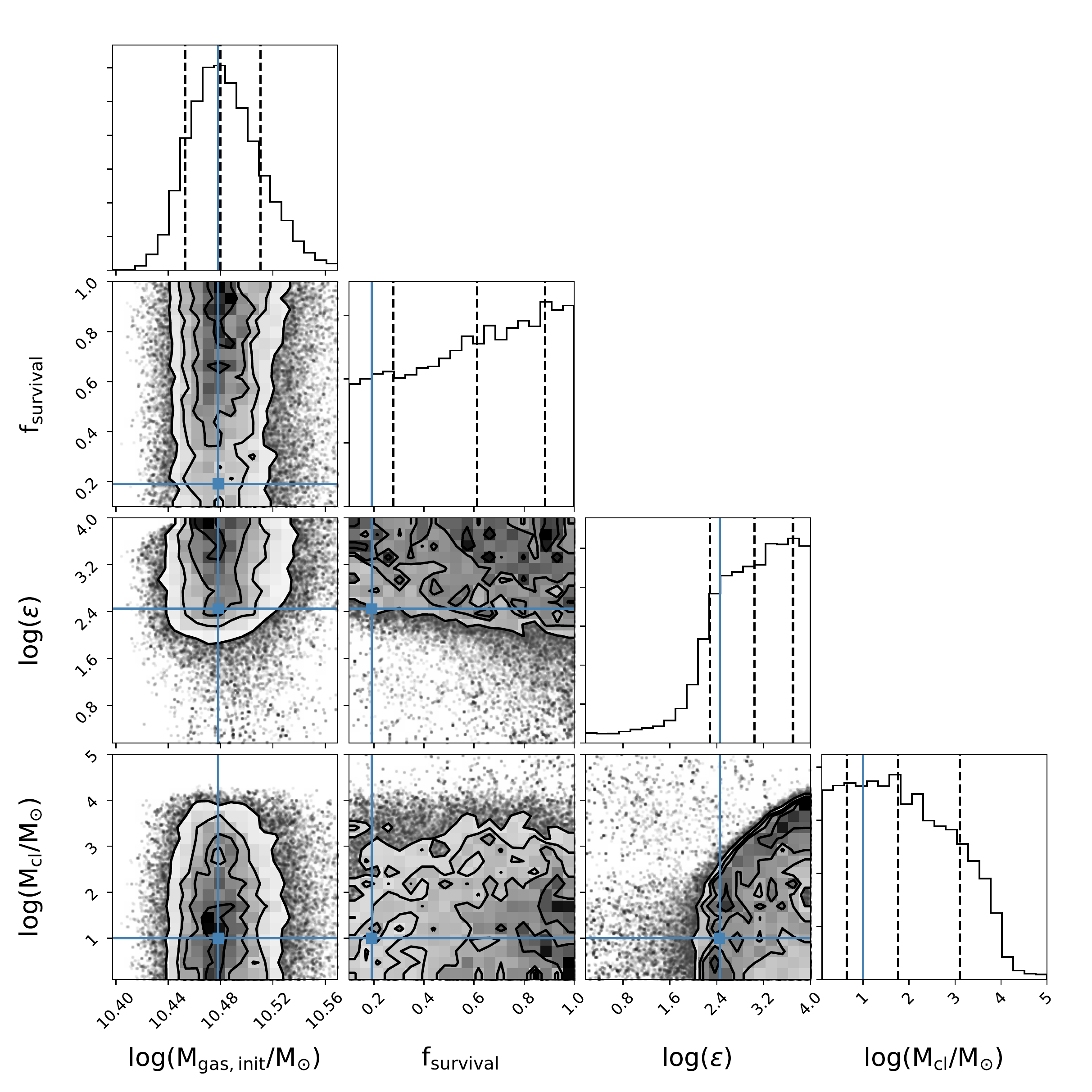}	
    \caption{Corner plot for Model I (i.e., closed-box models with a customised SFH) for galaxy bin 6. See caption of Fig. \ref{Image_cornerplot_ModelI_bin1} for more information.}     \label{Image_cornerplot_ModelI_bin6} 
\end{figure*}

\begin{figure*}
	\includegraphics[width=17.5cm]{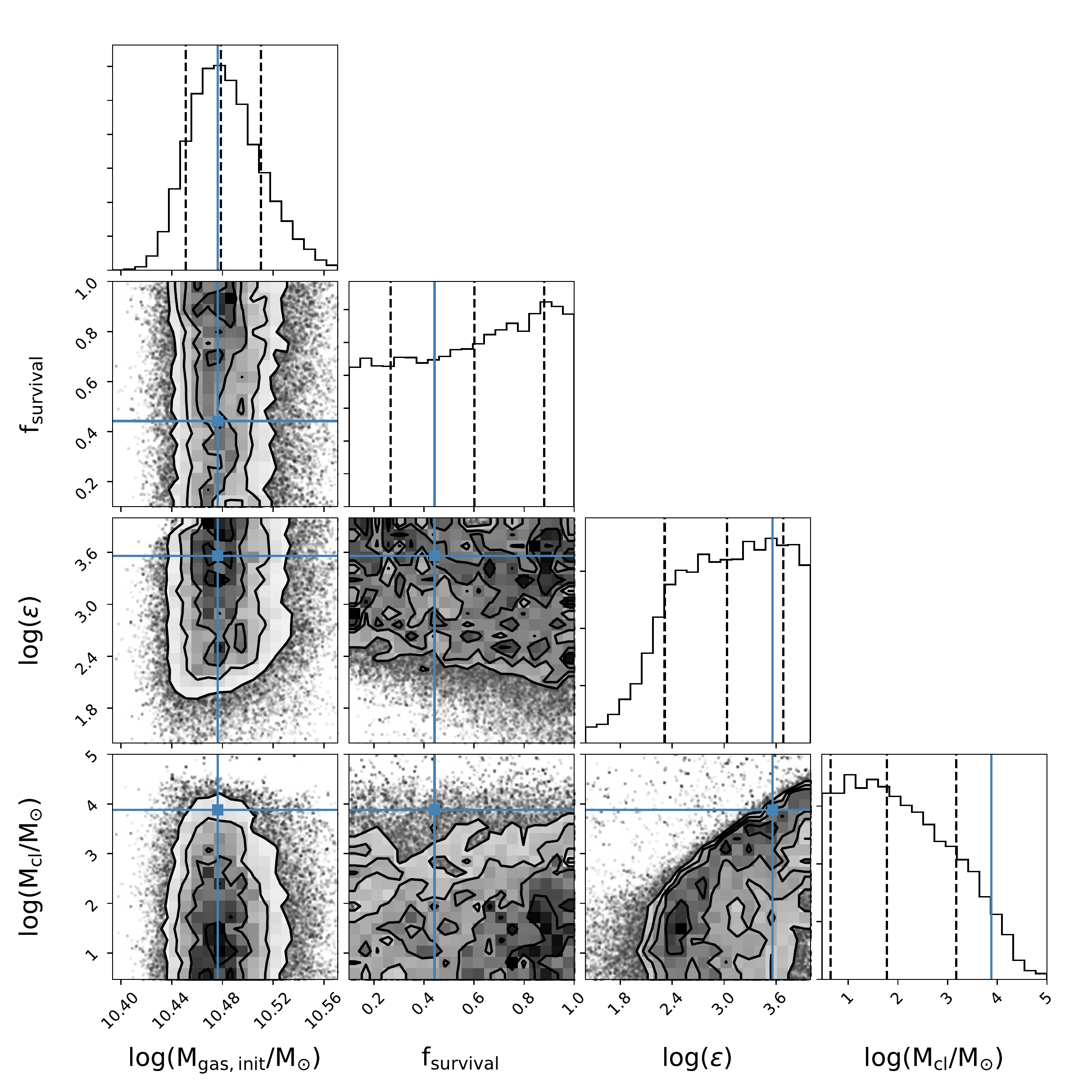}	
    \caption{Corner plot for Model II (i.e., closed-box models with a delayed SFH) for galaxy bin 6. See caption of Fig. \ref{Image_cornerplot_ModelI_bin1} for more information.}     \label{Image_cornerplot_ModelII_bin6} 
\end{figure*}

\begin{figure*}
	\includegraphics[width=17.5cm]{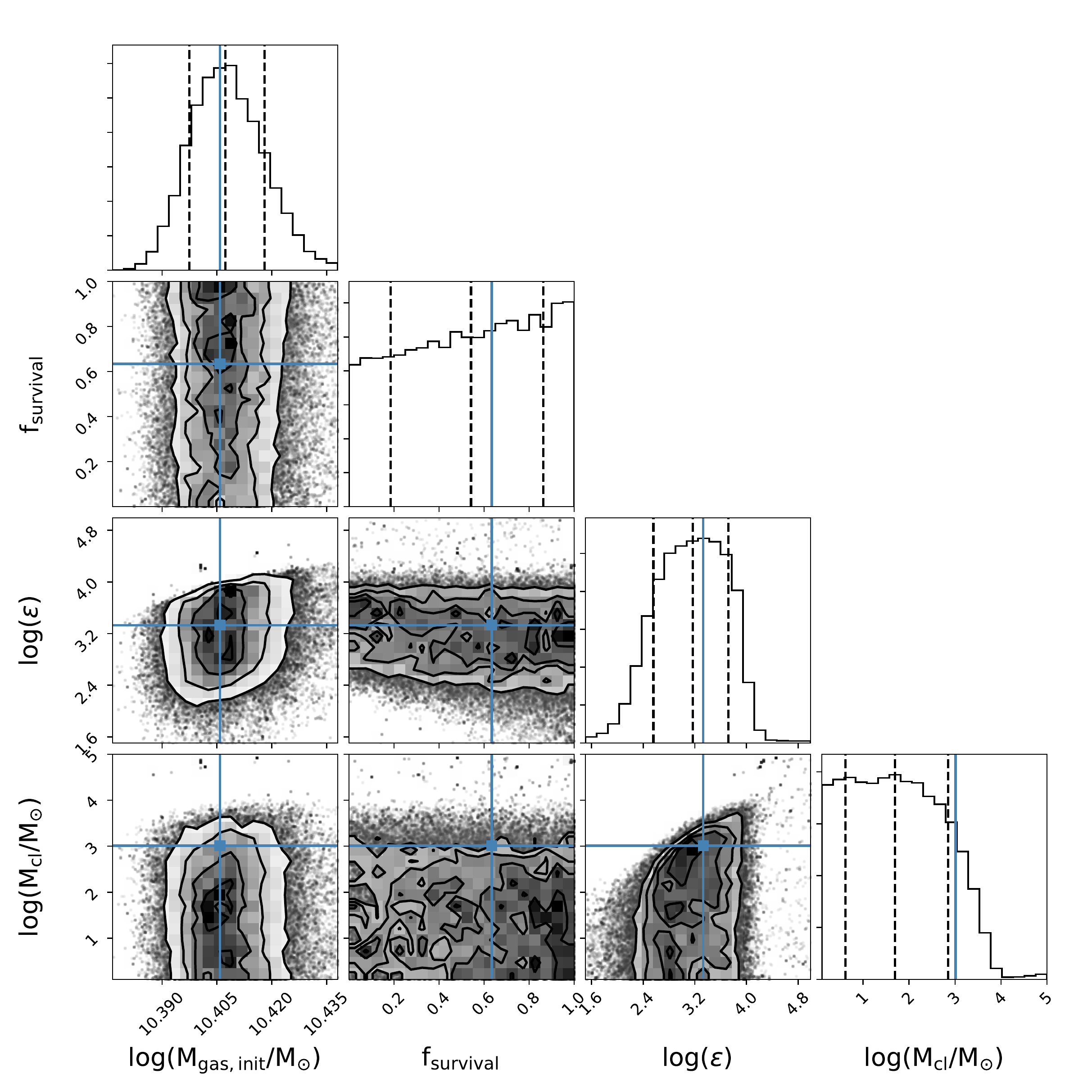}	
    \caption{Corner plot for Model III (i.e., models with gaseous flows and a customised SFH) for galaxy bin 6. See caption of Fig. \ref{Image_cornerplot_ModelI_bin1} for more information.}     \label{Image_cornerplot_ModelIII_bin6} 
\end{figure*}


%%%%%%%%%%%%%%%%%%%%%%%%%%%%%%%%%%%%%%%%%%%%%%%%%%

% Don't change these lines
\bsp	% typesetting comment
\label{lastpage}
\end{document}